\documentclass[12pt,a4paper]{article}

\usepackage[left=2cm,top=2cm,right=2 cm,bottom=2.5cm]{geometry}

\usepackage{amsfonts}
\usepackage{amssymb}
\usepackage{graphicx}

\usepackage{color}

\begin{document}

\centerline{\bf\Large Non-minimally coupled curvaton}

\bigskip

\centerline{\bf\large
 Lei-Hua Liu$^{a}$~\footnote{liuleihua8899@hotmail.com}
and Tomislav Prokopec$^{b}$~\footnote{{t.prokopec@uu.nl}}}
%\email{liuleihua8899@hotmail.com}
%\author{Tomislav Prokopec$^{2}$}
%\email{t.prokopec@uu.nl}
\bigskip\bigskip

\centerline{$^a$ Department of Physics, College of Physics,  Mechanical and Electrical Engineering},
\centerline{ Jishou University, Jishou 416000, China}\\
%\affiliation{$^1$ Department of Physics, College of Physics,  Mechanical and Electrical Engineering, Jishou University, Jishou 416000, China}

%\medskip
\centerline{$^b$Institute for Theoretical Physics, Spinoza Institute}
\centerline{and the Center for Extreme Matter and Emergent Phenomena (EMME$\Phi$), }
\centerline{Utrecht University, Buys Ballot Building, Princetonplein 5,}
\centerline{ 3584 CC Utrecht, the Netherlands}
%\affiliation{$^2$Institute for Theoretical Physics, Spinoza Institute and the Center for Extreme Matter and Emergent Phenomena (EMME$\Phi$), Utrecht University, Buys Ballot Building, Princetonplein 5, 3584 CC Utrecht, the Netherlands}

\bigskip\bigskip\bigskip\bigskip

%\begin{abstract}

\centerline{\bf\large Abstract}
\bigskip

We investigate two-field inflationary models in which scalar cosmological pertubations are generated {\it via} 
a {\it spectator field nonminimally
coupled to gravity}, with the particular emphasis on 
{\it curvaton} scenarios. 
The principal advantage of these models is in the possibility to tune
the spectator spectral index {\it via} the nonminimal coupling.
Our models naturally yield red spectrum of the adiabatic perturbation demanded by observations.
We study how the nonminimal coupling affects the spectrum of the curvature perturbation
generated in the curvaton scenarios. In particular we find that for small, negative nonminimal couplings 
the spectral index gets a contribution that is negative and linear in the nonminimal coupling.
Since  in this way the curvature spectrum becomes redder, some of curvaton scenarios can 
be saved, which would otherwise be ruled out.
In the power law inflation we find that a large nonminimal coupling is excluded since 
it gives the principal slow-roll parameter that is of the order of unity. 
Finally, we point out that nonminimal coupling can affect
the postinflationary growth of the spectator perturbation, 
and in this way the effectiveness of the curvaton mechanism.

%\end{abstract}

%\maketitle

\bigskip\bigskip\bigskip\bigskip\bigskip

\eject

%%%%%%%%%%%%%%%%%%%%%%%%%%%%%%%%%%%%%%%%%%%%%%%%%%%%%
%%%%%%%%%%%%%%%%  I N T R O D U C T I O N  %%%%%%%%%%%%%%%%%%%%%%%%%
%%%%%%%%%%%%%%%%%%%%%%%%%%%%%%%%%%%%%%%%%%%%%%%%%%%%%

\section{Introduction}	

In most of inflationary models the inflaton (which drives inflation) is the origin of the curvature perturbation
that sources the principal part of the CMB temperature fluctuations.
However, viable alternatives exist
in which the curvature perturbation is predominantly generated by another scalar field,
whose energy density is subdominant during inflation. 
These models are known 
as multifield inflationary models
with spectator fields, an important 
class of which was dubbed
curvaton scenarios~\cite{Enqvist:2001zp, Moroi:2001ct, Lyth:2001nq}.

In curvaton scenarios, the standard relation for the curvaton spectral index, $n_{\chi}=-2\epsilon+ 2m_{\chi}^2/(3H^2)$, where
 $\epsilon=-\dot H/H^2$ is the principal slow-roll parameter, 
$m_{\chi}$ the curvaton mass and
 $H$ the Hubble parameter, yields {\it via} its post-inflationary decay a curvature perturbation with a spectral index
given by, $n_s= 1+n_{\chi}$. 
Here we consider a simple modified spectator field 
model in which
the spectator condensate 
$\bar\chi=\langle\hat\chi\rangle$ couples 
nonminimally to gravity
and observe that the spectator spectral index $n_\chi$ 
acquires an additional contribution from its nonminimal coupling 
$\xi$ of the form, 
$\delta n_{\chi}\sim \xi$,
where $\xi$ is the nonminimal coupling.  
Since that contribution is $\propto \xi$, it can be used to tune the spectral index of the spectator field
-- and thus also {\it via} the curvaton mechanism 
that of the curvature perturbation
--
which be of the crucial 
importance for viability of the curvaton model. 
This simple observation is the principal result of this work.

The nonminimal coupling is not only important during 
inflation~\cite{Linde:1996gt,Takahashi:2020car,Feng:2013pba},
but it can also play an important role
for the post-inflationary curvaton decay, which we investigate as well.
In most of curvaton scenarios the curvaton decays predominantly perturbatively~\cite{Bartolo:2002vf} significantly
after the end of inflation. Roughly speaking the decay occurs when the curvaton decay rate
$\Gamma_\chi$ becomes comparable to the expansion rate of the Universe $H(t)$, {\it i.e.} when $\Gamma_\chi\sim H$.
When the assumption that the curvaton condensate dominates over its perturbations is relaxed,
the decay process can produce large local non-Gaussianities~\cite{Lyth:2002my}.
Current observations~\cite{Ade:2015ava} severely constrain these models however, as
 (local) non-Gaussianity $f_{NL}$ cannot be too large ($|f_{NL}|< 10$),
thereby ruling out curvaton models that produce large non-Gaussianities. 

 There are situations where the inflaton does not decay perturbatively, but instead non-perturbative decay channels, such
as parametric resonant or tachyonic decay
channels~\cite{Traschen:1990sw,Kofman:1994rk,Shtanov:1994ce,Prokopec:1996rr,Greene:1997ge,Kofman:1997yn,Greene:1997fu}
are more efficient. The possibility that the curvaton may decay non-perturbatively has also been
envisaged~\cite{Liu:2019xhn,Enqvist:2008be}. 
Furthermore, it is known that one can produce a significant amount
of gravitational waves during preheating~\cite{Figueroa:2017vfa}.
If the curvaton lives longer, it can couple to the Higgs field in which case the mass of the curvaton can vary significantly~\cite{Enqvist:2012tc,Enqvist:2013gwf}. In this work we provide a 
preliminary analysis
of post-inflationary dynamics of two fields after and leave  
a more complete account of it for future work.

The paper is organized as follows. In section~\ref{model} we introduce our inflationary model with two scalar fields,
one being the inflaton and the other the spectator nonminimally coupled to gravity. In section~\ref{Power spectrum} we make use of 
the gauge-invariant two-field formalism to 
calculate the spectra of the curvature perturbation and 
entropy perturbation by making use of the general slow-roll analysis.
We pay a particular attention to the role of the nonminimal coupling.
In section~\ref{Post-inflationary dynamics}
and in appendix~B we study how 
nonminimal coupling influences post-inflationary 
dynamics and the corresponding spectra of the 
curvature and entropy perturbations.
In section~\ref{Conclusion} 
we summarize our main results and discuss some 
possible future lines of research.

We work in natural units in which $c=1=\hbar$, but retain the Newton constant $G$.

%%%%%%%%%%%%%%%%%%%%%%%%%%%%%%%%%%%%%%%%%%%%%%%%%%%%%%%%
%%%%%%%%%%%%%%%%%%%    T H E     M O D E L   %%%%%%%%%%%%%%%%%%%%%%%%%%
%%%%%%%%%%%%%%%%%%%%%%%%%%%%%%%%%%%%%%%%%%%%%%%%%%%%%%%%

\section{The model}
\label{model}

In this section we consider an inflationary model consisting of two scalar fields, in which one scalar ($\phi$) is 
the inflaton and the other ($\chi$) is the spectator field nonminimally coupled to gravity.~\footnote{Even though we are mainly interested 
here in a class of two field models of inflation,
one of which is the inflaton and the other the 
non-minimally coupled curvaton field,
the formalism we develop applies to the more general
situations in which the second field is a spectator field.
This more general approach is dictated by the nonminimal 
coupling, as many of the standard formulas developed 
in the context of curvaton scenarios do not apply in 
this more general setting.
%that couples to the inflaton with an arbitrary strenght.
}
The action in Jordan frame (denoted by subscript $J$) is,
\begin{equation}
S_J=\int d^4x\sqrt{-g_J}\left(\frac{1}{2}F(\chi_J)R_J-\frac{1}{2}g^{\mu\nu}_J(\partial_\mu\phi_J)(\partial_\nu\phi_J)-\frac{1}{2}g^{\mu\nu}_J(\partial_\mu\chi_J)(\partial_\nu\chi_J)-V_J(\phi_J,\chi_J)\right)
\,,
\label{curvaton model in Jordan frame}
\end{equation}
where in this work $F(\chi_J)$ and potential $V(\phi_J,\chi_J)$ are given by, 
 \begin{eqnarray}
 F(\chi_J)&=&M_{\rm P}^2-\xi\chi_J^2,
\label{nonminimal coupling function}\\
 V_J(\phi_J,\chi_J) &=&  V_J(\phi_J)+ V_J(\chi_J)
\,,\qquad 
 V_J(\chi_J)=\frac{1}{2}m_{\chi}^2\chi_J^2 +\frac{\lambda_\chi}{4!}\chi_J^4
\,,
\label{potential}
\end{eqnarray}
where $M_{\rm P}=1/\sqrt{8\pi G} \simeq 2.45\times 10^{18}~{\rm GeV}$ is the reduced Planck mass,
$V_J(\phi_J)$ is the inflaton potential,
$m_\chi$ is the spectator mass, $\xi$ is its nonminimal coupling
and, unless stated otherwise, 
we take the spectator self-coupling $\lambda_\chi=0$.
Next, for simplicity we assume no direct coupling between the inflaton 
and the spectator. 
That significantly simplifies our analysis but -- unless the mutual coupling is quite strong -- 
in no essential way affects the main results of this work.
Furthermore, in this work we work with a simple potential for the inflaton, 
\begin{equation}
V_J(\phi_J)= V_0 \exp\left(-\lambda \frac{\phi_J}{M_{\rm P}}\right)
%\frac12m_\phi^2\phi_J^2+\frac{\lambda_\phi}{4!}\phi_J^4
\,,
\label{quartic selfinteraction inflaton}
\end{equation}
even though the precise form of the potential is not important for the purposes of this paper.
The exponential potential in~(\ref{quartic selfinteraction inflaton}) is particularly useful since 
 the single field inflationary model  in its attractor mode
leads to particularly simple slow-roll parameters, $\epsilon\equiv \epsilon_\phi=\lambda^2/2$,
and all other slow-roll parameters are exactly zero (in the attractor mode of the theory), $\epsilon_i=0$ ($i=2,3,\cdots$).  
Of course, it is important to study other types of 
inflaton potentials and its interactions with 
other matter fields, and we leave that for future work.
Namely, our main interest here is to study the effects of the nonminimal coupling 
of the spectator field $\chi$, and therefore in this work we shall not complicate that by including 
more complex interactions and further couplings to gravity such as the inflaton nonminimal
coupling or the kinetic coupling to the Einstein tensor.

\begin{figure}[h!]
 \centering
  \includegraphics[width=0.99\textwidth]{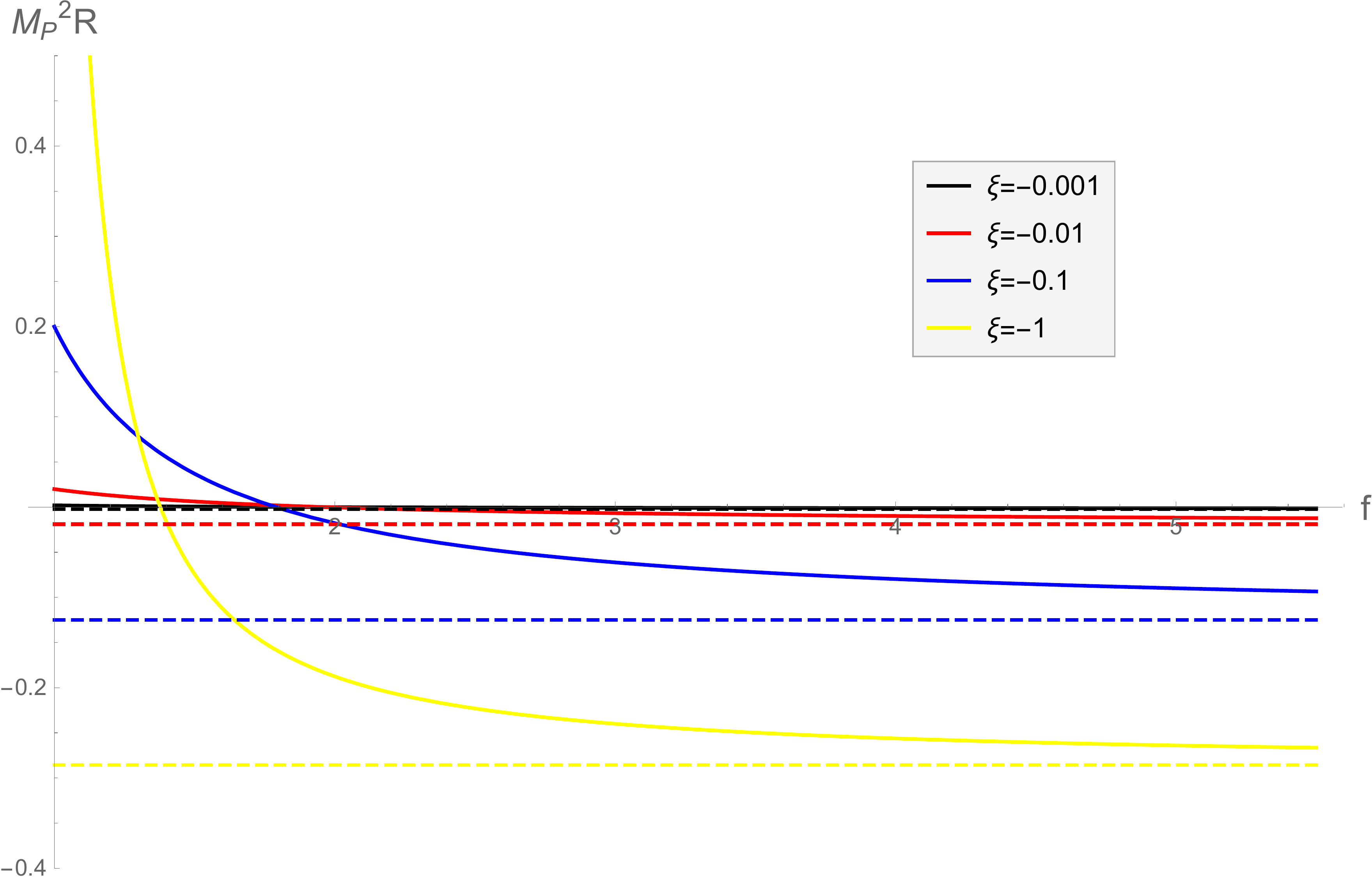}
\vskip -0.4cm 
 \caption{The configuration space scalar Ricci curvature~(\ref{configuration space curvature})
 $M_{\rm P}^2{\cal R}$ as a function of $f=1\!-\!\xi\chi^2/M_{\rm P}^2$ 
 for negative nonminimal couplings: $\xi=-1$ (bright yelow curve asymptoting $\approx -0.28$),
  $\xi=-0.1$ (light blue curve asymptoting $\approx -0.12$),
   $\xi=-0.01$ (solid red curve asymptoting $\approx -0.02$)
and  $\xi=-0.001$ (solid black curve asymptoting almost zero).
When the coupling is very large and negative,  $M_{\rm P}^2{\cal R}$
asymptotes $-1/3$, which is a hyperbolic space ${\mathbf H}^2$ of constant curvature.
}
 \label{figure: Ricci curvature}
 \end{figure}

 It turns out that a particularly useful frame is the one in which gravity is transformed into Einstein frame, 
while the inflaton and curvaton are kept in Jordan frame,
 \begin{eqnarray}
 g_{\mu\nu}&=&\frac{F(\chi_J)}{M_{\rm P}^2}g_{\mu\nu}^J,
 \label{metric transformaiton}\\
 \phi &=&\phi_J,
 \label{inflaton transformation}\\
 \chi&=&\chi_J\,, \quad 
 f(\chi) = \frac{F}{M_{\rm P}^2} = 1 - \xi\frac{\chi^2}{M_{\rm P}^2}
   \label{curvaton transformtion}
\,.
\label{conformal transformation}
\end{eqnarray}
After the above transformations are exacted,
 the action~(\ref{curvaton model in Jordan frame}) 
 becomes,
\begin{eqnarray}
S[g_{\alpha\beta},\chi,\phi]
&=&\int d^4x\sqrt{-g}\Biggl\{\frac{M_{\rm P}^2}{2}R
-\frac{1}{2f(\chi)}g^{\mu\nu}(\partial_\mu\phi)(\partial_\nu\phi)
\nonumber\\
&& \hskip 2cm
 -\frac{1}{2f}\left[1+\frac{3M_{\rm P}^2}2\frac{f'^2}{f}\right]g^{\mu\nu}(\partial_\mu\chi)(\partial_\nu\chi)
     -V(\phi,\chi)\Biggr\}
\,,\quad
\label{curvaton model in Einstein frame}
\end{eqnarray}
where the transformed potential equals,
\begin{equation}
V(\phi,\chi)=\frac{V_J(\phi,\chi) }{f^2(\chi)}
 \equiv \frac{V_\phi(\phi)}{f^2} +\frac{V_\chi(\chi)}{f^2}
 \,,\qquad
 \frac{V_\chi(\chi)}{f^2}=\frac{M_{\rm P}^2m_\chi^2}{-2\xi}\left(\frac1f-\frac1{f^2}\right)
   +\frac{\lambda_\chi M_{\rm P}^4}{24\xi^2}\left(1-\frac{1}{f}\right)^2
\,.
\label{potential in Einstein frame}
\end{equation}
Note that in the limit of the minimal coupling, $\xi\rightarrow 0$, we have $f\rightarrow 1$ 
and the two fields in~(\ref{curvaton model in Einstein frame}--\ref{potential in Einstein frame})
decouple, implying that  
this two-field inflationary model reduces to a single field inflation driven by the inflaton $\phi$
which can be treated within the standard slow-roll inflationary framework.
 For small curvaton condensates the curvaton part of the potential
 in~(\ref{potential in Einstein frame}) can be expanded as, 
\begin{equation}
\frac{V_\chi(\chi)}{f^2}\simeq\frac{1}{2}m_{\chi}^2\chi^2\left[1+2\xi\frac{\chi^2}{M_{\rm P}^2} + {\cal O}(\chi^4)\right]
\qquad \big(|\xi\chi^2|\ll M_{\rm P}^2\big)
\,.
\label{potential in Einstein frame curvaton1}
\end{equation}
In order to facilitate the analysis, it is convenient to introduce 
the covariant multifield 
formalism~\cite{Gong:2011uw,Kaiser:2012ak,Elliston:2012ab}, in 
which~(\ref{curvaton model in Einstein frame}) can be recast as,

\begin{eqnarray}
S[g_{\alpha\beta},\chi,\phi]
&=&\int d^4x\sqrt{-g}\Biggl\{\frac{M_{\rm P}^2}{2}R
-\frac{1}{2}{\cal G}_{AB}g^{\mu\nu}(\partial_\mu\phi^A)(\partial_\nu\phi^B) - V(\phi^A)\Biggr\}
\,,\quad
\label{curvaton model: covariant formalism}
\end{eqnarray}
where ${\cal G}_{AB}$ is the 
{\it configuration (field) space metric}
in~(\ref{curvaton model in Einstein frame})
which, in the field space coordinates $\phi^A=(\phi,f)$, reads
\begin{equation}
{\cal G}_{AB} 
= {\rm diag}\left(\frac1{f},
 \frac{M_{\rm P}^2}{-4\xi}\frac{6\xi+(1\!-\!6\xi)f}{f^2(f\!-\!1)}\right)
 \,.
 \label{configuration space metric}
\end{equation}
Note that the field space metric is diagonal. Since the corresponding 
configuration space curvature tensor does not vanish,
\begin{equation}
{\cal R}_{ABCD}=\frac{{\cal R}}{2}
 \left({\cal G}_{AC}{\cal G}_{BD}-{\cal G}_{AD}{\cal G}_{BC}\right)
\,,\qquad
{\cal R} = \frac{-2\xi}{M_{\rm P}^2}
                \frac{6\xi\!+\!2(1\!-\!6\xi)f\!-\!(1\!-\!6\xi)f^2}
                        {[6\xi\!+\!(1\!-\!6\xi)f]^2}
\,,
\label{configuration space curvature}
\end{equation}
 the kinetic terms in the 
 action~(\ref{curvaton model: covariant formalism}) cannot be brought
 into the canonical form.   It is in this sense that the Einstein frame for 
 the fields does not exist.
 The dependence of ${\cal R}$ on $\xi$ and the field $f$ is illustrated in figure~\ref{figure: Ricci curvature},
 from which we see that, in the limit of large and negative $\xi$, the configuration space curvature 
 asymptotes to a negative constant $-1/(3M_{\rm P}^2)$,
and thus belongs to the class of models 
with a negative configuration space curvature. 
These models have gained in popularity, and notable 
examples are the super-gravity inspired
$\alpha$-attractors~\cite{Kallosh:2013yoa,Galante:2014ifa,Carrasco:2015uma} 
and the Weyl symmetric 
models~\cite{Kallosh:2013hoa,Barnaveli:2018dxo}.

In the following section we discuss how to calculate the curvature power spectrum under the assumption that the inflaton contribution dominates.
While the small field expansion~(\ref{potential in Einstein frame curvaton1}) often suffices 
for rough estimates, it is in general not enough to provide accurate answers for the curvaton dynamics 
and the respective spectrum of its quantum fluctuations
during inflation. For this reason, in what follows we present the analysis by using the full 
potential~(\ref{potential in Einstein frame}).

%%%%%%%%%%%%%%%%%%%%%%%%%%%%%%%%%%%%%%%%%%%%%%%%%%%
%%%%%%%%%%%%%%%%    P O W E R    S P E C T R U M   %%%%%%%%%%%%%%%%%%%%
%%%%%%%%%%%%%%%%%%%%%%%%%%%%%%%%%%%%%%%%%%%%%%%%%%%

\section{Power spectrum}
\label{Power spectrum}

We work in spatially flat cosmological space-times whose background metric is of the form, 
%(and adopt a metric signature, ${\rm sign}(-,+,+,+)$),
\begin{equation}
g_{\mu\nu}(x)={\rm diag} \left(-N^2(t),a^2(t),a^2(t),a^2(t)\right)
%\,,\qquad \eta_{\mu\nu}={\rm diag}(-,+,+,+)
\,,
\label{FRLW metric}
\end{equation}
where $a=a(t)$ denotes the scale factor and $N=N(t)$ is the lapse function. The expansion of the Universe is driven by field condensates, $\chi(t)=\langle\hat \chi\rangle$ and 
 $\phi(t)=\langle\hat \phi\rangle$, and it is governed by Friedmann equations, 
\begin{eqnarray}
H^2(t) &=& \frac{1}{3M_{\rm P}^2}\rho
\,,\quad \rho= \frac12{\cal G}_{AB}\dot\phi^A\dot\phi^B+V(\phi^A)
\label{Friedmann 1}
\\
\dot H &=& 
-\frac{1}{2M_{\rm P}^2}\left(\rho+{\cal P}\right)
\,,\quad
\rho+{\cal P}={\cal G}_{AB}\dot\phi^A\dot\phi^B
\,,
\label{Friedmann 2}
\end{eqnarray}
where $H=d\ln(a)/(Ndt)$ is the Hubble rate and 
$\dot H=dH/(Ndt)$ and $\dot\phi^A=d\phi^A/(Ndt)$ denotes 
the time reparametrization invariant time derivative and
$\phi^A=(\phi(t),f(t))$ are the background fields,
which obey~\cite{Kaiser:2012ak}~\footnote{One can easily show 
that Eq.~(\ref{background field equations}) is not independent,
as it can be  derived from
covariant conservation of the background stress-energy tensor, 
$\dot\rho + 3H(\rho+{\cal P})=0$ 
and the Friedmann equation~(\ref{Friedmann 1}).}
\begin{equation}
 D_t^2\phi^A + 3HD_t \phi^A + \partial^AV(\phi^B) = 0
 \label{background field equations}
 \end{equation}
 where $D_t = \dot \phi^B\nabla_B$ is the covariant derivative on the field 
 space. Since the background fields are coordinates on the configuration space manifold, the covariant derivative $D_t$ 
 acts simply on the background fields, 
 \begin{equation}
 D_t \phi^A = \dot \phi^A\,,\quad 
 D_t^2\phi^A = \ddot \phi^A 
         +\Gamma^A_{BC}\dot \phi^B\dot \phi^C
         \,,
\label{background field covariant derivatives}
\end{equation}
where $\Gamma^A_{BC}$ are the Christoffel symbols of the field space. We therefore see that the background fields 
obey a geodesic equation~(\ref{background field equations})
in presence of a time dependent (Hubble) friction $\gamma=3H$ and an external force $F^A=-\partial^AV$.~\footnote{If one understands
$\dot H=dH/[N(t)dt]$, $H(t)=d\ln(a)/[N(t)dt]$ and 
$D_t \phi^A$ as $\dot \phi^A=d\phi^A/[N(t)dt]$, 
the background equations~(\ref{Friedmann 1}--\ref{Friedmann 2})
and~(\ref{background field equations}) become 
time reparametrization invariant, and thus can be easily converted 
to any other definition of time, {\it e.g.} conformal time for which 
$N(t)=a(\tau)$ and $dt=d\tau$. }

In order to obtain the curvature power spectrum, one ought to solve the operator 
equation of motion for the curvaton and inflaton, which can be 
obtained by varying the 
action~(\ref{curvaton model in Einstein frame}).
When gravitational constraints are solved
the linearized equations for field perturbations $Q^A(x)$
in the zero curvature gauge~\footnote{Since this is 
a fully fixed gauge,
the equations~(\ref{EOM 1}) are identical to the equations of motion for
the corresponding gauge invariant variables in the zero curvature gauge, 
in which $\hat Q^A(x)$ reduces to the field perturbations 
$\delta\hat \phi^A(x)$. 
For a more detailed discussion of this issue see, for example,
 Refs.~\cite{Prokopec:2012ug,Prokopec:2013zya}.
}
 are~\cite{Kaiser:2012ak}, 
\begin{eqnarray}
&&\left[D_t^2\!+\!3HD_t\!-\!\frac{\nabla^2}{a^2}\right]\hat Q^A(x)
 \!+\! \nabla_B\nabla^A V(\phi^I)\hat Q^B
 \!-\! {\cal R}^A_{\;IJB}\dot\phi^I\dot\phi^J\hat Q^B
\label{EOM 1}\\
&& \hskip 7.8cm
 -\frac{1}{M_{\rm P}^2a^3}D_t\left(\frac{a^3}{H}\dot \phi^A\dot \phi_B\right)\hat Q^B \!+\! {\cal O}\Big(\hat Q^C\hat Q^E\Big) = 0
\,,\quad
\nonumber
\end{eqnarray}
where 
$\nabla^2\equiv \sum_{i=1}^3\partial_i^2$,
${\cal R}^A_{\;IJB}={\cal G}^{AC}{\cal R}_{CIJB}$,
the Riemann tensor 
${\cal R}_{CIJB}$ is given in~(\ref{configuration space curvature}) 
and 
\begin{eqnarray}
 D_t \hat Q^A &=& \dot{\hat Q}^A
                     +\Gamma^A_{\;BC}\dot \phi^B\hat Q^C
\label{derivative of QA}
\\
D_t^2 \hat Q^A &=& \ddot{\hat Q}^A 
 + \left(\partial_D\Gamma^A_{\;BC}+\Gamma^A_{\;DE}\Gamma^E_{BC}\right)\dot \phi^D\dot \phi^B\hat Q^C
 + \Gamma^A_{\;BC}\ddot \phi^B\hat Q^C 
 + 2\Gamma^A_{\;BC}\dot \phi^B\dot{\hat Q}^C
 \,,
\label{2 derivatives of QA}
\end{eqnarray}
such that 
\begin{eqnarray}
\left[D_t^2+3HD_t\right]\hat Q^A
   -{\cal R}^A_{\;ABCD}\dot\phi^B\dot\phi^C\hat Q^D
    &=& \ddot{\hat Q}^A +3H\dot{\hat Q}^A
      + 2\Gamma^A_{\;BC}\dot \phi^B\dot{\hat Q}^C
\nonumber\\
     &-&\Gamma^A_{\;BC}(\partial^BV)\hat Q^C
       + (\partial_D\Gamma^A_{\;BC})\dot \phi^B\dot \phi^C\hat Q^D
\,, 
\label{part of perturbations}
\end{eqnarray}
where we made use of Eq.~(\ref{background field equations}) and of,
\begin{equation}
{\cal R}^A_{\;BCD}=\partial_C\Gamma^A_{\;BD}
  + \Gamma^A_{\;CE}\Gamma^E_{\;DB}
  - \partial_D\Gamma^A_{\;BC}
   - \Gamma^A_{\;DE}\Gamma^E_{\;CB}
   \,.
\label{Riemann tensor}
\end{equation}
The nonvanishing Christoffel symbols are, 
\begin{eqnarray}
\Gamma^\phi_{\phi f} &=& \Gamma^\phi_{f\phi}=-\frac{1}{2f}
\nonumber\\
\Gamma^f_{\phi\phi} &=& \frac{-2\xi}{M_{\rm P}^2}
          \frac{f\!-\!1}{6\xi\!+\!(1\!-\!6\xi)f}
\nonumber\\
\Gamma^f_{ff} &=& \frac{1}{2}\left[
          \frac{1\!-\!6\xi}{6\xi\!+\!(1\!-\!6\xi)f}
           -\frac{2}{f}-\frac{1}{f\!-\!1}
          \right]
\,.
\label{Christoffel symbols}
\end{eqnarray}

\bigskip

The next natural step is canonical quantisation, according to which 
the fields $\phi^A$ and their canonical momenta 
\begin{equation}
\pi_A(x)=\frac{\delta S}{\delta \partial_0\phi^A(x)}
    = a^3{\cal G}_{AB}\frac{\partial_0\phi^B}{N(t)}
\label{canonical momenta AB}
\end{equation}
satisfy canonical commutation relations, 
\begin{equation}
 [\hat\phi^A(t,\vec{x}),\hat\pi_{B}(\tau,\vec{x}')]
% [\delta\hat\phi^A(t,\vec{x}),\delta\hat\pi_{B}(\tau,\vec{x}')]
   = i \hbar \delta^A_{\;B}\delta^{3}(\vec{x}\!-\!\vec{x}')
\,,
\label{canonical quantisation AB}
\end{equation}
while the fields and their canonical momenta mutually commute. 
Perturbations around the field condensates satisfy identical
commutation relations as in~(\ref{canonical quantisation AB}). 
This can be seen by expanding 
the action~(\ref{curvaton model in Einstein frame}) 
to the quadratic order in perturbations, including the effect of coupling to 
the gravitational perturbations.
The resulting action can be found {\it e.g.} in Ref.~\cite{Gong:2016qmq}
(see Eq.~(4.19)) 
(for a discussion of more general multifield Lagrangians see Ref.~\cite{Langlois:2008mn}), 
from where it is clear that 
coupling to gravity does not change the structure of 
the canonical kinetic term, such that the canonical quantization
relation~(\ref{canonical quantisation AB})  holds also for the perturbations,
\begin{equation}
% [\hat\phi^A(t,\vec{x}),\hat\pi_{B}(\tau,\vec{x}')]
 [\hat Q^A(t,\vec{x}),\hat\Pi_{B}(\tau,\vec{x}')]
   = i \hbar \delta^A_{\;B}\delta^{3}(\vec{x}\!-\!\vec{x}')
\,,
\label{canonical quantisation AB 2}
\end{equation}
where $\hat\Pi_{B}(x) = a^3 {\cal G}_{AB}(\partial_0 Q^B(x))/N(t)$.

Since the procedure for studying
the dynamics of quantized linear curvature perturbations
 is standard~\cite{Sasaki:1995aw},
here we outline just its main steps.
The quantum fields that exhibit kinetic and potential mixing
(which are both evident from~(\ref{curvaton model in Einstein frame}--\ref{potential in Einstein frame}) 
and~(\ref{EOM 1}))
can be decomposed into spatial momentum modes as, 
\begin{equation}
\hat Q^A(x) = \int\frac{d^3k}{(2\pi)^3}\sum_{\alpha=1,2} 
\left(e^{i\vec k\cdot\vec x}q^A_{\;\alpha}(t,k)\hat a_\alpha(\vec k)
+e^{-i\vec k\cdot\vec x}[q^A_{\;\alpha}(t,k)]^*\hat a_\alpha^\dagger(\vec k)
\right) 
\,,
\label{Fourier decomposition}
\end{equation}
where $q^A_{\;\alpha}(t,k)$ are matrix valued mode functions, 
 $\hat a^\dagger_\alpha(\vec k\,)$ and  $\hat a_\alpha(\vec k\,)$ $(\alpha=1,2)$ are the creation 
and annihilation operators in the instantaneously diagonal basis
denoted by the intex $\alpha$, which obey, 
\begin{equation}
[\hat a_\alpha(\vec k\,),\hat a^\dagger_{\alpha'}(\vec k'\,)]=(2\pi)^3\delta_{\alpha\alpha'}\delta^3(\vec k-\vec k')
\,,\quad [\hat a_\alpha(\vec k\,),\hat a_{\alpha'}(\vec k'\,)]=0
\,,\quad [\hat a^\dagger_\alpha(\vec k\,),\hat a^\dagger_{\alpha'}(\vec k'\,)]=0
\,.
\label{canonical commutation relations: momentum space}
\end{equation}
The spectra of different field components can be then defined as, 
\begin{equation}
 P_A(t,k) = \frac{k^3}{2\pi^2}\sum_{\alpha=1,2}|q^A_\alpha(t,k)|^2
 \,,\qquad
  P_{AB}(t,k) = \frac{k^3}{2\pi^2}\sum_{\alpha=1,2}
    q^A_\alpha(t,k){q^B}^*_\alpha(t,k)
\label{field spectra}
\end{equation}
where the normalization of the modes $q^A_\alpha(t,k)$ can be determined from the Wronskian,
\begin{equation}
\sum_\alpha \Big[q^A_{\,\alpha}(t,k)\pi^\alpha_{\,B}(t,k)^* 
                 - q^A_{\,\alpha}(t,k)^*\pi^\alpha_{\,B}(t,k)\Big]
                             =\imath\delta^A_{\;B}
\,,
\label{Wronskian matrix}
\end{equation}
where $\pi^\alpha_B(t,k)
=a^3{\cal G}_{AB}(\partial_0q_\alpha^B(t,k))/N(t)$ 
%are the mode function and 
are the canonical momenta associated with the mode functions 
$q_\alpha^B(t,k)$. 

These quantities are, however, not directly observable. In order to reach observable spectra,
it is convenient to define the 
{\it curvature and entropy} directions in the field space as follows,
\begin{equation}
 \|\dot \phi^A\| \equiv \dot \sigma = \sqrt{{\cal G}_{AB}\dot\phi^A\dot\phi^B}
 \,,\qquad \hat \sigma^A \equiv \frac{\dot\phi^A}{\dot\sigma}
 \,,
 \label{adiabatic and entropic}
\end{equation}
such that the norm of the entropy vector is unity, $\|\hat \sigma^A\|=1$. In terms of these 
quantities the background Friedmann equations~(\ref{Friedmann 1}--\ref{Friedmann 2}) simplify to,
\begin{eqnarray}
H^2(t) &=& \frac{1}{3M_{\rm P}^2}\left(\frac{\dot\sigma^2}{2}+V(\phi^A)\right)
\label{Friedmann 1B}
\\
\dot H &=& 
-\frac{\dot\sigma^2}{2M_{\rm P}^2}
\,.
\label{Friedmann 2B}
\end{eqnarray}
and the background field equation~(\ref{background field equations})
 for the adiabatic mode $\sigma$ becomes identical to that of one field inflation,
\begin{equation}
\ddot \sigma + 3H\dot \sigma +\partial_\sigma V = 0
\,,
\label{field equation for sigma}
\end{equation}
where $\partial_\sigma V =\hat\sigma^A\partial_A V(\phi^B)$. Note that, just as in the one field case, 
Eq.~(\ref{field equation for sigma}) can be derived from 
Eqs.~(\ref{Friedmann 1B}--\ref{Friedmann 2B}) by taking a time derivative of~(\ref{Friedmann 1B}).
Eqs.~(\ref{Friedmann 2B}) suggest the following definition of the principal geometric slow-roll parameter,
\begin{equation}
\epsilon = -\frac{\dot H}{H^2} = \frac{\dot \sigma^2}{2M_{\rm P}^2H^2} 
\,.
\label{epsilon}
\end{equation}
By projecting Eq.~(\ref{EOM 1}) onto $\sigma_I$, one can then show that the equation of motion 
for $Q_\sigma=\hat\sigma_A Q^A$ reads, 
\begin{equation}
\ddot Q_\sigma + 3H\dot Q_\sigma + \left[-\frac{\nabla^2}{a^2}+{\cal M}^2_{\sigma\sigma} - \omega^2
   - \frac{1}{M_{\rm P}^2a^3}\frac{d}{dt}\left(\frac{a^3\dot\sigma^2}{H}\right)\right]Q_\sigma
     = 2\left[\frac{\partial}{\partial t}-\frac{\partial_\sigma V}{\dot\sigma}+\epsilon H\right]\left(\omega Q_s\right)
\,,
\label{evolution adiabatic}
\end{equation}
where 
\begin{equation}
 {\cal M}^2_{\sigma\sigma} = \hat \sigma^A\hat \sigma^B\nabla_A\nabla_B V(\phi^C)
\label{adiabatic mass Msigmasigma}
\end{equation}
is the mass term of the adiabatic perturbation and 
\begin{eqnarray}
 \omega^A= {\cal D}_t \hat\sigma^A 
  = \frac{\hat\sigma^A\hat\sigma^B\partial_B V-\partial^A V}{\dot\sigma}
\nonumber\\
\omega = \|\omega^A\|=\sqrt{{\cal G}_{AB}\omega^A\omega^B}
   = \frac{\sqrt{(\partial^A V)(\partial_A V)-(\hat\sigma^B\partial_B V)^2}}{\dot\sigma}
\label{turning rate}
\end{eqnarray}
defines the {\it turning rate}, which is by definition 
orthogonal to $\hat \sigma^A$, 
$\omega_A\hat\sigma^A=0$, and can be used to define convenient orthonormal basis vectors for the perturbations. 
Indeed, we can define a unit turning vector as, 
\begin{equation}
 \hat \omega^A = \frac{\omega^A}{\omega} 
  = \frac{\hat\sigma^A\hat\sigma^B\partial_B V-\partial^A V}{\sqrt{(\partial^A V)(\partial_A V)-(\hat\sigma^B\partial_B V)^2}}
 \,,\quad  \|\hat \omega^A\|=1  \quad\&\quad \hat\sigma^A \hat\omega_A=0
 \,,\quad
\label{unit turning vector}
\end{equation}
which can be used to project out the entropy perturbation,
\begin{equation}
 Q_s =  \hat \omega_B Q^B 
 \,,
\label{entropy perturbation}
\end{equation}
Note that since  $\hat \omega_A\hat \sigma^A=0$ are orthogonal, 
the adiabatic and entropy (or isocurvature) perturbations, $\{Q_\sigma,Q_s\}$,
denote the two orthogonal perturbations
(since we have only 
two fields, this completes the procedure of diagonalization of the perturbations).

 Since the projection vectors $\hat\sigma^A$ and 
$\hat\omega ^A$ are orthonormal, from~(\ref{canonical quantisation AB 2})
one can infer that the nonvanishing canonical commutation
relations for $\hat Q_\sigma$ and $\hat Q_s$ are,
\begin{equation}
 [\hat Q_\sigma(t,\vec{x}),\hat\Pi_{\sigma}(\tau,\vec{x}^{\,\prime})]
   = i \hbar \delta^{3}(\vec{x}\!-\!\vec{x}^{\,\prime})
\,,\qquad
 [\hat Q_s(t,\vec{x}),\hat\Pi_{s}(\tau,\vec{x}^{\,\prime})]
   = i \hbar \delta^{3}(\vec{x}\!-\!\vec{x}^{\,\prime})
\,,
\label{canonical quantisation sigma s}
\end{equation}
where $\hat \Pi_\sigma = \hat\sigma^A\hat \Pi_A
 = a^3\left(\dot{\hat Q}_\sigma-\omega \hat Q_s\right)$ and 
$\hat \Pi_s = \hat\omega^A\hat \Pi_A
  = a^3\left(\dot{\hat Q}_s+\omega \hat Q_\sigma\right)$.
Notice also that, as a consequence of orthogonality, 
$\hat\sigma^A\hat\omega_A=0$, 
the remaning commutation relations vanish, 
{\it e.g.} 
$ [\hat Q_\sigma(t,\vec{x}),\hat\Pi_{s}(\tau,\vec{x}^{\,\prime})]
   = 0
 = [\hat Q_s(t,\vec{x}),\hat\Pi_{\sigma}(\tau,\vec{x}^{\,\prime})]$, 
 which is true in spite of 
 the mixing of perturbations in~(\ref{evolution adiabatic}).

The effective mass in the evolution equation~(\ref{evolution adiabatic}) 
for the adiabatic perturbations $Q_\sigma$ 
does not depend on the configuration space curvature (which drops out due to a Bianchi identity),
but it acquires a negative contribution from the turning rate $\omega^2$. 
Note also that the source on the right hand side of~(\ref{evolution adiabatic}) is entirely 
due to the entropy perturbation $Q_s$. 

The entropy perturbation obeys,~\footnote{Since we are considering here only the two fields case, 
our equation for the entropy perturbation~(\ref{evolution entropy}) is simpler than the more general one 
presented in~\cite{Kaiser:2012ak}, which holds for general multifield case.} 
\begin{equation}
\ddot Q_s + 3H\dot Q_s + \left[-\frac{\nabla^2}{a^2}+{\cal M}^2_{ss} + 3\omega^2 
            - \frac12\dot\sigma^2{\cal R}
  \right]Q_s
     = -4M_{\rm P}^2\frac{\omega}{\dot\sigma}\frac{\nabla^2}{a^2}\Psi
\,,
\label{evolution entropy}
\end{equation}
where the mass term and the curvature contributions read,
\begin{equation}
 {\cal M}^2_{ss} = \hat \omega^A\hat \omega^B\nabla_A\nabla_B V (\phi^C)
 \,,\qquad
{\cal R}_{s\sigma\sigma s}=\hat \omega^A\hat \omega^B{\cal R}_{AIJB}\hat \sigma^I\hat\sigma^J 
 =\frac{{\cal R}}{2}
\label{mass of the entropy mode}
\end{equation}
and $\Psi$ is the Bardeen's spatial (gauge invariant) potential. Note that 
-- unlike in the case of the adiabatic perturbation -- the turning rate contributes positively to the mass term of 
the entropy mode (see, however, Eq.~(\ref{evolution entropy:2}) below). 
Furthermore,  while in the adiabatic mode equation 
the configuration space curvature does not contribute, it does contribute to 
the mass of the entropy perturbations as, 
$-\dot\sigma^2 {\cal R}/2 = -\epsilon M_{\rm P}^2H^2{\cal R}$, which is positive (negative) for a negatively 
(positively) curved configuration space manifold.  

The equations for the perturbations~(\ref{evolution adiabatic}) and~(\ref{evolution entropy})
can be closed by making use of the relation between the Bardeen potential and the curvature and entropy perturbations, 
\begin{equation}
 \frac{\nabla^2}{a^2}\Psi =\epsilon H\left[\dot {\cal R}_c + 2\omega {\cal S}\right] 
    =     \frac{\sqrt{2\epsilon} H}{2M_{\rm P}}\left[\dot Q_\sigma - \frac12\epsilon_2HQ_\sigma+2\omega Q_s\right]                                 
 \label{Bardeen potential}
\end{equation}
where, in the last step, we used the following relations between
 the curvature perturbation ${\cal R}_c$ and the entropy perturbation ${\cal S}$ 
 and the variables $Q_\sigma$ and $Q_s$,
\begin{eqnarray} 
{\cal R}_c &=&\frac{H}{\dot \sigma}Q_\sigma 
  =\frac{1}{\sqrt{2\epsilon}M_{\rm P}}Q_\sigma  
\label{relation between curvature perturbation and Qsigma}
\\
{\cal S} &=& \frac{H}{\dot \sigma}Q_s
  =\frac{1}{\sqrt{2\epsilon}M_{\rm P}}Q_s 
\,,
\label{relation between entropy perturbation and Qs} 
\end{eqnarray}
and we have introduced the second geometric slow-roll parameter, 
\begin{equation}
\epsilon_2 = \frac{\dot \epsilon}{\epsilon H}
\,.
\label{epsilon 2}
\end{equation}

Next, it is convenient to introduce the directional curvature and entropy covariant derivatives as, 
\begin{equation}
  D_\sigma \equiv \hat \sigma^A\nabla_A
  \,,\qquad
  D_s \equiv \hat \omega^A\nabla_A
\label{directional derivatives}
\end{equation}
Of course, if $D_\sigma$ and $D_s$ act on a scalar quantity $\phi$ once, they act as ordinary derivatives,
 and we shall denote them to indicate that,
 {\it i.e.} $D_\sigma \phi= \hat \sigma^A\partial_A \phi \equiv \partial_\sigma \phi$
and $D_s \phi= \hat \omega^A\partial_A \phi \equiv \partial_s \phi$.
Armed with these, one can show that 
${\cal M}^2_{\sigma\sigma}$ in~(\ref{adiabatic mass Msigmasigma}) 
can be rewritten as,
\begin{equation}
 {\cal M}^2_{\sigma\sigma} = \hat \sigma^A\nabla_A\hat \sigma^B\nabla_B V(\phi^C)
                                        - \frac{1}{\dot\sigma}(D_t\hat \sigma^B)\nabla_B V(\phi^C)
                 = D_s^2 V -\frac{\omega}{\dot\sigma}\partial_sV
                 = D_s^2 V +\omega^2
                 \,,
\label{adiabatic mass Msigmasigma:2}
\end{equation}
where we made use of~(\ref{turning rate}--\ref{unit turning vector}) and of 
\begin{equation}
 \partial_sV=-\omega\dot\sigma
 \,.
 \label{definition of omega}
 \end{equation}
This equality follows from~(\ref{turning rate}) and can be used 
to determine the sign of $\omega$.

Upon making use of~(\ref{mass of the entropy mode}--\ref{Bardeen potential})
and~(\ref{adiabatic mass Msigmasigma:2})
in~(\ref{evolution adiabatic}) and~(\ref{evolution entropy})
we obtain the following equations,
\begin{eqnarray}
\ddot Q_\sigma \!+\! 3H\dot Q_\sigma \!+\! \left[-\frac{\nabla^2}{a^2}
       \!-\!   \frac12\epsilon_2  \left(3-\!\epsilon\!+\!\frac12\epsilon_2\!+\!\epsilon_3\right) H^2
\right]Q_\sigma
     &=& 2\left[D_t\!+\!\Big(3\!+\!\frac{\epsilon_2}2\Big)H\right]\left(\omega Q_s\right)
\,,\qquad
\label{evolution adiabatic:2}
\\
\ddot Q_s + 3H\dot Q_s + \left[-\frac{\nabla^2}{a^2}
+{\cal M}^2_{ss} - \omega^2 
            - \epsilon H^2M_{\rm P}^2{\cal R}
  \right]Q_s
     &=& -2\omega\left[\dot Q_\sigma -\frac12\epsilon_2 HQ_\sigma\right]
\,,
\label{evolution entropy:2}
\end{eqnarray}
where we made use of,
\begin{equation}
\partial_\sigma V = -3H\dot\sigma-\ddot\sigma 
= -\sqrt{2\epsilon}\left(3-\epsilon+\frac{\epsilon_2}{2}\right)M_{\rm P}H^2 
\,,\quad \dot \sigma = \sqrt{2\epsilon}M_{\rm P}H
\label{partial_sigma V}
\end{equation}
and its directional derivative, $D_\sigma = (1/\dot\sigma)D_t$,~\footnote{Since $\partial_\sigma V$ is a scalar quantity,
we then have $D_\sigma (\partial_\sigma V) = \partial^2_\sigma V$.} 
\begin{equation}
D_\sigma^2V=\partial_\sigma^2V= \left(6\epsilon-\frac32\epsilon_2+\frac52\epsilon\epsilon_2-2\epsilon^2-\frac14\epsilon_2^2
              -\frac12\epsilon_2\epsilon_3\right)H^2
\,,\qquad 
\epsilon_3\equiv \frac{\dot\epsilon_2}{\epsilon_2 H}
\,.
\nonumber
\end{equation}
and we have converted, when possible, to slow-roll parameters.
%In addition, we have defined one new (potential) slow-roll parameter, 
 %
 %\begin{equation}
 %\eta_{ss} = \frac{{\cal M}_{ss}}{H^2}
 %\,.
 %\label{eta sigma sigma eta ss}
 %\end{equation}
 % 
 Notice that the form of the equation for the curvature 
 perturbation~(\ref{evolution adiabatic:2}) is such that 
  the only difference between the corresponding one field equation and~(\ref{evolution adiabatic:2})
  is that, in the multi-field case, the curvature perturbation 
  is sourced 
by the entropy perturbation, whose precise form is shown on the right hand side of~(\ref{evolution adiabatic:2}).
The structure of 
 Eqs.~(\ref{evolution adiabatic:2}--\ref{evolution entropy:2}) reveals that 
 $Q_\sigma$ and $Q_s$ decouple when the turning rate $\omega=0$.
 From Eq.~(\ref{partial_sigma V}) we see that this will be the case if the directional 
 dervative of 
 $V$ along $\hat \omega^A$ vanishes. Alternatively, 
 $\omega$ will vanish if $\hat \sigma^A=\dot\phi^A/\dot\sigma$ is time independent,
 which will be approxuimately the case if 
 $\dot\sigma = \sqrt{{\cal G}_{AB}\dot\phi^A\dot \phi^B}$ 
 is dominated by the inflaton kinetic energy, {\it i.e.} if $\dot\phi^2\gg \dot\chi^2$.
 In practice that will be the case if $\xi=0$ and if the curvaton condensate 
 is small enough, $\chi\simeq 0$.

 \medskip

In view of~(\ref{relation between curvature perturbation and Qsigma}--\ref{relation between entropy perturbation and Qs}),  equations~(\ref{evolution adiabatic:2}--\ref{evolution entropy:2}) can be easily converted into
equations for ${\cal R}_c$ and ${\cal S}$, 
\begin{eqnarray}
\ddot {\cal R}_c \!+\! (3+\epsilon_2)H\dot{\cal R}_c \!-\frac{\nabla^2}{a^2}{\cal R}_c
     &=& 2\left[D_t\!+\!\Big(3\!+\!\epsilon_2\Big)H\right]\left(\omega {\cal S}\right)
\qquad
\label{evolution adiabatic:3}
\\
\ddot {\cal S} \!+\!  (3\!+\!\epsilon_2)H\dot {\cal S}\!+\! \left[-\frac{\nabla^2}{a^2}
         \!+\!{\cal M}^2_{ss}\!-\! \omega^2\!+\!\big(\Delta_\epsilon\!-\!\epsilon M_{\rm P}^2{\cal R}\big) H^2 \right]{\cal S}
     &=& \!-\!2\omega\dot {\cal R}_c
\,,
\label{evolution entropy:3}
\end{eqnarray}
where 
\begin{equation}
\Delta_\epsilon=\frac{\epsilon_2}{2}\left(3-\epsilon+\frac12\epsilon_2+\epsilon_3\right)
\,.
\label{Delta epsilon}
\end{equation}
As one could have expected, the mass term ($\propto {\cal R}_c$) has completely disappeared from the equation for 
the curvature perturbation ${\cal R}_c$, which must be so also in the multifield case.
What is also interesting is that the same operator as it acts on $\dot {\cal R}_c$, acts in the source on $\omega {\cal S}$.
Upon rewriting~(\ref{evolution adiabatic:3}) as, 
\begin{equation}
\left[D_t +(3+\epsilon_2)H\right]\left(\dot{\cal R}_c - 2\omega{\cal S}\right) = \frac{\nabla^2}{a^2}{\cal R}_c
\,,
\label{evolution adiabatic:3b}
\end{equation}
we see that, on super-Hubble scales, on which $\|\nabla^2\|\ll (aH)^2$, the following quantity is conserved, 
\begin{equation}
  \exp\left(\int^n\big(3+\epsilon_2(n^\prime)\big)dn^\prime\right)
         \left[\dot{\cal R}_c(n,\vec x) - 2\omega(n){\cal S}(n,\vec x)\right]
    = {\rm const.}
    \,,
\label{conserved quantity}
\end{equation}
where we introduced the number of e-foldings, $n=\ln(a)$ ($\partial_t =H\partial_n$). 
This means that the (rate of change of the) curvature perturbation on super-Hubble scales is given by, 
\begin{equation}
 \dot{\cal R}_c(n,\vec x)=  \left[\dot{\cal R}_c(n_*,\vec x)- 2\omega(n_*){\cal S}(n_*,\vec x)\right]
                                     {\rm e}^{-\int_{n_*}^n\big(3+\epsilon_2(n^\prime)\big)dn^\prime}
                                      + 2\omega(n){\cal S}(n,\vec x)
\,,
\label{conserved quantity:2}
\end{equation}
where $n=n_*$ is chosen such that the gradient term on the right hand side of Eq.~(\ref{evolution adiabatic:3b})
can be neglected, which is the case when $\|\nabla^2\|/(aH)^2$ evaluated at $n=n_*$ is sufficiently small when
compared with unity. 
Hence, in order to calculate the spectrum of the curvature perturbation, we need to know how 
the entropy perturbation evolves in time. Since Eq.~(\ref{evolution entropy:3}) cannot be solved in general, 
we shall solve it in slow-roll approximation, which is what we do next.

%%%%%%%%%%%%%%%%%%%%%%%%%%%%%%%%%%%%%%%%%%%%%%%%%%%%
%%%%%%%%%%%%%    S L O W   R O L L    A N A L Y S I S   %%%%%%%%%%%%%%%%%%%%%
%%%%%%%%%%%%%%%%%%%%%%%%%%%%%%%%%%%%%%%%%%%%%%%%%%%%

\subsection{Slow-roll analysis}
\label{Slow-roll analysis}

Equations~(\ref{evolution adiabatic:3}--\ref{evolution entropy:3}) are easy enough such that they can be analyzed in 
{\it slow-roll approximation}. We shall perform our analysis in two steps. In {\it step 1} we determine the spectra
at a scale close to the Hubble scale. This analysis can be done at the leading (zeroth) order in slow-roll parameters,
but the gradient operators must be kept.
In {\it step 2} we shall study the evolution of the curvature perturbation on super-Hubble scales induced by 
the entropy perturbation, again to the leading order in slow-roll approximation. 
While this analysis is guaranteed to correctly reproduce
the spectra and spectral indices to 
the leading order in slow-roll parameters,
because of the coupling between the fields, potentially 
interesting features may be hidden in the subleading results for the spectra.
For that reason in Appendix~A we analyze the spectra
at the subleading order in the slow-roll parameters and find 
that the curvature-entropy spectrum is generated at the 
next-to-leading order in the slow-roll parameters
(its amplitude being proportional to the turning rate $\omega$),
even if at the beginning
of inflation this correlator was set to zero.

{\bf Step 1.} Observe firstly that at the zeroth order in slow-roll, 
Eqs.~(\ref{evolution adiabatic:3}--\ref{evolution entropy:3}) simplify to,
\begin{eqnarray}
\ddot {\cal R}_c \!+\! 3H\dot{\cal R}_c \!-\frac{\nabla^2}{a^2}{\cal R}_c  &\approx& 0
\qquad
\label{evolution adiabatic: LO}
\\
\ddot {\cal S} \!+\!  3H\dot {\cal S}\!-\frac{\nabla^2}{a^2}{\cal S}
     &\approx& 0
\,,
\label{evolution entropy: LO}
\end{eqnarray}
where we made use of the fact that the coupling between the perturbations is suppressed by the turning rate, 
$\omega=\eta_\omega H$, which we assume to be suppressed in slow-roll approximation, {\it i.e.} $\eta_\omega\ll 1$. 

Since the perturbations decouple, it is easy to solve Eqs.~(\ref{evolution adiabatic: LO}--\ref{evolution entropy: LO}).
When written in terms of (conformal) time, $d\tau = (N/a)dt$,  Eqs.~(\ref{evolution adiabatic: LO}--\ref{evolution entropy: LO})
become, 
\begin{eqnarray}
{\cal R}_c^{\prime\prime} \!+\!2{\cal H}{\cal R}_c^{\prime} \!-\nabla^2{\cal R}_c  &\approx& 0
\qquad
\label{evolution adiabatic: LOa}
\\
{\cal S}^{\prime\prime} \!+\! 2{\cal H}{\cal S}^{\prime}\!-\nabla^2{\cal S}  &\approx& 0
\,,
\label{evolution entropy: LOa}
\end{eqnarray}
where ${\cal R}_c^{\prime}=\partial_\tau {\cal R}_c$ and ${\cal H} = aH=(d/d\tau)\ln[a(\tau)]$ is the conformal 
expansion rate, which in this approximation is simply, ${\cal H} \approx - 1/\tau$ ($\tau<0$).
In order to obtain the spectra on the sub-Hubble scales, one ought to solve the quantum version of 
(\ref{evolution adiabatic: LOa}--\ref{evolution entropy: LOa}), {\it i.e.} one ought to promote ${\cal R}_c$ and ${\cal S}$ 
to operators, ${\cal R}_c\rightarrow \hat {\cal R}_c$, ${\cal S}\rightarrow \hat {\cal S}$, which satisfy the following
canonical commutation relations ($\hbar =1$), 
\begin{equation}
 \left[ \hat {\cal R}_c(t,\vec x), \hat \Pi_{{\cal R}_c}(t,\vec x^{\,\prime})\right]
  = i \delta^3(\vec x-\vec x^{\,\prime})
\,,\quad 
\left[ \hat {\cal S}(t,\vec x), \hat \Pi_{\cal S}(t,\vec x^{\,\prime})\right] 
= i \delta^3(\vec x-\vec x^{\,\prime})
\label{canonical quantization}
\end{equation}
and all other commutators vanish. Here we have introduced canonical momenta, 
\begin{equation}
   \hat \Pi_{{\cal R}_c} 
   = 2\epsilon M_{\rm P}^2a^2 
   \left(\hat {\cal R}_c^\prime -a \omega\hat {\cal S}\right)
\,,\quad
   \hat \Pi_{\cal S} =
   2\epsilon M_{\rm P}^2  a^2 \left(\hat {\cal S}^\prime
              + a \omega\hat {\cal R}_c\right)
\,,
\label{canonical momenta}
\end{equation}
where these relations follow from the canonical
momenta $\hat\Pi_\sigma$ and $\hat\Pi_s$ in 
Eq.~(\ref{canonical quantisation sigma s}).
Upon transforming into the spatial momentum space ({\it cf.} Eq.~(\ref{Fourier decomposition})), 
from~(\ref{evolution entropy: LO}) we obtained the mode equations 
${\cal R}_c(\tau,k)$ and ${\cal S}(\tau,k)$ 
($k=\|\vec k\|$), 
\begin{equation}
 {\cal R}_c^{\prime\prime} \!-\!\frac{2}{\tau}{\cal R}_c^{\prime} \!+k^2\!{\cal R}_c  \approx 0
\,,\quad 
 {\cal S}^{\prime\prime} \!-\!\frac{2}{\tau}{\cal S}^{\prime} \!+k^2\!{\cal S} \approx 0
 \,,
\label{mode equations: LO}
\end{equation}
which can be solved in terms of the Hankel functions
with the index, $\nu=3/2$. 
The normalization can be determined 
(up to Bogolyubov transformations) from the Wronskian conditions ({\it cf.} Eq.~(\ref{Wronskian matrix})), 
\begin{equation}
{\cal R}_c(\tau,k)\Pi_{{\cal R}_c}^*(\tau,k)-{\cal R}_c^*(\tau,k)\Pi_{{\cal R}_c}(\tau,k)=i
\,,\quad
{\cal S}(\tau,k)\Pi_{\cal S}^*(\tau,k)-{\cal S}^*(\tau,k)\Pi_{\cal S}(\tau,k)=i
\,.
\label{Wronskian conditions: LO}
\end{equation}
Notice that here the mode functions are ordinary functions, which is to be contrasted with the general 
case~(\ref{Fourier decomposition}), in which they are matrix valued.  
Here we make the simplest -- positive frequency -- choice of the vacuum (also known as the Bunch-Davies or
Chernikov-Tagirov vacuum), and we obtain, 
\begin{equation}
{\cal R}_c(\tau,k) \approx \frac{H}{2\sqrt{\epsilon k^3}M_{\rm P}}
        \left(1+i k\tau\right){\rm e}^{-ik\tau}
\,,\quad
 {\cal S}(\tau,k) \approx \frac{H}{2\sqrt{\epsilon k^3}M_{\rm P}}
        \left(1+i k\tau\right){\rm e}^{-ik\tau}
\,.
\label{solutions mode functions: LO}
\end{equation}
These short-wavelength solutions can be inserted into the standard formulas for the spectra,
 \begin{eqnarray}
 {\cal P}_{\cal R}(\tau,k) = \frac{k^3}{2\pi^2}  |{\cal R}_c(\tau,k)|^2 
     ={\cal P}_{{\cal R}*}\left(\frac{k}{k_*}\right)^{n_{\cal R}-1}
 \label{spectrum adiabatic perturbation:LO}
 \\
  {\cal P}_{\cal S}(\tau,k) = \frac{k^3}{2\pi^2}  |{\cal S}(\tau,k)|^2 
    ={\cal P}_{{\cal S}*}\left(\frac{k}{k_*}\right)^{n_{\cal S}}
 \label{spectrum entropy perturbation:LO}
 \end{eqnarray}
which are valid up to mildly super-Hubble scales $\tau=\tau_*$ 
(on which $k/a_*\ll H_*$ with $a_*\equiv a(\tau_*)$ and $H_*\equiv H(\tau_*)$).
Upon inserting~(\ref{solutions mode functions: LO}) 
into~(\ref{spectrum adiabatic perturbation:LO}--\ref{spectrum entropy perturbation:LO}) we obtain, 
 \begin{eqnarray}
{\cal P}_{{\cal R}*}\approx \frac{H_*^2}{8\pi^2\epsilon_*M_{\rm P}^2}
\,,\qquad 
{\cal P}_{{\cal S}*}\approx \frac{H_*^2}{8\pi^2\epsilon_*M_{\rm P}^2}
\,,
 \label{spectra LO}
 \end{eqnarray}
where we neglected the conformal parts, which come as a multiplicative factor $1+k^2/(a_*H_*)^2$ in~(\ref{spectra LO}),
which is justified on super-Hubble scales. 
The solutions~(\ref{spectra LO}) are correct up to the leading 
order in slow-roll parameters. One can construct slow-roll corrections to
these solutions by solving the full equations~(\ref{evolution adiabatic:3}--\ref{evolution entropy:3}) iteratively in powers of slow-roll parameters,
{\it e.g.} by using the method of Green's functions. The resulting 
corrections are slow-roll suppressed when compared with
 the leading results~(\ref{spectra LO}).
 The details of such an analysis can be found in Appendix~A.
The spectral indices 
in~(\ref{spectrum adiabatic perturbation:LO}--\ref{spectrum entropy perturbation:LO})
are then obtained in the standard manner, by taking 
a derivative with respect to $\ln(k)$ and setting it to the Hubble crossing scale, $k=aH$.
The result is, to leading order in slow-roll,
\begin{equation}
 [n_{\cal R}-1]_1=-2\epsilon -\epsilon_2
 \,,\qquad 
   [n_{\cal S}]_1=-2\epsilon -\epsilon_2
\label{spectral indices LO}
\end{equation}
where all quantities are evaluated at a fiducial scale $k_*=\mu a_*H_*$,
with $\mu\lesssim 1$. 
This completes our analysis of short scales. 

{\bf Step 2.} As we have shown in Eq.~(\ref{conserved quantity:2}) above, 
in the two field case the curvature perturbation is not constant on super-Hubble scales,
but it is sourced by the entropy perturbation, which in turn can modify its spectrum.
In order to make progress, in what follows we shall solve the evolution 
equations~(\ref{evolution adiabatic:3}--\ref{evolution entropy:3})
on super-Hubble scales, but now keeping the linear slow-roll corrections.

Indeed, when the entropy field mass ${\cal M}^2_{ss}$~\footnote{The opposite limit, when
${\cal M}^2_{ss}\gg H^2$ is rather easy, since in this case one can use adiabatic approximation to solve for the mode  
functions of the entropy perturbation. Since in this case the effect of the entropy perturbation on the curvature perturbation 
is expected to be small on super-Hubble scales, this case is 
trivial and we do not consider it any further.}
and the turning rate $\omega$ are small, {\it i.e.} when, 
\begin{equation}
{\cal M}^2_{ss} \equiv \eta^2_{ss}H^2
\,,\qquad   \omega \equiv \eta_\omega H
  \,,\qquad {\rm with} \quad |\eta^2_{ss}|,\eta_\omega\ll 1
\label{turning rate slow roll}
\end{equation}
both satisfied, then  the source on 
the right hand side of~(\ref{evolution entropy:3}) can be approximated by, 
$-2\omega\dot{\cal R}_c\simeq -4\omega^2{\cal S}$, such that, on super-Hubble scales, 
Eq.~(\ref{evolution entropy:3}) simplifies to, 
\begin{equation}
\ddot {\cal S} \!+\!  (3\!+\!\epsilon_2)H\dot {\cal S}\!+\! 
        \big({\eta}^2_{ss}\!+\! 3\eta_\omega^2\!+\!\Delta_\epsilon\!-\!\epsilon M_{\rm P}^2{\cal R}\big) H^2{\cal S}
%     \simeq  \frac{\nabla^2}{a^2}{\cal S}
       \approx 0
\,.
\label{evolution entropy:4}
\end{equation}
Since the last term on the left hand side is suppressed by $\epsilon$, and $M_{\rm P}^2{\cal R}$ is typically 
of the order unity or smaller, all terms contributing to the effective mass of the entropy perturbation
are suppressed (at least linearly) by slow-roll parameters.
Notice next that the form~(\ref{evolution entropy:4}) of the equation for the entropy perturbation 
follows immediately from Eq.~(\ref{evolution entropy}), in which the source on the right 
hand side is suppressed by the Laplacian of the Bardeen potential, and hence can be neglected on super-Hubble scales.
Equation~(\ref{evolution entropy:4}) tells us that on super-Hubble scales 
${\cal S}$ approximately decouples from ${\cal R}_c$, implying that one can first solve~(\ref{evolution entropy:4}) 
for the entropy perturbation, and then insert the solution into the equation for the curvature 
perturbation~(\ref{conserved quantity:2}) to get the desired spectrum. 

To the leading order in slow-roll parameters and on super-Hubble scales Eq.~(\ref{evolution entropy:3}) simplifies to, 
\begin{eqnarray}
 \partial_n{\cal S}
  &=& -\frac{1}{3\!-\!\epsilon\!+\!\epsilon_2}\left[\left(\eta^2_{ss}\!+\!\Delta_\epsilon\!+\! 3\eta_\omega^2
                   \!-\!\epsilon M_{\rm P}^2{\cal R}\right) {\cal S}\!+\!\partial_n^2{\cal S}\right] 
\nonumber\\
 &\approx& -\frac{1}{3}\left(\eta^2_{ss}\!+\! 3\eta_\omega^2\!+\!\frac32\epsilon_2
                   \!-\!\epsilon M_{\rm P}^2{\cal R}\right) {\cal S}
\,,
\label{evolution entropy:superHubble}
\end{eqnarray}
 where $\partial_n$ is a derivative with respect to the number of e-foldings $n$ (defined by $dn=HN(t)dt$)
 and we kept only the leading (linear) order terms in slow-roll 
 (${\cal S}^{\prime\prime}$ is of higher (second) order in slow-roll). 
 Eq.~(\ref{evolution entropy:superHubble}) can be easily solved,  
\begin{equation}
 {\cal S}(n,\vec x)\approx {\cal S}(n_*,\vec x)\exp\left[-\frac13\int_{n_*}^n d\tilde n
                    \Big(\eta^2_{ss}\!+\! 3\eta_\omega^2\!+\!\frac32\epsilon_2\!-\!\epsilon M_{\rm P}^2{\cal R}\Big)
                                                     \right]
\,,
\label{solution entropy:superHubble}
\end{equation}
which tells us how ${\cal S}(n)$ evolves on very large scales, where $n>n_*$. This evolution 
results in an additional contribution to the spectral index $n_{\cal S}$ 
({\it cf.} Eq.~(\ref{spectral indices LO}))~\footnote{One can show 
 that the spectral index of the entropy perturbation $n_{\cal S}$
 is twice the derivative with respect to the Hubble crossing time, $\ln(aH)=n+\ln(H)$ 
 of the exponent of the solution given in~(\ref{solution entropy:superHubble}) which is, to leading order 
 in slow-roll, equal to the derivative with respect to $n$.} of the form,
\begin{equation}
 \left[n_{\cal S}\right]_2 = \frac23\eta^2_{ss}\!+\!2\eta_\omega^2\!+\!\epsilon_2\!-\!\frac23\epsilon M_{\rm P}^2{\cal R}
\,,
\label{spectrum of entropy perturbation}
\end{equation}
where all parameters in~(\ref{spectrum of entropy perturbation}) are evaluated at 
$n=n_*$. In fact, evaluating these quantities at a different time is permitted, since that would lead to a result that differs at 
higher order in slow-roll, and thus is immaterial for the present analysis. 

\medskip

We are now ready to consider the adiabatic perturbation. 
Integrating Eq.~(\ref{conserved quantity:2}) and neglecting the first term (which amounts to neglecting the 
decaying mode), we obtain,
\begin{eqnarray}
 {\cal R}_c(n,\vec x) &\approx&  {\cal R}_c(n_*,\vec x) 
      + 2\int_{n_*}^n \eta_\omega(n^\prime){\cal S}(n^\prime,\vec x)dn^\prime
\nonumber\\
&\approx&  {\cal R}_c(n_*,\vec x)
    \left[1+
           2\int_{n_*}^n \eta_\omega(n^\prime){\cal T}_{\cal S}(n^\prime,\vec x;n_*)dn^\prime
  \right]
\label{curvature perturbation: solution}
\end{eqnarray}
where, to get the last result, we made use of ${\cal R}_c(n_*,\vec x)\approx{\cal S}(n_*,\vec x)$ and 
we have introduced the transfer function for the entropy perturbation (see~(\ref{solution entropy:superHubble})) ,
\begin{equation}
{\cal T}_{\cal S}(n,\vec x;n_*)\equiv\frac{{\cal S}(n,\vec x)}{{\cal S}(n_*,\vec x)} 
 =\exp\left[-\frac13\int_{n_*}^n d\tilde n
                    \Big(\eta^2_{ss}\!+\! 3\eta_\omega^2\!+\!\frac32\epsilon_2\!-\!\epsilon M_{\rm P}^2{\cal R}\Big)
                                                     \right] 
\,.
\label{transfer function for entropy perturbation}
\end{equation}

Now upon taking derivative of the logarithm of~(\ref{curvature perturbation: solution}) with respect to $\ln(aH)\approx n$, multiplying by 2 and making use of~(\ref{transfer function for entropy perturbation}), 
we get the following contribution to the spectral index of the adiabatic perturbation due to its coupling to the 
entropy perturbation, 
 \begin{equation}
  \left[ n_{\cal R}(n)-1\right]_2 = 4\eta_\omega(n) 
  \frac{{\rm e}^{-\frac13\!\int_{n_*}^{n}\! 
                     \big(\eta^2_{ss}\!+\! 3\eta_\omega^2\!+\!\frac32\epsilon_2\!-\!\epsilon M_{\rm P}^2{\cal R}\big)d\tilde n
                                                   }}
       {1+2\int_{n_*}^n \eta_\omega(n^\prime){\cal T}_{\cal S}(n^\prime,\vec x;n_*)dn^\prime}
\,.
 \label{spectral index of the curvature perturbation}
 \end{equation}
 Several remarks are now in order.
The entropy perturbation can through 
 Eq.~(\ref{transfer function for entropy perturbation}) 
 contribute to  the curvature perturbation. 
Unless the transfer function ${\cal T}_{\cal S}$ is quite sizable, 
the contribution in the denominator
of~(\ref{spectral index of the curvature perturbation})
 can be neglected as it is suppressed by 
 the slow-roll parameter $|\eta_\omega|\ll 1$.
 Notice also that, even though the exponent of  the transfer function ${\cal T}_{\cal S}$ 
 in~(\ref{transfer function for entropy perturbation}) is suppressed by slow-roll parameters, it is not 
 necessarily small because of the integral, which produces an enhancement by a factor
 $\sim n-n_*$. For sufficiently late times $n-n_*\gg 1$,
 such that it can compensate the smallness 
 of the slow-roll parameters. 
 For that reason it is important to keep that term 
 in Eq.~(\ref{spectral index of the curvature perturbation}) even though na\^ively one would be 
 tempted to conclude that it contributes 
 at a higher order in slow-roll parameters. 
Furthermore, the sign of the exponent 
in~(\ref{transfer function for entropy perturbation}) is important.
Namely, if the sign of the integrand 
$\eta^2_{ss}\!+\! 3\eta_\omega^2\!+\!\frac32\epsilon_2\!-\!\epsilon M_{\rm P}^2{\cal R}$ 
is positive (negative), the transfer function ${\cal T}_{\cal S}(n,\vec x;n_*)$ decreases (increases) in time,
which in turn implies that the contribution of the entropy perturbation to the spectral index 
decreases (grows) in time, rendering the curvature spectrum 
bluer (redder).  

To conclude, the principal results of this section are  formulas~(\ref{spectra LO}) and 
(\ref{spectrum adiabatic perturbation:LO})
for the spectrum of of the adiabatic and entropy perturbation, with the spectral indices given in 
Eqs.~(\ref{spectral indices LO}), (\ref{spectrum of entropy perturbation}) 
and~(\ref{spectral index of the curvature perturbation}) which, when summed, yield, 
\begin{eqnarray}
 n_{\cal R} &=& 1-2\epsilon -\epsilon_2+4\eta_\omega(n) 
  \frac{{\rm e}^{-\frac13\!\int_{n_*}^{n}\! 
                     \big(\eta^2_{ss}\!+\! 3\eta_\omega^2\!+\!\frac32\epsilon_2\!-\!\epsilon M_{\rm P}^2{\cal R}\big)d\tilde n}}
       {1+2\int_{n_*}^n \eta_\omega(n^\prime){\cal T}_{\cal S}(n^\prime,\vec x;n_*)dn^\prime}
\label{spectral index: curvature perturbation}\\
   n_{\cal S} &=& -2\epsilon
    +\frac23\eta^2_{ss}\!+\!2\eta_\omega^2\!-\!\frac23\epsilon M_{\rm P}^2{\cal R}
\,.
\label{spectral index: entropy perturbation}
\end{eqnarray}
There is also the mixed correlator, 
$\langle \hat{\cal R}_c(t,\vec k){\cal S}(t,\vec x ')\rangle$,
whose amplitude is suppressed by the turning rate $\omega$,
see Eq.~(\ref{Appendix: spectrum RS}),
and since the turning rate is typically small (it is suppressed by 
the slow-roll
parameter $\eta_\omega$)
its amplitude is suppressed when compared with that of the curvature and entropy correlators. Its spectral index is simply,
$n_{\cal R\cal S} = (n_{\cal R}+n_{\cal S})/2$.
Unless either the turning rate  or the transfer function ${\cal T}_{\cal S}$ is rather large, 
the contribution of the ratio in~(\ref{spectral index: curvature perturbation}) can be approximated 
by unity. In this case the adiabatic spectral index simplifies to, 
\begin{eqnarray}
 n_{\cal R} \approx 1-2\epsilon -\epsilon_2+4\eta_\omega
\,,
\label{spectral index: curvature perturbation: simple}
\end{eqnarray}
such that the principal contribution of the entropy perturbation to the spectral index of the curvature perturbation,
$\delta_{\cal S}n_{\cal R} \approx 4\eta_\omega = 4\omega/H$, 
comes from the turning rate $\omega$ (expressed in units of $H$). When $\omega<0$ ($\omega>0$)
the coupling to the entropy perturbation reduces (enhances)
the spectral 
index in~(\ref{spectral index: curvature perturbation: simple}), 
such that the corresponding spectrum becomes redder (bluer).
While the curvature spectrum gets a correction from the entropy perturbation through the transfer function
${\cal T}_{\cal S}$ in~(\ref{curvature perturbation: solution}), 
this correction is typically small and can be neglected, unless either 
the turning rate $\omega$ or 
the transfer function ${\cal T}_{\cal S}$ is quite large.
More precisely, when $\omega{\cal T}_{\cal S}\gg H$
%If that is the case, 
and $\omega$ and ${\cal T}_{\cal S}$ change
adiabatically slowly in time, then
the spectrum of the curvature 
perturbation~(\ref{spectral index: curvature perturbation})
can be approximated by,
\begin{equation}
n_{\cal R}\simeq 1-2\epsilon-\epsilon_2+\frac{2}{n-n_*}
\,,
\label{curvature spectrum: large transfer function}
\end{equation}
which in the limit $n-n_*\gg 1$ 
approaches that of the single field inflation. 
What is also interesting in Eqs.~(\ref{spectral index: curvature perturbation: simple}) 
and~(\ref{spectral index: entropy perturbation}) is that, while the configuration space curvature contributes to the spectral index of the entropy perturbation, it does not contribute 
to the spectral index of the curvature perturbation.

 %%%%%%%%%%%%%%%%%%%%%%%%%%%%%%%%%%%%%%%%%%%%%%%%%%%%
%%%%%%%%%%%%%    S L O W    R O L L    P A R A M E T E R S  %%%%%%%%%%%%%%%%%%%
%%%%%%%%%%%%%%%%%%%%%%%%%%%%%%%%%%%%%%%%%%%%%%%%%%%%

 \subsection{Explicit form for slow-roll parameters}
 
 In this subsection we give explicit forms for the slow-roll parameters. 
The Hubble parameter and the principal slow-roll parameter are given by (to leading order in derivatives),
\begin{equation}  
 H^2 \approx \frac{V(\phi^A)}{3M_{\rm P}^2}
 \,,\qquad 
 \epsilon = \frac{\dot\sigma^2}{2M_{\rm P}^2H^2} \approx \frac{M_{\rm P}^2}{2}\frac{(\partial^AV)(\partial_AV)}{V^2} 
 \,.
 \label{H and epsilon}
 \end{equation}
 The higher order slow-roll parameters are, 
 \begin{eqnarray}
 \epsilon_2 &=& \partial_n\ln(\epsilon) = (\partial_n \phi^A)\partial_A\ln(\epsilon)
 = -M_{\rm P}^2\frac{\left(\partial^A\ln(V)\right)
            \partial_A\left[\left(\partial^B\ln(V)\right)\left(\partial_B\ln(V)\right)\right]
                   }
                  {\left(\partial^C\ln(V)\right)\left(\partial_C\ln(V)\right)}
    \label{epsilon2}\\
 \epsilon_3 &=& \partial_n\ln(\epsilon_2) 
   =  M_{\rm P}^2\left(\partial^A\ln(V)\right)\left[
    \frac{\partial_A
    \left\{
      \left(\partial^B\ln(V)\right)\partial_B\left[\left(\partial^C\ln(V)\right)\left(\partial_C\ln(V)\right)\right]\right\}
      }{\left(\partial^I\ln(V)\right)\partial_I\left[\left(\partial^J\ln(V)\right)\left(\partial_J\ln(V)\right)\right]}\right.
     \label{epsilon3}\\
      &&     \left.\hskip 5cm
 - \frac{\partial_A\left[\left(\partial^B\ln(V)\right)\left(\partial_B\ln(V)\right)\right]
                   }
                  {\left(\partial^C\ln(V)\right)\left(\partial_C\ln(V)\right)}\right]
   \,,
\nonumber
 \end{eqnarray}
  where the last term in~(\ref{epsilon3}) equals to $\epsilon_2$. 

\begin{figure}[h!]
 \centering
  \includegraphics[width=0.99\textwidth]{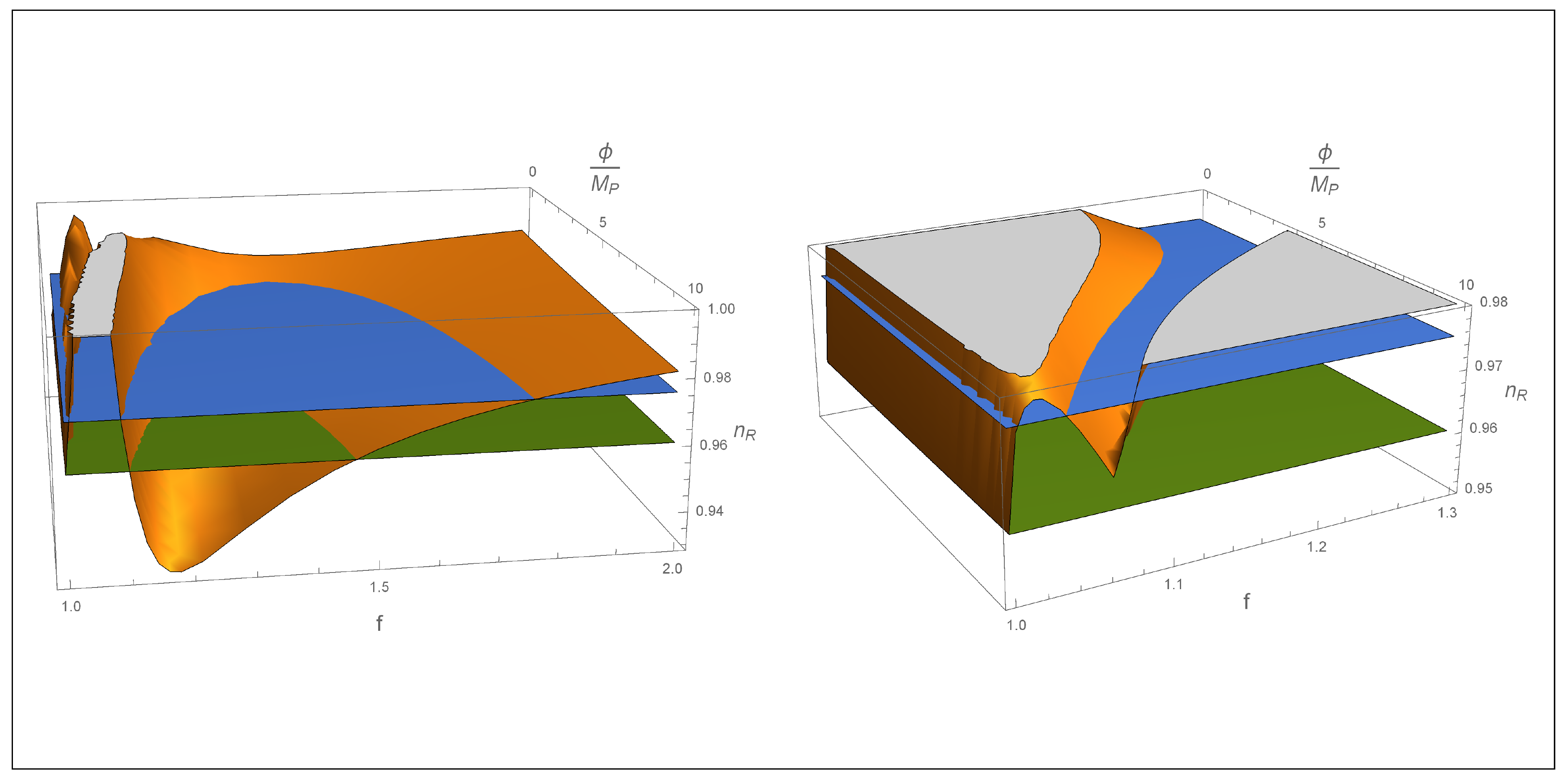}
\vskip -0.4cm 
 \caption{The spectral index $n_{\cal R}$ from Eq.~(\ref{spectral index: curvature perturbation: simple}) is shown 
  as a function of the curvaton field $f$ (horizontal axis) and the inflaton condensate $\phi/M_{\rm P}$
 for negative nonminimal couplings: $\xi=-0.001$ (left panel) and 
 $\xi=-0.01$ (right panel).
The values of the parameters are: 
$m_\chi=10^{-7}M_{\rm P}$, 
$\lambda=10^{-1}$, $\lambda_\chi=10^{-12}$ 
and $V_0=10^{-9}M_{\rm P}^4$. 
}
 \label{figure: scalar spectral index}
 \end{figure}
 Based on the above expressions in figure~(\ref{figure: scalar spectral index}) 
 we plot the spectral index of the curvature perturbation  (red surface)
 which, in the limit of a small transfer function ${\cal T}_{\cal S}$ ({\it i.e.} a small turning rate $\omega/H$), 
 can be well approximated by~(\ref{spectral index: curvature perturbation: simple}).  
 The lower and upper Planck collaboration limits on the scalar spectral index, $n_{\cal R}=0.9649\pm0.0042$, 
 are also shown
  (green and blue horizontal planes, respectively). The spectral index $n_{\cal R}$ is shown 
  as a function of $f$ (horizontal axis) and $\phi/M_{\rm P}$  (the axis pointing into the paper)
  and the noniminimal coupling.
  We see 
  that the values which are consistent with the observations typically corresponds to 
  $f$ in the range from 1 to 2 and rather small, negative nonminimal couplings. 
  The value of $\phi$ is not very relevant, since for the exponential inflaton potential we 
  consider in this work~(\ref{quartic selfinteraction inflaton}),  
  a shift in $\phi$ can be always compensated by a multiplicative change in $V_0$.

  \bigskip
  
\begin{figure}[h!]
 \centering
  \includegraphics[width=0.99\textwidth]{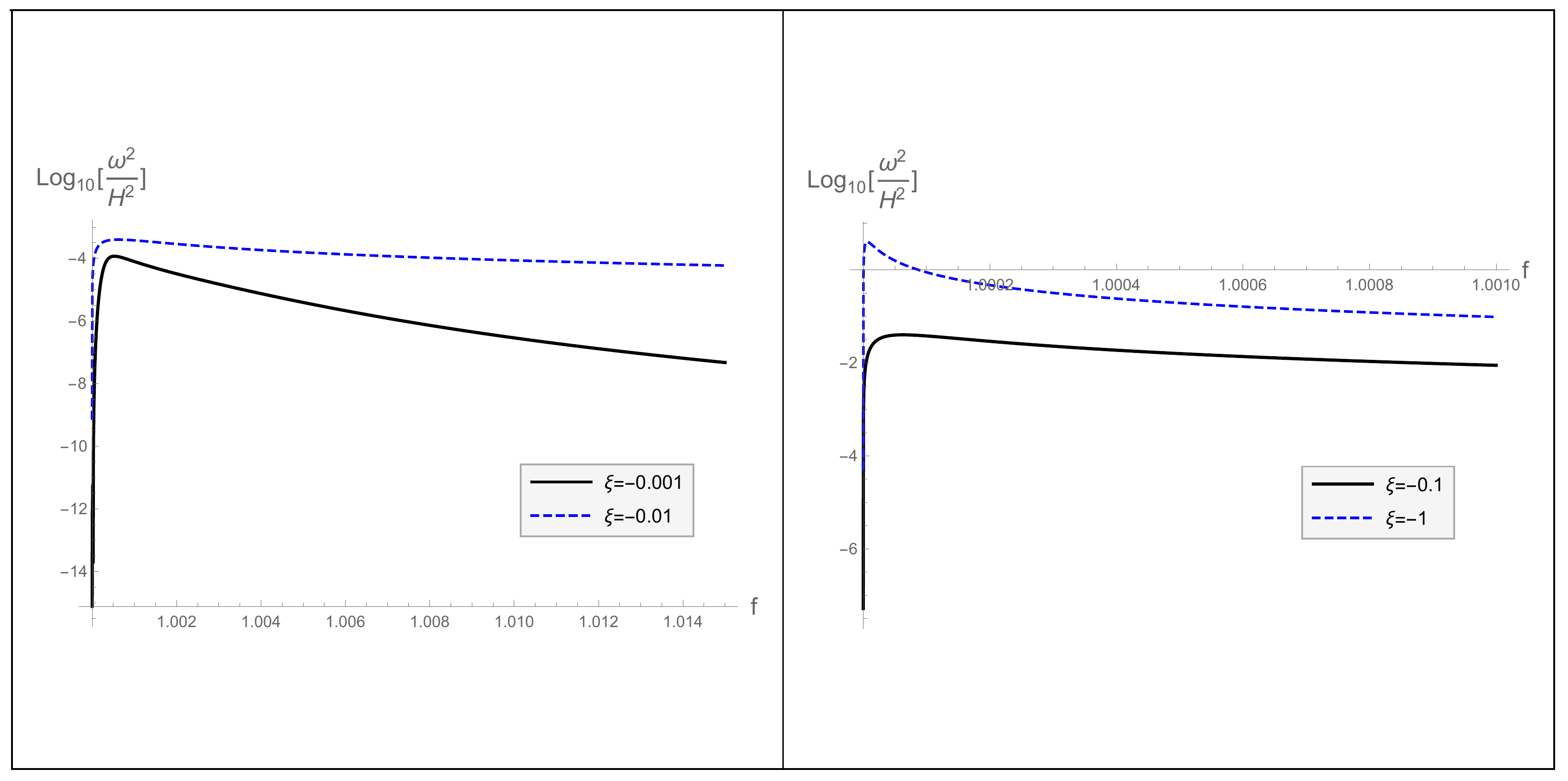}
\vskip -0.4cm 
 \caption{The dimensionless turning rate $\eta_\omega^2=\omega^2/H^2$ 
 calculated in slow-roll approximation~(\ref{turning rate: slow roll})
 as a function of the curvaton field $f$ and 
 for negative nonminimal couplings: $\xi=-0.01$ (blue dashed curve on the left panel),
 $\xi=-0.001$ (black solid curve on the left panel),
  $\xi=-0.1$ (solid black curve on the right panel),
   $\xi=-1$ (blue dashed curve on the right panel).
The scale at the vertical axes is logarithmic, while at the horizontal axes it is linear. 
 The values of other parameters are identical as in figure~2.
}
 \label{figure: turning rate}
 \end{figure}

  Next we need is the unit vectors $\hat\sigma^A$ and $\hat\omega^A$ 
  and the turning rate $\omega$~(\ref{turning rate}).
From~(\ref{adiabatic and entropic}) we know that $\hat\sigma^A=\dot\phi^A/\dot\sigma$, which in slow-roll 
approximation becomes, 
\begin{equation}
 \hat \sigma^A\approx -\frac{\partial^A V}{\|\nabla V\|}
 \,,\qquad \|\nabla V\| = \sqrt{(\partial^AV)(\partial_AV)}
 \,.
\label{sigma A: slow roll}
\end{equation}
To get the turning rate $\omega^A$, one inserts the slow-roll 
result~(\ref{sigma A: slow roll}) 
into the definition ~(\ref{turning rate}) to obtain,
\begin{equation}
\omega^2 = \|\omega^A\|^2  %=\sqrt{{\cal G}_{AB}\omega^A\omega^B}
   \approx  \frac{M_{\rm P}^2}{3V}
   \left\{\left(\partial^A \|\nabla V\|\right)\left(\partial_A \|\nabla V\|\right)
   -\frac{\left[\left(\partial^A V\right)\left(\partial_A \|\nabla V\|\right)\right]^2}{\|\nabla V\|^2}
   \right\}
   \,,
\label{turning rate: slow roll}
\end{equation}
where we made use of $\omega^A=\dot\sigma \hat\sigma^B\nabla_B\hat\sigma^A$ and 
$\hat\sigma^A=-(\partial^AV)/\|\nabla V\|$. Eq.~(\ref{turning rate: slow roll})
then immediately implies, 
\begin{equation}
\eta_\omega^2=\frac{\omega^2}{H^2}\approx \frac{3M_{\rm P}^2}{V}\omega^2
\,.
\label{eta omega 2}
\end{equation}
In figure~\ref{figure: turning rate} we illustrate how $\eta_\omega^2=\omega^2/H^2$  
defined in~(\ref{eta omega 2})
and~(\ref{turning rate: slow roll}) depends on the curvaton field condensate $f=1-\xi\chi^2/M_{P}^2$
and on the nonminimal coupling $\xi<0$. The generic trend is that the turning rate peaks at a 
rather small field value, $f\simeq 1$ ($-\xi\chi^2\ll M_{\rm P}^2$), and then decays as $f$ increases. 
Furthermore, the peak value of $\eta_\omega^2$ increases as $-\xi$ increases, which means that the coupling 
between the curvature and entropy perturbations becomes stronger, as can be seen 
from {\it e.g.} Eq.~(\ref{evolution adiabatic:3}). 

The unit turning vector is formally,
\begin{equation}
 \hat \omega^A =\frac{\dot\sigma}{\omega}\frac{\partial^BV}{\|\nabla V\|}\nabla_B\frac{\partial^AV}{\|\nabla V\|}
 \,,
\label{unit turning rate vector}
\end{equation}
where $\dot\sigma \approx M_{\rm P}\|\nabla V\|/\sqrt{3V}$.
While this expression is formally correct and can be used to construct $\hat\omega^A$, 
there is an easier way to proceed, namely to
use $\|\hat \omega^A\|=1$ and $ \hat \sigma^A\hat \omega_A=0$, which uniquely fix it to, 
\begin{equation}
\hat\omega^A  \approx  \left(\!\!\begin{array}{c}
                           \frac{\partial_2V}{\|\nabla V\|\sqrt{{\rm det}[{\cal G}_{AB}]}}\cr
                            -\frac{\partial_1V}{\|\nabla V\|\sqrt{{\rm det}[{\cal G}_{AB}]}}\cr
                          \end{array}
                   \right)
\,.
\label{unit turning rate vector:2}
\end{equation}
Even though this expression looks noncovariant, it is in fact covariant, as its covariant form 
is given by~(\ref{unit turning rate vector}). It is nevertheless simple, and thereby convenient to use in practical calculations.
   
Next, we need a slow-roll expression for ${\cal M}^2_{ss}$ defined in~(\ref{mass of the entropy mode}), 
or equivalently the corresponding slow-roll parameter $\eta^2_{ss}={\cal M}^2_{ss}/H^2$. 
Making use of~(\ref{unit turning rate vector}) and~(\ref{turning rate: slow roll}), after some algebra, one gets, 
\begin{eqnarray}
 {\cal M}^2_{ss} &=&\left[\|\nabla V\|^2\left(\partial^A \|\nabla V\|\right)\left(\partial_A \|\nabla V\|\right)
   -\left[\left(\partial^A V\right)\left(\partial_A \|\nabla V\|\right)\right]^2\right]^{-1}
   \nonumber\\
   &\times&\Bigg\{
 \|\nabla V\|^2\left(\partial^A\|\nabla V\|\right)\left(\partial^B\|\nabla V\|\right)\nabla_A\nabla_BV
\label{Mss: slow roll}\\
 \hskip 0cm &&
     -\,2\|\nabla V\|\left(\partial^AV\right)\left(\partial_A\|\nabla V\|\right)
                         \left(\partial^B\|\nabla V\|\right)\left(\partial_B\|\nabla V\|\right)
                         +\frac{\left[\left(\partial^AV\right)\left(\partial_A\|\nabla V\|\right)\right]^3}{\|\nabla V\|} 
   \Bigg\}
\,,
\nonumber
\end{eqnarray}
where the denominator comes from multiplying by $\dot\sigma^2/\omega^2$.

Next, from equation~(\ref{spectral index: entropy perturbation}) we also need contribution
from the configuration space curvature, 
\begin{equation}
-\epsilon M_{\rm P}^2{\cal R} \approx (-\xi)M_{\rm P}^2\frac{\|\nabla V\|^2}{V^2}
  \frac{-6\xi\!-\!2(1\!-\!6\xi)f\!+\!(1\!-\!6\xi)f^2}{[6\xi\!+\!(1\!-\!6\xi)f]^2}
\,,
\label{minus epsilon R}
\end{equation}
where we made use of ~(\ref{H and epsilon}) and~(\ref{configuration space curvature}).
When $\xi<0$ (for which  $f>1$),
 the curvature term~(\ref{minus epsilon R}) contributes positively to the  spectral index of the entropy perturbation. 

\begin{figure}[h!]
 \centering
  \includegraphics[width=0.99\textwidth]{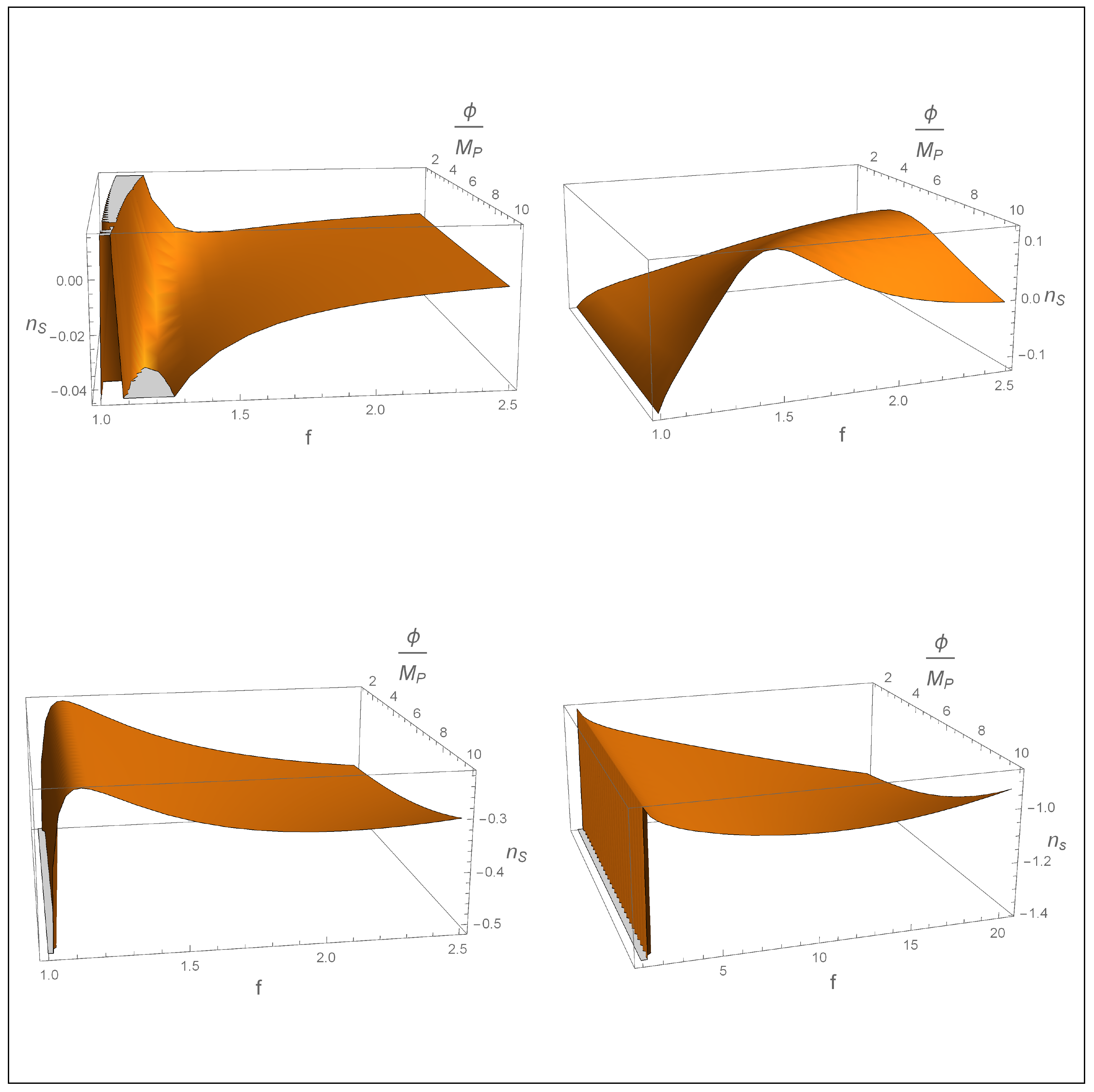}
\vskip -0.4cm 
 \caption{The spectral index of the entropy perturbation 
 $n_{\cal S}$ defined in Eq.~(\ref{spectral index: entropy perturbation}) as a function of the fields 
 $f$ and $\phi/M_{\rm P}$. Each of the panels represents one nonminimal coupling. On the top left panel  
 $\xi=-0.001$, on the top right panel $\xi=-0.01$, on the bottom left panel  
 $\xi=-0.1$, and finally on the bottom right panel $\xi=-1$.
The general trend is that $n_{\cal S}$ becomes more and more negative as $|\xi|$ increases,
{\it i.e.} the spectrum becomes more red and thus its fluctuations grow on very large scales. 
The values of the parameters are: $m_\chi=10^{-7}M_{\rm P}$,
 $\lambda=10^{-1}$, $\lambda_\chi=10^{-12}$ and $V_0=10^{-9}M_{\rm P}^4$. 
}
 \label{figure: entropy spectral index}
 \end{figure}
Figure~\ref{figure: entropy spectral index} shows four panels illustrating the spectral index $n_{\cal S}$ 
of the entropy perturbation
defined in Eq.~(\ref{spectral index: entropy perturbation}) as a function of the curvaton and inflaton condensates. 
The four panels illustrate the dependence of the spectral index $n_{\cal S}$ on the nonminimal coupling $\xi<0$. 
The general trend is that the spectral index  $n_{\cal S}$ becomes more negative as 
$\xi$ becomes more negative, indicating 
that the spectrum of the entropy perturbation grows faster on very large scales, which 
means that the entropy perturbation dominates over the curvatre spectrum of the single field adiabatic model.
When combined with the observation that also the turning 
rate~(\ref{turning rate: slow roll})
 grows with increasing $-\xi$ (see figure~\ref{figure: turning rate}), 
this suggests that, as $\xi$ becomes more and more negative,  the energy between the entropy and the curvature perturbations
gets more efficiently transferred.
%suggesting that the curvaton 
%mechanism is more effective for larger $|\xi|$'s, 
However, there are limitations on
how large $|\xi|$ can be since -- as we argue below -- 
$|\xi|$ cannot be much larger than unity. 
A detailed study of the precise consequences of these crude observations we leave for future work.

 Finally, in order to evaluate the transfer function ${\cal T}_{\cal S}$~(\ref{transfer function for entropy perturbation})
 (see also~(\ref{spectral index of the curvature perturbation})), we need to integrate 
 slow-roll suppressed terms over the number of e-foldings $n$. 
 The number of e-foldings can be expressed in slow-roll as,
 \begin{equation}
 n(\phi^A) = \int^t Hdt \approx \frac{1}{M_{\rm P}^2}\int ^\sigma  \frac{V}{\|\nabla V\|}d\sigma^\prime
 \approx  \frac{1}{M_{\rm P}}\int ^\sigma  \frac{d\sigma^\prime}
         {\sqrt{2\epsilon}}
 \,,
 \label{number of efoldings}
 \end{equation}
 where $d\sigma=\dot\sigma dt = (\nabla_A\sigma)d\phi^A$ is used as the clock during inflation.
Inflation ends at a point when, 
\begin{equation}
\epsilon(\phi_{\rm e}^A)\equiv \epsilon_{\rm e} =1
\,.
\label{end of inflation:epsilon}
\end{equation}
However, these formulas are not useful as long as we do not have an explicit expression on how the fields depend 
on the number of e-foldings, $\phi^A=\phi^A(n)$. In what follows we use the slow-roll relation,
$d\phi^A\approx -(\partial^AV)/(3H^2)dn$, to obtain,
\begin{equation}
 dn^2 = \frac{V^2}{\|\nabla V\|^2M_{\rm P}^4}{\cal G}_{AB}d\phi^A d\phi^B
          = \frac{1}{2\epsilon M_{\rm P}^2}{\cal G}_{AB}d\phi^A d\phi^B
\,.
\label{slow roll metric}
\end{equation}
This expression can be, at least in principle, used to obtain the functional dependence of how $n$ depends on 
the fields $\phi^A$. 
For completeness, we shall also derive a relationship that allows to express $s=s(\phi^A)$. Starting with 
$\hat \omega^A=\frac{\dot\sigma}{\omega}\frac{\partial^CV}{\|\nabla V\|}\nabla_C \frac{\partial^AV}{\|\nabla V\|}$
and $ds=\hat \omega_Ad\phi^A$, one can quite straightforwardly derive the desired expression, 
%
%\begin{eqnarray}
% ds^2 &=& \frac{1}{3V\omega^2}
%  \left(\!\partial_A\|\nabla V\| - \frac{(\partial_A V)(\partial^CV)(\partial_C\|\nabla V\|)}{\|\nabla V\|^2}
%  \right)\left(\partial_B\|\nabla V\|\! -\! \frac{(\partial_B V)(\partial^DV)(\partial_D\|\nabla V\|)}{\|\nabla V\|^2}
%  \!\right)
%  \nonumber\\
%  &\times&d\phi^Ad\phi^B
%  \,,
%  \label{ds2}
%\end{eqnarray}
% 
%where $\omega^2$ is given in Eq.~(\ref{turning rate: slow roll}).
%
\begin{eqnarray}
 ds^2 &=& 
\frac{\left(\!\partial_A\|\nabla V\| - \frac{(\partial_A V)(\partial^CV)(\partial_C\|\nabla V\|)}{\|\nabla V\|^2}
  \right)\left(\partial_B\|\nabla V\|\! -\! \frac{(\partial_B V)(\partial^DV)(\partial_D\|\nabla V\|)}{\|\nabla V\|^2}
  \!\right)}
{\left(\partial^A \|\nabla V\|\right)\left(\partial_A \|\nabla V\|\right)
   -\frac{\left[\left(\partial^A V\right)\left(\partial_A \|\nabla V\|\right)\right]^2}{\|\nabla V\|^2}}
    \times\frac{d\phi^Ad\phi^B}{M_{\rm P}^2}
\,,\qquad
\label{ds2}
\end{eqnarray}
which is rather complicated.
Together with~(\ref{slow roll metric}), this equation allows for construction of the curvilinear coordinates 
$(n,s)$ from the original field coordinates $(\phi^A)$. The advantage of using $(n,s)$ (or equivalently ($\sigma,s$)) 
is that these coordinates have a direct physical interpretation: 
the number of e-foldings $n$ can be used to measure time in inflation and (from $\epsilon(n_e)=1$) to signal the end of inflation,
while $s$ can be used to 
measures the distance between neighboring inflaton trajectories. Finally, in the context of 
$\Delta N$ formalism, $n=n(\phi^A)$ can be used to study the spectrum of cosmological perturbations.

The expression~(\ref{slow roll metric}) 
defines the metric along which the fields move with the number of e-foldings chosen as proper time, 
{\it i.e.} it defines $\phi^A=\phi^A(n)$ in slow-roll 
approximation. In general, it is hard to integrate~(\ref{slow roll metric}), as $\epsilon$ mixes $\phi$ and $f$, 
\begin{equation}
\epsilon \approx   \frac{\lambda^2 V_\phi^2}{2V^2f^3}  
    + (-8\xi)\frac{f\!-\!1}{6\xi\!+\!(1\!-\!6\xi)f}
         \left(1-\frac{m_\chi^2M_{\rm P}^2}{-4\xi}\frac1{Vf}-\frac{\lambda_\chi M_{\rm P}^4}{24\xi^2}\frac{f\!-\!1}{Vf}\right)^2
\,, 
\label{epsilon: explicit}
\end{equation}
where $V(\phi,\chi)=[V_\phi+V_\chi]/f^2$ 
is the potential in Einstein frame~(\ref{potential in Einstein frame}) and we made use of 
\begin{equation}
\|\nabla V\|^2 =\frac{\lambda^2 V_\phi^2}{M_{\rm P}^2f^3}
    + \frac{-16\xi}{M_{\rm P}^2}\frac{f\!-\!1}{6\xi\!+\!(1\!-\!6\xi)f}
         \left(V-\frac{m_\chi^2M_{\rm P}^2}{-4\xi}\frac1f-\frac{\lambda_\chi M_{\rm P}^4}{24\xi^2}\frac{(f\!-\!1)}{f}\right)^2
\,.
\label{nabla V 2}
\end{equation}
To make progress, it is useful to expand~(\ref{epsilon: explicit}) for small and for large $\xi$.
First recall that we are interested in the {\it curvaton model}, 
in which $V_\phi\gg V_\chi$,
such that upon making use of, 
\begin{equation}
V\approx \frac{V_\phi}{f^2}\left[1+\frac{V_\chi}{V_\phi}\right]
\,,
\label{expansion of V}
\end{equation}
we get the following simplified expression for $\epsilon$~(\ref{epsilon: explicit}), 
\begin{equation}
\epsilon \approx   \frac{\lambda^2}{2}f\left[1-2\frac{V_\chi}{V_\phi}\right]  
    + (-8\xi)\frac{f\!-\!1}{6\xi\!+\!(1\!-\!6\xi)f}
\,.
\label{epsilon: simple}
\end{equation}
This expression can be further simplified by taking the $|\xi|\ll 1$ limit and assuming $f={\cal O}(1)$, 
\begin{equation}
\epsilon \approx   \frac{\lambda^2}{2}f
    -8\xi\frac{f\!-\!1}{f}
    %\approx \frac{\lambda^2}{2}f \qquad (|\xi|\ll 1)
\,.
\label{epsilon: simple:2}
\end{equation}
When in addition, $16|\xi|\ll \lambda^2$ is satisfied, $\epsilon$ further simplifies to $\epsilon\approx (\lambda^2/2)f$. 
When combined with
the constraint on the tensor-to-scalar ratio, and 
assuming that the one-field inflation relationship holds, 
 $r\simeq 16\epsilon<0.06$, this gives an upper limit 
 on the coupling, $\lambda< 0.1$ 
 (valid if $\xi<0$). 
Inserting~(\ref{epsilon: simple:2}) into~(\ref{slow roll metric}) gives, 
\begin{equation}
 dn^2= \frac{1}{\lambda^2 M_{\rm P}^2}
        \left(\frac{ d\phi^2}{f^2-\frac{16\xi}{\lambda^2}(f-1)} 
         + \frac{M_{\rm P}^2}{-4\xi}\frac{df^2}{\left(f^2-\frac{16\xi}{\lambda^2}(f-1)\right)(f-1)}
        \right)
\,,
\label{slow roll metric: small xi}
\end{equation}
where we used the expanded configuration space metric~(\ref{configuration space metric}), 
\begin{equation}
{\cal G}_{AB} 
\approx {\rm diag}\left(\frac{1}{f},
 \frac{M_{\rm P}^2}{-4\xi}\frac{1}{f(f-1)}+{\cal O}(\xi^0)\right)
 \,.
 \label{configuration space metric: small xi}
\end{equation}
One can think of the problem of finding $n=n(\phi^A)$  as being equivalent to a 
curved space-time with a configuration space metric, 
\begin{equation}
 {\cal H}_{AB}={\rm diag}\left(\frac{1}{\lambda^2 M_{\rm P}^2}\frac{1}{f^2-\frac{16\xi}{\lambda^2}(f-1)},
  \frac{1}{\lambda^2(-4\xi)}\frac{1}{\left[f^2-\frac{16\xi}{\lambda^2}(f-1)\right](f-1)}
 \right)
 \,.
\label{slow roll metric:H}
\end{equation}
We shall refer to $ {\cal H}_{AB}$ as a {\it slow-roll metric}. 
Since~(\ref{slow roll metric:H}) does not depend on $\phi$, 
$\partial_\phi$ is a Killing vector, implying a conserved configuration space momentum in the $\phi$ 
direction,~\footnote{Notice that the na\^ive configuration space momentum, ${\cal G}_{\phi\phi}d\phi/dn= (d\phi/dn)/f$ 
is not conserved due to the nonvanishing gradient of  
the potential $\partial_A V$ and the 
configuration space curvature.}
\begin{equation}
P_\phi = \left(\partial_\phi\right)_A \frac{d\phi^A}{dn} 
= \frac{1}{\lambda^2 M_{\rm P}^2} \frac{1}{f^2-\frac{16\xi}{\lambda^2}(f-1)} \frac{d\phi}{dn}
 \,,
 \label{conserved phi momentum: small xi}
\end{equation}
where the sign of $P_\phi$ is chosen such that $P_\phi>0$ when 
$d\phi^A/dn$ is positive. This is a convenient choice when early in inflation
both of the field condensates are positive. For definiteness, 
here we assume that to be the case. Since the potential and the nonminimal
coupling function $f(\chi)$ are symmetric under the exchange of the field sign, the results for the other three cases can be easily obtained from the 
case studied here.
This conserved momentum  can be used to convert~(\ref{slow roll metric: small xi}) 
into an ordinary differential equation for $f(n)$, 
\begin{equation}
 \frac{df}{\sqrt{\left[f^2-\frac{16\xi}{\lambda^2}(f-1)\right]
        \left[1 - \lambda^2M_{\rm P}^2P_\phi^2\left(f^2-\frac{16\xi}{\lambda^2}(f-1)\right)\right](f\!-\!1)}} 
          =  - 2\lambda\sqrt{-\xi}dn
\,,
\label{slow roll metric: small xi:2}
\end{equation}
where the sign is chosen such that, as $f$ decreases, $n$ increases. 
Since during inflation $f$ decreases, $n$ will increase.
From Eq.~(\ref{slow roll metric: small xi}) we see 
that the opposite sign choice is allowed, and it corresponds to $n$ 
measuring the time lapse from the end of inflation backwards in time.  
Since Eq.~(\ref{slow roll metric: small xi:2}) is hard to integrate in general, 
we shall make simplifying assumptions, namely that $|\xi|$ is sufficiently
small to satisfy, 
\begin{equation}
 -8\xi\ll \frac{\lambda^2}{2} \equiv \epsilon_\phi
 \,,
\label{very small xi}
\end{equation}
and that the conserved momentum is small, {\it i.e.} 
\begin{equation}
 \tilde P_\phi\equiv \sqrt{2\epsilon_\phi} M_{\rm P}P_\phi  \ll\frac1f<1
 \label{small P}
 \end{equation}
 To estimate $\tilde P_\phi$, recall first that in the limit 
 when $V_\phi\gg V_\chi$ and  $|\xi|\ll 1$, the principal slow-roll 
 parameter~(\ref{epsilon: simple:2}) can be well approximated by,
 $\epsilon\simeq \epsilon_\phi
 \left[f-\frac{8\xi}{\epsilon_\phi}\left(1-\frac{1}{f}\right)\right]$
 and $d\phi/dn=\dot\phi/H =\lambda M_{\rm P}f/(1+V_\chi/V_\phi)$. 
 When this is inserted into~(\ref{conserved phi momentum: small xi})
 one obtains, 
 \begin{equation}
 \tilde P_\phi  \sim \frac{1}
     {\left[f-\frac{8\xi}{\epsilon_\phi}\big(1-\frac{1}{f}\big)\right]}
\,,
 \label{size of conserved P phi}
 \end{equation}
 where we neglected the factor $(1+V_\chi/V_\phi)^{-1}\simeq 1$. 
 This tells us that inequality~(\ref{small P}) is {\it marginally}
 satisfied.  At a first sight it seems strange that our 
estimate~(\ref{size of conserved P phi}) of  $\tilde P_\phi$
%the conserved configuration space slow-roll momentum 
depends on $f$ and therefore it does not seem conserved.
The explanation is in our inexact (slow-roll) estimate of $d\phi/dn$.

When keeping only the linear quantities in the perturbations $\tilde P_\phi^2$
and $-8\xi/\epsilon_\phi$, the integral of~(\ref{slow roll metric: small xi:2}) simplifies to,
\begin{eqnarray}
 \int\frac{df}{f\sqrt{f\!-\!1}} \left(1 \!+\!\frac{4\xi}{\epsilon_\phi}\frac{f\!-\!1}{f^2}\!+\!\frac{\tilde P_\phi^2}{2}f^2\right)
 &=&\Bigg\{
  2\left(1\!+\!\frac{\xi}{2\epsilon_\phi}\right){\rm Arctan}\left[\sqrt{f\!-\!1}\right]
    \!-\!\frac{\xi}{\epsilon_\phi}\frac{2\!-\!f}{f^2}\sqrt{f\!-\!1}
\nonumber\\
  &+&\frac{\tilde P_\phi^2}{3}(f\!+\!2)\sqrt{f\!-\!1}
     \Bigg\}\Bigg|_{f_*}^f 
          \approx  -\sqrt{-8\xi\epsilon_\phi}(n\!-\!n_*)
\,,
\label{slow roll metric: small xi:2B}
\end{eqnarray}
where we neglected  the terms that are quadratic and higher order in the perturbations~(\ref{very small xi}--\ref{small P}). 
This then implies the following expression for $n=n(f)$,
\begin{eqnarray}
 n(f)&\approx& n_*\!-\frac{1}{\sqrt{-2\xi\epsilon_\phi}}\!\bigg\{\!
           \biggl[ \left(\!1\!+\!\frac{\xi}{2\epsilon_\phi}\right){\rm Arctan}\left[\sqrt{f\!-\!1}\right]
            \!-\!\frac{\xi}{2\epsilon_\phi}\frac{2\!-\!f}{f^2}\sqrt{f\!-\!1}
           \!+\!\frac{\tilde P_\phi^2}{6}(f\!+\!2)\sqrt{f\!-\!1}\biggr]
\nonumber\\
 &&          -\,{\cal N}(f_*)\!
 \bigg\}
,
\label{n as function of f}
\end{eqnarray}
where ${\cal N}(f_*)$ is the same function of $f$ as given in the 
square brackets 
in~(\ref{n as function of f}) with $f\rightarrow f_*$.

In what follows, for simplicity we give expressions to the leading (zeroth) order in the perturbations 
$-\xi/\epsilon_\phi$ and $\tilde P_\phi^2=\lambda^2M_{\rm P}^2 P_\phi^2$. If needed, one can always 
go back to~(\ref{n as function of f}) to iteratively include the linear (and if desired higher) order corrections 
in the perturbations. 
Eq,~(\ref{n as function of f}) can be inverted to yield~\footnote{By making use of 
$\tan(\alpha+\beta)=[\tan(\alpha)+\tan(\beta)]/[1-\tan(\alpha)\tan(\beta)]$,
$f_*$ can be pulled out of the tangent in Eq.~(\ref{f of n: small xi}) to arrive at, 
\begin{equation}
  f(n)=1 +\left\{\frac{\sqrt{f_*-1}
     \!-\!\tan\left[\sqrt{-2\xi\epsilon_\phi}(n-n_*)\right]}
 {1\!+\!\sqrt{f_*-1}\tan\left[\sqrt{-2\xi\epsilon_\phi}(n-n_*)\right]}\right\}^2
\,.
\label{f of n: small xi:footnote}
\end{equation}
}
\begin{equation}
 f(n)\approx \frac{1}
            {\cos^2\left[{\rm Arctan}\left(\sqrt{f_*-1}\right)
                         -\sqrt{-2\xi\epsilon_\phi}(n-n_*)\right]}
\,,
\label{f of n: small xi}
\end{equation}
Since $0\leq\cos^2(x)<1$, $f>1$, as it should be. 
From~(\ref{conserved phi momentum: small xi}) we can get  $\phi(n)$,
\begin{eqnarray}
 \phi(n) &=& \phi_* +\sqrt{2\epsilon_\phi}M_{\rm P}\tilde P_\phi 
 \int_{n_*}^n\left[ f^2(n^\prime)
    \!-\!\frac{8\xi}{\epsilon_\phi}\Big(f(n^\prime)\!-\!1\Big)\right]dn^\prime
\label{phi of n: small xi}\\
 &\approx&\phi_* \!+\!
 \frac{\sqrt{2\epsilon_\phi}M_{\rm P}\tilde P_\phi}{\sqrt{-\xi}}
 \Bigg\{\!
  \frac{2\!+\!f_*}{3}\sqrt{f_*\!-\!1}\!-\!(f_*\!-\!1)
 \label{phi of n: small xi:X}\\
 &&\hskip 0cm
 -\,\tan\left({\rm Arctan}\Big(\!\sqrt{f_*\!-\!1}\Big)
  \!-\!\sqrt{\!-2\xi\epsilon_\phi}(n\!-\!n_*)\right)
 \nonumber\\
 &&\hskip 0.5cm \times \left[\frac13
  \!\tan^2\left({\rm Arctan}\Big(\!\sqrt{f_*\!-\!1}\Big)
  \!-\!\sqrt{\!-2\xi\epsilon_\phi}(n\!-\!n_*)\right)\!-\!\sqrt{f_*\!-\!1}
  \!+\!1\right]
 \Bigg\}
  .\;\;
\nonumber
\end{eqnarray}
Notice that, if $\tilde P_\phi\approx 0$, then $\phi(n) \approx \phi_*$ is constant, implying that $\tilde P_\phi$ is the principal
cause for $\phi$ to roll. It is interesting to observe that such a trajectory exists. In order to get a better idea on understanding of what the solutions~(\ref{f of n: small xi}) 
and~(\ref{phi of n: small xi:X}) convey, 
it is useful to consider the following two limits, 
early times when $\sqrt{-2\epsilon\xi}(n-n_*)\ll 1$, and 
late times (close to the end of inflation) when 
${\rm Arctan}(\sqrt{f_*-1})-\sqrt{-2\epsilon\xi}(n-n_*)\ll 1$ which, in 
the  limit when $\sqrt{f_*-1}\gg1$, reduces to 
$(n\!-\!n_*)\lesssim \pi/[2\sqrt{-2\epsilon\xi}]$.
In the former case the following approximation
can be used for the tangents in~(\ref{f of n: small xi}) 
and~(\ref{phi of n: small xi:X}),
\begin{equation}
\tan\left[{\rm Arctan}(\sqrt{f_*-1})-\sqrt{-2\epsilon\xi}(n-n_*)\right]\approx \frac{\sqrt{f_*-1}-\sqrt{-2\epsilon\xi}(n-n_*)}
{1+\sqrt{f_*-1}\sqrt{-2\epsilon\xi}(n-n_*)}
\,,
\label{tangent approximation}
\end{equation}
with the help of which one obtains,
\begin{eqnarray}
f(n)&\approx&f_*-2\sqrt{f_*\!-\!1}\sqrt{-2\xi\epsilon_\phi}(n\!-\!n_*)+{\cal O}\big((-2\epsilon\xi)(n\!-\!n_*)^2\big)
\label{small argument expansion of f(n)}
\\
\phi(n)&\approx& \phi(n_*)+2\epsilon_\phi M_{\rm P}\tilde P_\phi
     f_*\Big(f_*\!+\!2-\sqrt{f_*\!-\!1}\Big)(n\!-\!n_*)
%-\frac{\tilde P_\phi}{3\sqrt{-\xi}}
%\left(1\!-\!\frac{8\xi}{\epsilon_\phi}\right) 
+{\cal O}\big((-2\epsilon\xi)(n\!-\!n_*)^2\big)
\,.
\label{small argument expansion of phi(n)}
\end{eqnarray}
In the latter limit (when 
${\rm Arctan}(\sqrt{f_*-1})-\sqrt{-2\epsilon\xi}(n-n_*)\ll 1$)
one gets,
\begin{eqnarray}
f(n)&\approx&1 + \left({\rm Arctan}(\sqrt{f_*\!-\!1})-
          \sqrt{-2\xi\epsilon_\phi}(n\!-\!n_*)\right)^2
\label{small argument expansion of f(n):2}
\\
\phi(n)&\approx& \phi(n_*)+\sqrt{\frac{2\epsilon_\phi}{-\xi}}
 \frac{M_{\rm P}\tilde P_\phi}{3}
     \bigg[(f_*-1)^{3/2}\!-\!3(f_*\!-\!1)\!+\!3\sqrt{f_*\!-\!1}
\nonumber\\     
  && \hskip 0cm  +3\,\Big(\sqrt{f_*\!-\!1}\!-\!1\Big)
      \Big({\rm Arctan}(\sqrt{f_*\!-\!1})
          \!-\!\sqrt{-2\xi\epsilon_\phi}(n\!-\!n_*)\Big)
      \bigg]
\,.
\label{small argument expansion of phi(n):2}
\end{eqnarray}
This suggests that inflation ends when $f=1$,
at which point the number of e-folds reaches, 
$n\approx n_* +{\rm Arctan}(\sqrt{f_*\!-\!1})/
[\sqrt{-2\xi\epsilon_\phi}(n\!-\!n_*)]$ and $\phi$ reaches 
the value given by the first line in 
Eq.~(\ref{small argument expansion of phi(n):2}).
 The more accurate
statement is that this two field slow-roll inflation ends   
${\cal O}(1)$ e-foldings earlier than that, at which point 
the $\chi$ field enters a fast roll regime and oscillates around 
$\chi=0$ ($f\gtrsim 1$). The two field inflation ends there and 
one enters an approximately one field inflation, during which 
the slow-roll parameters approach,
 $\epsilon\rightarrow \epsilon_\phi=\lambda^2/2$ and all higher 
 ones vanish, $\epsilon_i\rightarrow 0$ ($i\geq 2$), such that 
 inflation never ends.
One way to terminate inflation is to add a small mass term 
for $\phi$, which creates a local minimum in the potential of 
$\phi$, such that the field starts oscillating around that minimum, 
thus ending inflation.

The solutions~(\ref{f of n: small xi}) and~(\ref{phi of n: small xi}) (or their improved versions
that include corrections to some order in the perturbations
$\xi/\epsilon_\phi$ and $\tilde P_\phi^2$) can be used 
in~(\ref{transfer function for entropy perturbation}) and~(\ref{curvature perturbation: solution})
to obtain the transfer function for the entropy perturbation as well as 
the evolution of the adiabatic perturbation on super-Hubble scales due to its coupling to 
the entropy perturbation. Furthermore, one can obtain
the spectral index for the curvature perturbation~(\ref{spectral index of the curvature perturbation}).
Since in general such evaluations involve complicated integrals that cannot be dealt with analytically,
we leave these for future study.

\bigskip

 Let us now consider large $\xi$ limit. In this case Eq.~(\ref{epsilon: simple}) reduces to, 
%(\ref{n as function of f}) and
\begin{equation}
\epsilon \simeq \frac 43 + \frac{\lambda^2}{2}f +{\cal O}(\xi^{-1})
\label{epsilon: large xi}
\end{equation}
where we again assumed the inflaton dominance, $V_\phi\gg V_\chi$.
Note also that, in this limit
 the Ricci curvature scalar of the configuration space~(\ref{configuration space curvature}) is negative and 
 constant,~\footnote{One can easily show that the configuration space metric in the limit of $\xi\rightarrow \infty$
 can be reduced to that of the Poincar\'e plane,
\begin{equation}
dS^2 = \frac{6M_{\rm P}^2}{g^2}\left[d\phi^2+dg^2\right]
\nonumber
\end{equation}
where $g=M_{\rm P}\sqrt{6f}$. The corresponding Ricci scalar is equal to $-1/(3M_{\rm P}^2)$, and thus constant 
and negative. 
 }
\begin{equation}
 {\cal R} \approx -\frac{1}{3M_{\rm P}^2} + {\cal O}(\xi^{-1})
\,.
\label{curvature configuration space large xi}
\end{equation}
This means that negative configuration space curvature prevents inflation from happening
(it makes the inflaton potential too steep). This can be rigorously shown 
by transforming to the frame in which the configuration space metric
is of the form, $(d\tilde\psi)^2 + \sinh^2(\psi/\sqrt{6M_{\rm P}})(d\tilde\chi)^2$
(see Ref.~\cite{Barnaveli:2018dxo}  for details). The leading order behavior in the potential
for large field values is then $\propto \exp(\sqrt{8/3}\tilde\psi/M_{\rm P})$, which corresponds to 
a coupling $\lambda_0 = \sqrt{8/3}$, and the corresponding $\epsilon_0=\lambda_0^2/2=4/3$. 
It is interesting that the same asymptotically late time $\epsilon$ is reached in 
inflationary models in Einstein frame 
driven by a cosmological constant and studied in~\cite{Glavan:2015aqa}. 

To summarize, we have analyzed the two-field inflationary model
with a nonminimally coupled spectator field and found 
that, a large nonminimal coupling induces a large negative curvature of the configuration space manifold, 
such that the principal slow-roll parameter $\epsilon\simeq 4/3$, which is too large to be of interest for inflationary 
model building. On the other hand, the model is viable when the nonminimal coupling $\xi$ is small.
In what follows we analyze this model in a post-inflationary setting. In particular we discuss how 
cosmological perturbations evolve in radiation and matter era.

%%%%%%%%%%%%%%%%%%%%%%%%%%%%%%
%%%%  POSTINFLATIONARY DYNAMICS  %%%%%%%%%%
%%%%%%%%%%%%%%%%%%%%%%%%%%%%%%

\section{Post-inflationary dynamics}
\label{Post-inflationary dynamics}

Here we consider several plausible scenarios for the
evolution of scalar cosmological perturbations after inflation,
with the principal goal to clarify the role of the 
{\it nonminimal coupling}. Not all of the scenarios 
correspond to the traditional curvaton scenario, but instead 
some belong 
to the more general class of two-field inflationary models.
Recall that post-inflationary dynamics 
is very weakly constrained by the current data, leaving us with
numerous theoretical possibilities. For simplicity, here 
we shall keep track of post-inflationary evolution of 
the scalar cosmological perturbations 
on super-Hubble scales only. At 
linear order the graviton and scalar pertubations decouple, 
such that tensor cosmological perturbations obey the usual 
post-inflationary dynamics (their amplitude remains 
frozen on super-Hubble scales), and we shall not discuss 
them any further here. 

In the former section we analyzed slow-roll dynamics of 
scalar cosmological perturbations in a two-field inflationary model. 
We transformed to the more natural basis spanned by 
the comoving curvature perturbation ${\cal R}_c$ 
and the isocurvature (entropy) perturbation ${\cal S}$. Our principal results are given in 
Eqs.~(\ref{spectrum adiabatic perturbation:LO}--\ref{spectra LO}) and~(\ref{spectral index: curvature perturbation}--\ref{spectral index: entropy perturbation}). 
Assuming that the fields (and the corresponding spectra) 
at the end of inflation 
(more precisely, at the end of the slow-roll regime) are known
and given by ${\cal S}( n_e,\vec x)={\cal S}_e(\vec x)$,
${\cal R}_c(n_e,\vec x)={\cal R}_{e}(\vec x)$
(${\cal P}_{{\cal R}e}(k)$, ${\cal P}_{{\cal S}e}(k)$ with
$k\ll aH$),
in what follows we discuss how to evolve them through
 post-inflationary epochs. Before we begin our analysis, 
 we note that, at the end of inflation, one of the two different 
 possibilities can be realized: 
 \begin{enumerate}
 \item[$(a)$] {\bf Weak coupling regime.} In this 
 regime, the curvature and entropy perturbation
 couple weakly during inflation (in the sense that 
 the turning rate $\omega$ is small), 
 such that during inflation there is no significant power transfer between the 
 entropy and curvature perturbations, 
 resulting in the usual one-field 
 estimate for the spectra, 
 $n_{{\cal R}\,e}\approx 1-2\epsilon-\epsilon_2$,
 $n_{{\rm S}\,e}\approx 
 - 2\epsilon +\frac23\eta^2_{ss}+2\eta_\omega^2 -\frac23\epsilon M_{\rm P}{\cal R}$, see 
 Eqs.~(\ref{spectral index: curvature perturbation}--\ref{spectral index: entropy perturbation}). The last term in 
 $n_{\cal S}$ comes from non-canonical structure 
 of the kinetic terms, and can be conveniently expressed 
 by making use of the configuration space curvature ${\cal R}$.
 
 \item[$(b)$] {\bf Strong coupling regime.} The entropy
perturbation couples {\it strongly} to the curvature perturbation,
 such that power 
 transfer between the two spectra is efficient during inflation, 
 resulting in a modified spectrum of the curvature perturbation 
 given in Eq.~(\ref{spectral index: curvature perturbation}).
 \end{enumerate} 
Precisely these two regimes can be identified also 
 during preheating and the subsequent 
 radiation and matter epochs, the only difference being that 
 the power transfer between the entropy and curvature 
 perturbations continues after inflation, and thus ought to be taken into account in the final estimate of 
 the curvature spectrum. The latter scenario -- 
 the one in which 
 a significant power gets transferred from ${\cal S}$ 
 onto ${\cal R}_c$ after inflation -- corresponds to the 
usual {\it curvaton scenario}, in which the late time 
comoving curvature perturbation inherits the spectrum 
of the entropy perturbation. We emphasize that our analysis goes 
beyond that of the standard curvaton scenario,
%{\color{red}
and since in the multi-field scenario considered in this work
the curvature perturbation is not generally conserved on super-Hubble 
scales, the effects of the coupling between the fields must be carefully
studied in order to get a reliable prediction for the spectrum of
scalar cosmological perturbations observed in the late time Universe.
While the general problem is beyond the scope of this work, 
in what follows we make a crude analysis of the postinflationary evolution
in this model.
%} 
 
Since the perturbation spectra 
 are frozen on super-Hubble scales, the 
spectral indices of both entropy and curvature perturbations 
are inherited from the end of inflation. 
The relevant equations of motion in the postinflationary regime 
are therefore (\ref{evolution adiabatic:3b}) 
and~(\ref{evolution entropy:4}),
which after inflation simplify to,
\begin{eqnarray}
\big(D_t +3H\big)\left(\dot{\cal R}_c - 2\omega{\cal S}\right)  %= \frac{\nabla^2}{a^2}{\cal R}_c
 &\approx& 0
\,,
\label{evolution adiabatic:postinflation}
\\
\ddot {\cal S} \!+\!  3H\dot {\cal S}\!+\! 
        \big({\eta}^2_{ss}\!+\! 3\eta_\omega^2\!-\!\epsilon M_{\rm P}^2{\cal R}\big) H^2{\cal S}
       &\approx& 0
\,,
\label{evolution entropic:postinflation}
\end{eqnarray}
with initial conditions given by,
\begin{equation}
{\cal R}_{c}(n_e,\vec x)={\cal R}_{c\,e}(\vec x)
\,,\qquad 
{\cal S}(n_e,\vec x)={\cal S}_{e}(\vec x)
\,.
 \label{initial condition in radiation}
 \end{equation}
When writing Eqs.~(\ref{evolution adiabatic:postinflation})
and~(\ref{evolution entropic:postinflation})  we assumed that $\dot\epsilon = 0$,
which is approximately true during 
preheating (during which $\epsilon\approx 3/2$ or $2$)
and in radiation era, in which $\epsilon=2$,
after preheating is completed. In this work we shall not take into account the time dependence in $\epsilon$, which can be significant during preheating after inflation. To see that, recall
that during preheating,
$\epsilon\approx \dot\sigma^2/(2M_{\rm P}^2 H^2)$,
where the kinetic energy in the adiabatic mode
$\dot\sigma^2/2$ can exhibit a relatively strong dependence on time.  

For the modes of physical interest the decaying mode 
of $\dot {\cal R}_c-\omega{\cal S}$ has decayed by the end of 
inflation, such that~(\ref{evolution adiabatic:postinflation}) 
can be easily solved ({\it cf.} Eq.~(\ref{conserved quantity:2})), 
\begin{equation}
 {\cal R}_c(n,\vec x) = {\cal R}_{c}(n_e,\vec x) 
      + 2\int_{n_e}^n \eta_\omega(n'){\cal S}(n',\vec x)dn' 
\,.
\label{solution for curvature perturbation in radiation}
\end{equation}
Even though Eq.~(\ref{evolution entropic:postinflation})
cannot be solved exactly, one can make progress assuming 
that $\epsilon\approx {\rm const.}$ and that 
the parameters $\eta^2_{ss}$, $\eta_\omega^2$ and 
$\!-\epsilon M_{\rm P}^2{\cal R}$ evolve adiabatically 
in time (in the sense that they do not change much 
during one expansion time). In this case 
Eq.~(\ref{evolution entropic:postinflation}) is approximately
solved by, 
\begin{equation}
{\cal S}(n,\vec x)\approx {\cal S}(n_e,\vec x)
  {\rm e}^{-\int_{n_e}^n \iota(n')dn'}
\,,
\label{S approx solution: radiation}
\end{equation}
where 
\begin{equation}
\iota = \frac{3\!-\!\epsilon}{2}
         -\sqrt{\left(\frac{3\!-\!\epsilon}{2}\right)^2
             -\mu^2_{\cal S}}
\,,\qquad 
 \mu_{\cal S}^2 \equiv \eta^2_{ss}+3\eta_\omega^2
       \!-\!\epsilon M_{\rm P}^2{\cal R}
\,,
\label{growth factor of S}
\end{equation}
where we ignored corrections of the order 
$\partial_n \iota\sim \partial_n \mu_{\cal S}^2$, 
which are of a higher adiabatic order. 
When writing (\ref{growth factor of S}) we have assumed that 
$\partial_n {\cal S}$ is small at the end of inflation 
($n=n_e$), such that its contribution to the postinflationary 
${\cal S}$ can be neglected.

In order to see whether there can be an appreciable
transfer of power between ${\cal S}$ and ${\cal R}_c$, 
we need the conditions for 
${\cal S}$ not to decay (significantly) after inflation.
From~(\ref{growth factor of S}) we see that 
that is the case when $\iota\leq 0$, or equivalently when 
$\mu_{\cal S}^2<0$; in other words when ${\cal R}>0$, 
or $\eta_{ss}<0$, or both.
For example, if $|\mu_{\cal S}^2|\ll 1$, 
then $\iota \approx \mu_{\cal S}^2/(3-\epsilon)$
and Eq.~(\ref{S approx solution: radiation}) reduces to,
${\cal S}\approx {\cal S}_e \exp\big(-\int_{n_e}^n  \frac{\mu_{\cal S}^2(n')}{3-\epsilon}
   dn' \big)$, which is valid both when $\mu_{\cal S}^2$
is positive or negative. When $\mu^2_{\cal S}\ll -1$,
$\iota\simeq -|\mu_{\cal S}|$ and 
${\cal S}\simeq {\cal S}_e \exp\left(\int_{n_e}^n  
|\mu_{\cal S}(n')|dn' \right)$ and ${\cal S}$ can grow 
very fast.
Finally, when $\mu_{\cal S}^2\gg  1$, $\iota$ is complex and 
there are two oscillatory solutions with the frequency 
$\simeq \mu_{\cal S}$ and their enveloping amplitude decays 
as, $\propto{\exp}\left[{-\frac{3-\epsilon}{2}(n-n_e)}\right]$,
such that ${\cal R}_c$ rapidly decouples from ${\cal S}$ .

\bigskip 

{\bf Configuration space curvature.}
The contribution to $\mu_{\cal S}^2$ 
in~(\ref{growth factor of S}), which is simplest 
to estimate, is that from the configuration space curvature,
\begin{equation}
% \big[\mu_{\cal S}^2\big]_1 =
-\epsilon M_{\rm P}^2{\cal R}\simeq 2\epsilon\xi
  \left[1-2(1\!-\!6\xi)(f\!-\!1)\right] +
 {\cal O}\big((f\!-\!1)^2\big) <0
\,,
 \label{curvature contribution}
 \end{equation}
where we assumed that after inflation the $\chi$ condensate 
is small when compared with the Planck scale, 
such that $f-1\ll1$ is a good expansion parameter. 
This contribution is negative, and  in radiation (matter) era
it equals {\it four} ({\it three})
times $(-\xi)$. If this contribution
dominates ${\cal S}$ will grow, thus providing an efficient mechanism for the transfer of 
power between ${\cal S}$ and ${\cal R}_c$, 
underlying the importance of {\it nonminimal coupling} for 
the curvaton mechanism.  
There are two other potentially important 
contributions to $\mu_{\cal S}^2$
in~(\ref{growth factor of S}), and before we reach 
any conclusion regarding effectiveness of 
the curvaton mechanism after inflation, we ought to estimate
them. 

{\bf The turning rate $\omega$} is defined 
in~(\ref{definition of omega})
in terms of the directional derivative of $V$,
 $\omega = - \partial_s V/\dot \sigma$, where 
\begin{equation}
 \dot\sigma^2 = {\cal G}_{AB}\dot\phi^A\dot\phi^B
 \,.
\label{kinetic energy}
\end{equation}
To make our analysis simpler, in what follows we assume that 
after inflation the field condensates are sub-Planckian, 
\begin{equation}
 |\phi|\ll M_{\rm P}
 \,,\qquad  |\chi|\ll M_{\rm P}, \frac{M_{\rm P}}{\sqrt{|\xi|}}
 \,. 
\label{small condensates}
\end{equation}
Our analysis in appendix~B shows that after inflation
the fields oscillate with a decaying envelope such that, 
if~(\ref{small condensates}) is met at the end of inflation, it will
be satisfied throughout the postinflationary epochs. 
The approximate solutions to the equations of motion 
are given in Eqs.~(\ref{Appendix: phi early times after inflation}--\ref{Appendix: chi early times after inflation}) and 
(\ref{Appendix: solution for phi}--\ref{Appendix: solution for chi}). From these, one  can easily obtain the two basis vectors
({\it cf.} Eqs.~(\ref{adiabatic and entropic}) 
and~(\ref{turning rate})),
\begin{equation}
\hat\sigma^A = \frac{1}{\dot\sigma}
\left(\dot\phi\atop\dot\chi\right)
\,,\qquad
\hat\omega^A = \frac{1}{\dot\sigma}
\left(-\dot\chi\atop\dot\phi\right)
\,,
\label{basis vectors after inflation}
\end{equation}
where $\dot\sigma=\sqrt{\dot\phi^2+\dot\chi^2}$.
These then imply the following expression for the turning rate 
\begin{equation}
\omega = \frac{\dot\chi\partial_\phi V-\dot\phi\partial_\chi V}
                     {\dot\sigma^2}
\,.
\label{omega after inflation}
\end{equation}
To get a better idea on how large $\omega$ is, 
we shall calculate it in two limits: (a) when $\phi\gg\chi$ and 
(b) when $\chi\gg\phi$. In the former case, provided there is 
no large hierarchy between the masses of the two fields,  
$\dot\sigma\simeq \dot\phi$, $\dot\chi\simeq 0$,
such that $\hat\omega^A=(0\;\; 1)^T$ and
\begin{equation}
 \omega\approx -\frac{\partial_\chi V}{\dot\phi}
  \simeq -\frac{m_{\chi\,\rm eff}^2\chi}{\dot\phi}
\,,
\label{omega small chi}
\end{equation}
where $m_{\chi\,\rm eff}^2$ is defined 
in~(\ref{Appendix: EOM for chi:2}) and 
\begin{equation}
\dot\phi\simeq 
-\frac{\phi_0{\rm e}^{-\Gamma_\phi t/2}}{a^{3/2}}\left[\Omega_\phi\sin\left(\int_0^t\Omega_\phi(t')dt'\right)
   +\frac{\Gamma_\phi}{2}
           \cos\left(\int_0^t\Omega_\phi(t')dt'\right)
\right]\left[1
+{\cal O}\left(\frac{H}{\Omega_\phi}\right)\right]
\,, \quad
\label{dot phi after inflation}
\end{equation}
where $\Omega_\phi\gg H$ is assumed.   From these expressions we see that $\omega$ oscillates around zero with a typical
amplitude,
\begin{equation}
   |\omega| \sim  \frac{\chi_0}{\phi_0} 
            \times \frac{m_{\chi\rm eff}^2}{m_{\phi\rm eff}}
\,,
\label{typical turning rate}
\end{equation}
such that it is suppressed by $\chi_0/\phi_0\ll 1$. 
Even though it is suppressed, the turning rate may be significant
if $|\omega|$ in~(\ref{typical turning rate}) 
is larger than the expansion rate $H$. 
From~(\ref{dot phi after inflation}) we also see that 
$\omega$ blows up when $\dot\phi=0$, which 
is when $\int_0^t\Omega_\phi(t')dt'+{\rm Arctan}
 \left(\frac{\Gamma_\phi}{2\Omega_\phi}\right)$ is an integer.
 This is however an artifact of the approximations we used to
 get to~(\ref{omega small chi}). Indeed, by using the more 
 accurate formula~(\ref{omega after inflation}) one sees 
 that at these special points $\omega$ grows at most to a value,
$|\omega|\sim 
 (m_{\phi\,\rm eff}^2\phi_0)/(m_{\chi\,\rm eff}\chi_0)$, which is large, but (at least it is) finite. During the short intervals
 when $|\omega|$ is relatively large, 
 the coupling between the two fields
 becomes strong, and may lead to an efficient transfer of 
 power between the entropy and curvature perturbation. 
We intend to explore the physical ramifications of this interesting 
effect in the future. 
For now we just note that the typical turning rate after inflation~(\ref{typical turning rate}) is small, and thus 
the transfer of power between the two fields will typically 
be not inefficient.

Before continuing our analysis, note that 
there is a simple approximation for 
$\dot\sigma=\sqrt{2\epsilon}M_{\rm P}H$,
which acquires a particularly simple form 
when $\epsilon=2$ or $\epsilon=3/2$. 
However, a constant value of $\epsilon$ is attained only 
when the condensates have decayed sufficiently such 
that  $\dot\sigma$ is dominated by the plasma contribution,
in which case, $\dot\sigma^2\simeq {\cal P}+\rho$,
where ${\cal P}$ and $\rho$ are the pressure and energy density.
But as long as the condensates dominate the energy density and 
pressure, the pressure (and therefore also $\epsilon$) oscillates, which in turn causes short periods of strong coupling 
between the curvature and entropy perturbation.

{\bf The mass parameter ${\cal M}_{ss}^2$} for the 
entropy perturbation~(\ref{mass of the entropy mode})
can be approximated by,
\begin{equation}
 {\cal M}_{ss}^2 =\frac{1}{\dot\sigma^2}
 \left(\dot\chi^2\partial_\phi^2
 -2\dot\phi\dot\chi\partial_\phi\partial_\chi
 +\dot\phi^2\partial_\chi^2\right)V(\phi,\chi)
\simeq \frac{m_{\phi\,\rm eff}^2\dot\chi^2
 +m_{\chi\,\rm eff}^2\dot\phi^2}
 {\dot\sigma^2}
 \,,
\label{mass of entropy perturbation after inflation}
\end{equation}
where we used the fact that the condensate amplitudes are 
much smaller than the Planck scale. If in addition 
$|\chi|\ll |\phi|$,
one easily gets, 
\begin{equation}
 {\cal M}_{ss}^2 \approx  m_{\chi\,\rm eff}^2
 \,.
\label{mass of entropy perturbation after inflation:2}
\end{equation}

\bigskip

Upon combining~(\ref{curvature contribution}),
(\ref{omega after inflation}--\ref{dot phi after inflation})
and~(\ref{mass of entropy perturbation after inflation:2})
we get for the mass parameter 
%$\mu_{\cal S}^2H^2$
(\ref{growth factor of S}) of the entropy 
perturbation 
({\it cf.} Eq.~(\ref{evolution entropic:postinflation})),
\begin{equation}
\mu_{\cal S}^2H^2 \simeq  m_{\chi\,\rm eff}^2
                        +3\omega^2(t)  + 2\epsilon\xi H^2
\,,
\label{effective mass of cal S}
\end{equation}
where we kept $\omega^2$ unspecified.
We did that because, even though $\omega^2$ is typically
small~(\ref{typical turning rate}), it can vary a lot with time, 
as can be seen 
from Eqs.~(\ref{omega small chi}--\ref{dot phi after inflation}). 
From Eqs.~(\ref{S approx solution: radiation}--\ref{growth factor of S}) we know that 
the entropy perturbation ${\cal S}$ can grow after inflation
only if $\mu_{\cal S}^2<0$, 
which implies the following condition on $\xi$,
\begin{equation}
\xi<- \frac{m_{\chi\,\rm eff}^2+3\omega^2}{2\epsilon H^2}
\,.
\label{condition on xi negative S mass}
\end{equation}
Since $\omega$ oscillates, at some instances when 
it goes through zero, 
the condition~(\ref{effective mass of cal S})
becomes weaker, 
\begin{equation}
|\xi|> \frac{m_{\chi\,\rm eff}^2}{2\epsilon H^2}
\,,\qquad ({\rm when}\;\; \omega=0)
\,.
\label{condition on xi negative S mass: omega zero}
\end{equation}
% 
%which is a milder and minimum condition on $\xi$. 
When the condition~(\ref{condition on xi negative S mass})
is satisfied, ${\cal S}$ will grow,~\footnote{The entropy 
perturbation ${\cal S}$ will likely grow also when the milder condition~(\ref{condition on xi negative S mass: omega zero})
is satisfied, but to establish that rigorously one would have to 
analyze Eq.~(\ref{evolution entropic:postinflation}) with 
$\omega^2(t)$ 
varying non-adiabatically in time, in which case ${\cal S}$
may exhibit a resonant growth. A complete analysis of this 
case is beyond the scope of this work.
 }
 such that,  when it later decays, 
 it can imprint its power onto
the curvature spectrum with the spectral index given 
by~(\ref{spectral index: entropy perturbation}). 
If that contribution 
dominates its amplitude, the curvaton mechanism will 
be effective, and the observed scalar spectrum will be of 
the form~(\ref{spectral index: entropy perturbation}).
In the simple case when the amplitude 
of $\chi$ is small enough, {\it i.e.} $f\simeq 1$, 
the spectral index of ${\cal S}$ can be approximated by,
\begin{equation}
n_{{\rm S}\,e}\approx 
 - 2\epsilon +\frac23\mu_{\cal S}^2
 \,,
 \label{spectrum of S}
\end{equation}
such that $n_{{\rm S}\,e}<-2\epsilon$ when 
(\ref{condition on xi negative S mass}) is satisfied,
which is what is required by the observations. One can see this by recalling the observational bound on the tensor-to-scalar ratio $r$, which in the case 
when the one-field inflation consistency relation
approximately applies, 
$2\epsilon \simeq r/8<0.01$, from which one
sees that the $-2\epsilon$ contribution in~(\ref{spectrum of S})
is not large enough to explain the spectral index in the 
observed scalar spectrum, whose deviation from scale 
invariance is approximately, 
$n_{s}-1\simeq 0.035$. 
However, in many curvaton models 
the tensor-to-scalar ratio is suppressed due to the 
additional curvaton contribution to the curvature 
perturbation, such that the observed limit $r<0.06$
represents a weaker constraint on $\epsilon$, 
which can also help to make the curvaton
perturbation~(\ref{spectrum of S}) redder,
underlining yet another advantage of curvaton models.
Notice further that adding a large enough 
negative nonminimal coupling which 
satisfies~(\ref{condition on xi negative S mass: omega zero})
can make the spectrum of the entropy perturbation 
red enough to be consistent with the observed value of $n_{s}-1$.
When this is combined with 
condition~(\ref{condition on xi negative S mass}),
 one sees that, the same condition which reddens the spectrum of 
 the entropy perturbation, induces its postinflationary 
 growth. When combined with an efficient transfer
 of power (which can be facilitated either by a strong enough 
 linear coupling $\omega$ or by the decay of 
 entropy perturbations typically mediated by some quantum
 loop process), one can activate the curvaton 
 mechanism. 
 
Since the analysis performed in this section concerns the deep
infrared, super-Hubble modes, they will all be amplified
by an equal factor (with the caveat that each of the two fields 
can get amplified by a different factor). 
That means that the spectra inherited from 
the end of inflation will not change by the postinflationary 
evolution. What can happen is that, as the result of the coupling
between the perturbations, the spectrum of 
the entropy perturbations gets imprinted onto the curvature
perturbation, such that its spectral index becomes that
of the curvature perturbation from the end of inflation.
Therefore, this scenario constitutes the two-field 
variant of the curvaton scenario. 

In the above analysis we did not invoke any 
formulas from the standard curvaton scenario, since 
they assume that the $\Delta N$ formalism applies, 
in which the coupling between the curvaton and inflaton is small
and plays no role such that not significant coupling between the two fields is allowed. However, nonminimal couplings
induce a coupling between the field perturbations, 
which in turn can induce a non-negligible transfer 
of power between the perturbations, 
invalidating the $\Delta N$ formalism.

To conclude, the analysis performed in this section shows that  
a nonminiml coupling of the curvaton field can 
have a  profound impact on the validity 
of curvaton scenarios. In particular, 
it can make viable those curvaton models which 
in their minimal coupling incarnation would be ruled out.

\vskip 2cm

%%%%%%%%%%%%%%%%%%%%%%%%%%%%%%%%%%%%%%%%%%%%%%%%%%%%%
%%%%%%%%%%%%%%%%   C O N C L U S I O N   %%%%%%%%%%%%%%%%%%%%%%%%%%
%%%%%%%%%%%%%%%%%%%%%%%%%%%%%%%%%%%%%%%%%%%%%%%%%%%%%

\section{Conclusion}
\label{Conclusion}

In this paper we investigate in some detail 
a two-field model of inflation in which one of the 
fields is nonminimally coupled. A particular attention is 
devoted to the case when the nonminimally coupled field is the 
{\it curvaton}. Models with nonminimally coupled fields are 
important to study, 
especially when one recalls that nonminimally coupled scalars are 
ubiquitous. This is so because, even if nonminimal couplings are 
set to zero at the classical level, they will be generated by quantum effects, as they
induce a running of the nonminimal coupling 
as long as it is different from its conformal value,
 $\xi=\xi_c=1/6$.

 The principal finding of this work is the observation that
a nonminimal coupling can help to keep alive a class of curvaton models that otherwise would be ruled out by observations. 
Namely, a small negative (positive) nonminimal coupling makes the spectral index of 
scalar cosmological perturbations {\it redder} (bluer), 
thus counteracting a positive  mass term that alone would 
make the curvaton mass term positive.

 In this work we adapt the two field formalism of 
 references~\cite{Gong:2011uw,Kaiser:2012ak}
 (there are earlier references as well), in which 
one studies scalar cosmological perturbations by rotating into the 
frame of curvature and entropy perturbation, 
where the curvature perturbation is the one that directly 
(at the linear level) 
couples to gravitational perturbations. The two fields couple 
by the parameter known as the {\it turning rate} $\omega$
defined in~(\ref{turning rate}) and~(\ref{definition of omega}). 
In section~\ref{Power spectrum} we perform the general 
analysis of our two field model in {\it slow-roll regime}. Our results for the spectral indices for the curvature and entropy
perturbations are given in 
Eqs.~(\ref{spectral index: curvature perturbation}) and~(\ref{spectral index: entropy perturbation}), 
from where one 
sees that the spectral index of the curvature perturbation is influenced by the {\it turning rate} $\omega$ and by the {\it transfer function}
(defined as the ratio of the entropy and curvature perturbation),
such that for positive (negative) $\omega$ the spectral index 
becomes bluer (redder). The spectral index of 
the entropy perturbation is determined not just by its mass term,
but also by the turning rate squared, but also by the {\it curvature 
of the field configuration space} ${\cal R}$
(generated due to the noncanonical
form of the kinetic terms), such that a negative (positive) 
curvature implies a bluer (redder) spectral index of the entropy 
pertubation.

For the analysis particularly useful is 
the slow-roll metric $n=n(\phi,f)$~(\ref{slow roll metric}), 
which tells us how the number of e-foldings $n$ depends on 
the two fields $\phi$ and $f$ (or equivalently $\chi$).
For 
our simple two-field model and in the limit when 
the inflaton potential dominates, the slow-roll metric reduces 
to~(\ref{slow roll metric: small xi}),  which 
admits a conserved quantity~(\ref{conserved phi momentum: small xi}), 
which we named the {\it configuration space momentum}
 $P_{\phi}$ (not to be confused with the canonical momentum of $\phi$). 
Consequently, $n$ becomes a function of only one field
and the trajectory in the field space becomes unique,
signifying an attractor regime of the two-field model.
We then 
analyze the model by expanding in powers of the conserved momentum. This expansion is valid only when $P_\phi$ is 
small enough ($\tilde P=\lambda M_{\rm P}P_\phi\ll 1$), 
and a more general (presumably numerial) analysis 
ought to be performed for larger momenta $P_\phi$.

When the nonminimal coupling is large ($|\xi|\gg 1$) and the field rolls down along the inflaton direction, 
then the principal slow-roll parameter $\epsilon\simeq 4/3$, which is not compatible with 
inflation (in which $\epsilon<1$). 
This result resonates with the late (postinflationary) behavior of inflation 
driven by a cosmological constant studied in Ref.~\cite{Glavan:2015aqa}.

In section~\ref{Post-inflationary dynamics} and in Appendix~B
we present a simple analysis of post-inflationary dynamics of deep
infrared modes of the curvature and entropy perturbation
in the two field model under consideration. In order to make 
analytical progress, we assume that both condensate 
fields are massive and oscillate with sub-Planckian amplitudes.
Then the dynamics of the field condensates 
and the corresponding perturbations simplifies, 
and the only way the 
non-trivial structure of the kinetic terms affects 
evolution of the linear perturbations is through a negative 
contribution to the mass of the entropy 
perturbation~(\ref{curvature contribution}).
The two perturbations couple {\it via} the turning rate $\omega$,
which is typically small and oscillates around zero, but 
has large short spikes. These spikes can induce 
brief periods of strong couplings between the perturbations,
whose implications are 
%proper analysis is 
beyond the scope of this paper. 
Our post-inflationary analysis of the entropy perturbation ${\cal S}$
shows
that its noninimal coupling can play an essential role in 
determining whether ${\cal S}$ grows or decays after inflation.
More specifically, if condition~(\ref{condition on xi negative S mass}) is satisfies, ${\cal S}$ will grow; otherwise it will decay.
Strictly speaking this conclusion is valid if 
$\omega$ varies adiabatically in time. Consequently, a more 
refined analysis is required to establish when ${\cal S}$ 
grows/decays if $\omega^2$ varies nonadiabatically. 
The principal results of section~\ref{Post-inflationary dynamics}
are as follows.
If the entropy perturbation decays after inflation, then 
there will not be any important transfer of power between 
the entropy and curvature perturbation. In this case 
the cuvaton mechanism is inoperative, and the observed scalar 
spectrum will be that from the end of 
inflation. 
If, on the other hand, the entropy perturbation grows
and if it couples significantly to the curvature perturbation 
(either through $\omega$ or through quantum loop decays), 
then the post-inflationary transfer of power between the perturbations will be efficient and the curvaton mechanism operative
such that the observed scalar power spectrum will be dominated by that of the inflationary entropy perturbation. 

The importance of nonminimal coupling of the curvaton field 
is twofold. Firstly, it can determine whether the curvaton mechanism is operative after inflation. Secondly, it can make the spectrum of the 
entropy perturbation redder, thus making 
the spectrum of the entropy perturbation consistent with 
observations. For these reasons, a nonminimal curvaton coupling 
presents a window of opportunity 
for a large class of curvaton models.

 In this work we present a preliminary analysis of nonminimally coupled curvaton models.
Our analysis is encouraging and it invites for a more thorough investigation of this class of models. 
In particular a better understanding is needed of questions such as in precisely what ways 
the inflaton and curvaton mix in the curvature and entropy perturbation. 
The question of what is the effect of the nonminimal coupling on non-Guassianities in the curvaton 
scenario is particularly intriguing, since many curvaton scenarios produce too large non-Gaussianties to 
be compatible with observations.
Furthermore, it would be worthwhile investigating how 
the form of the inflaton and curvaton potentials, 
as well as their interactions with matter fields, 
influence the results presented in this work.

\section*{Acknowledgments}

The authors acknowledge Tomo Takahashi and Anja 
Marunovi\'{c}  for their contributions to the early stages of 
this project and Tomo Takahashi for a critical reading of the 
manuscript and for many useful comments.
Furthermore, we would like to thank Marieke Postma 
and Diederick Roest for 
a critical reading of the manuscript and for suggestions on 
how to improve it. This work is part of the research program 
of the Foundation for Fundamental Research on Matter (FOM), which is part of the Netherlands Organization for Scientific Research (NWO). This work is in part supported by the D-ITP consortium, a program of the Netherlands Organization for Scientific Research (NWO) that is funded by the Dutch Ministry of Education, Culture and Science (OCW).  Lei-Hua Liu was funded by the Chinese Scholarschip Council (CSC) and by the Funding of Hunan Provincial Department of Education, no 19B464.

\section*{Appendix A: The amplitude of the curvature and entropy perturbation}

In this appendix we evaluate the amplitude of the curvature 
and entropy perturbation at near super-Hubble scales at the leading 
and subleading order in slow-roll parameters. These amplitudes are important 
as they determine the subsequent inflationary and postinflationary 
evolution of cosmological perturbations. Our analysis in the main text
is based on the asumption that one can match to the leading order evolution
equations in slow-roll 
parameters~(\ref{evolution adiabatic: LOa}--\ref{evolution entropy: LOa}), 
whose solutions are given in  Eqs.~(\ref{solutions mode functions: LO}).
Here we go one step further, and solve 
Eqs.~(\ref{evolution adiabatic:3}--\ref{evolution entropy:3}) 
to the next-to-leading order
in slow-roll parameters. Let us begin our analysis by rewriting
Eqs.~(\ref{evolution adiabatic:3}--\ref{evolution entropy:3})
in conformal time, $ad\tau = dt$,
\begin{eqnarray}
\hat{\cal R}_c'' \!+\! (2+\epsilon_2){\cal H}\hat{\cal R}'_c \!-\frac{\nabla^2}{a^2}\hat{\cal R}_c
     - 2{\cal H}\left[\partial_\tau
     \!+\!\Big(3\!-\epsilon\!+\!\!\epsilon_2\Big){\cal H}
\right]\left(\eta_\omega \hat {\cal S}\right) &=& 0
\qquad
\label{Appendix: evolution adiabatic}
\\
\hat {\cal S}'' \!+\!  (2\!+\!\epsilon_2){\cal H}\hat{\cal S}'
 \!+\! \left[-\frac{\nabla^2}{a^2}
         \!+\!\big(\eta^2_{ss}\!-\! \eta_\omega^2\!+\!\Delta_\epsilon\!-\!\epsilon M_{\rm P}^2{\cal R}\big) {\cal H}^2 \right]\hat {\cal S}
     + 2\eta_\omega{\cal H}\hat {\cal R}'_c &=& 0
\,,
\label{Appendix: evolution entropy}
\end{eqnarray}
Because the field perturbations mix, it is convenient to introduce 
vector perturbations, 
$\hat{\cal R}(\tau,\vec x)\equiv \left(\hat {\cal R}_c,\hat{\cal S}\right)^T$
and 
$\hat\Pi_{\cal R}(\tau,\vec x)
 \equiv \left(\hat \Pi_{{\cal R}_c},\Pi_{\cal S}\right)^T$,
where 
\begin{equation}
\hat\Pi_{\cal R}
= \left(\begin{array}{c}
  2\epsilon M_{\rm P}^2a^2(\partial_\tau\hat{\cal R}_c-a\omega\hat{\cal R}_c) \cr
 2\epsilon M_{\rm P}^2a^2(\partial_\tau\hat{\cal S}+a\omega\hat{\cal R}_c)\cr
 \end{array}
 \right)
\,,
\label{appendix: canonical momenta}
\end{equation}
which obey the following canonical commutation relations,
\begin{equation}
\left[\hat{\cal R}(\tau,\vec x),\hat\Pi_{\cal R}(\tau,\vec x^{\,\prime})\right]
   =i\hbar{1\!\!1}
  \delta^3(\vec x\!-\!\vec x\,')
\,,
\label{Appendix: canonical commutation}
\end{equation}
and the other two commutators vanish. This commutator should be 
understood as a direct product of two vectors, so the result is a tensor,
{\it i.e.} a matrix.
The symbol ${1\!\!1}$ in (\ref{Appendix: canonical commutation})
denotes the $2\times 2$ dimensional unity matrix. 
The following step is to Fourier decompose the fields,
\begin{eqnarray}
\hat{\cal R}^i(\tau,\vec x) 
\!\!&=&\!\! \int \frac{d^3k}{(2\pi)^3}e^{i\vec k\cdot\vec x}
\left[{\cal R}^{ij}(\tau,k)\hat a_j\big(\vec k\,\big)
     \!+\!{{\cal R}^{*}}^{ij}(\tau,k)
   \hat a_j^+\big(\!-\!\vec k\,\big) \right]
\quad (i,j=1,2)
\nonumber\\
\hat\Pi_{\cal R}^i(\tau,\vec x) 
\!\!&=&\!\! \int \frac{d^3k}{(2\pi)^3}e^{i\vec k\cdot\vec x}
\left[\Pi_{\cal R}^{ij}(\tau,k)\hat a_j\big(\vec k\,\big)
       \!+\!{\Pi_{\cal R}^{*}}^{ij}(\tau,k)
   \hat a_j^+\big(\!-\!\vec k\,\big) \right]
\,,
\label{Appendix: Fourier decomp}
\end{eqnarray}
where $\hat a_j\big(\vec k\,\big)$ and $\hat a_j^+\big(\vec k\,\big)$
are the annihilation and creation operators of the instantaneous 
diagonal basis ($\hat a_j$ annihilate the vacuum $|\Omega\rangle$,
$\hat a_j\big(\vec k\big)|\Omega\rangle=0$), with
\begin{equation}
\left[\hat a_i\big(\vec k\,\big),\hat a_j^+\big(\vec k^{\,\prime}\,\big)\right]
  = (2\pi)^3\delta_{ij}\delta^3\Big(\vec k\!-\!\vec k^{\,\prime}\Big)
\,,
\label{Appendix: creation annihilation commutation}
\end{equation}
and the other commutators vanish. 
It is not hard to show that consistency of 
(\ref{Appendix: canonical commutation}) 
and~(\ref{Appendix: creation annihilation commutation}) implies 
the following normalization condition, which in a matrix form, 
\begin{equation}
{\cal R}(\tau,k)\cdot\Pi_{\cal R}^+(\tau,k)
-\Pi_{\cal R}(\tau,k)\cdot{\cal R}^+(\tau,k)=i{1\!\!1}
\,.
\label{Appendix: normalization}
\end{equation}
The diagonal basis is different from the curvature-entropy basis
due to the coupling between the two fields introduced by 
the turning rate $\omega$. 
It is still convenient to work in the $\{{\cal R}_c,{\cal S}\}$ basis, 
because it is 
singled out by gravity as the interaction basis of gravity (not unlike 
the flavour basis of neutrinos), and therefore it is the basis in which 
we make observations.

Perturbing the fields 
 in powers of slow-roll parameters, 
\begin{equation}
\hat{\cal R} =\hat{\cal R}^{(0)}+\hat{\cal R}^{(1)}+\cdots
\,,\qquad
\hat\Pi_{\cal R} =\hat\Pi_{\cal R}^{(0)}+\hat\Pi_{\cal R}^{(1)}+\cdots
\label{Appendix: perturbing R and Pi}
\end{equation}
the zeroth order equations are solved in the main text and 
from~(\ref{Appendix: evolution adiabatic}--\ref{Appendix: evolution entropy})
 we get the first order equations for the mode matrices,
\begin{eqnarray}
\left(\!\begin{array}{cc}
           \partial_\tau^2-\frac{2}{\tau}\partial_\tau+k^2 & 0\cr
        0 & \partial_\tau^2-\frac{2}{\tau}\partial_\tau+k^2 \cr        
\end{array}
\!\right)\!{\cal R}^{(1)} \!\!&\simeq&\!\! 
\left(\!\begin{array}{cc}
       \frac{2\epsilon+\epsilon_2}{\tau}\partial_\tau & 
      -\frac{2\eta_\omega}{\tau}\left(\partial_\tau-\frac{3}{\tau}\right)\cr
        \frac{2\eta_\omega}{\tau}\partial_\tau & 
\frac{2\epsilon+\epsilon_2}{\tau}\partial_\tau 
   \!-\!\frac{\eta^2_{ss}\!-\! \eta_\omega^2\!+\!\Delta_\epsilon
                \!-\!\epsilon M_{\rm P}^2{\cal R}}{\tau^2}\cr        
\end{array}
\!\right)\!{\cal R}^{(0)}
\equiv {\cal P}
,
\qquad\;
\label{Appendix: perturbative equation}
\end{eqnarray}
where
\begin{equation}
{\cal R}^{(0)}(\tau,k) ={\cal R}_0(\tau,k)\;{1\!\!1}
\,,\qquad 
{\cal R}_0(\tau,k) = \frac{H}{2\sqrt{\epsilon k^3}M_{\rm P}}
    \left(1+ik\tau\right)e^{-ik\tau}  
\;%{1\!\!1}
\label{Appendix: leading order solution}
\end{equation}
where we expanded ${\cal H}=-1/[(1-\epsilon)\tau]$  to first order in 
$\epsilon$, and for 
completeness we kept the mass term of the entropy perturbation.
Let us now check that the normalization condition 
(Wronskian)~(\ref{Appendix: normalization})
is satisfied by the leading order solution~(\ref{Appendix: leading order solution}).
\begin{equation}
2\epsilon M_{\rm P}^2a^2\!\left[\!{\cal R}_0{1\!\!1}\! \cdot \! 
\left(\! \begin{array}{cc}\!
       \partial_\tau {\cal R}_0^* \!-\! a\omega {\cal R}_0^* &  0 \cr
       0 &\partial_\tau {\cal R}_0^* \!+\! a\omega {\cal R}_0^*\! \cr 
       \end{array}
       \right)
\!-\!  \left(\begin{array}{cc}\!
      \partial_\tau {\cal R}_0 \!-\! a\omega {\cal R}_0 &  0 \cr
       0 &  \partial_\tau {\cal R}_0 \!+\! a\omega {\cal R}_0\! \cr 
       \end{array}\! 
       \right)\! \cdot\!  {1\!\!1}{\cal R}^*_0\right]\!=i{1\!\!1}
,\quad
\label{Appendix: normalization 2}
\end{equation}
where we made use of~(\ref{appendix: canonical momenta}).
To see that this equality is satisfied, note first that the terms $\propto\omega$ 
cancel out (as both terms contribute as $\pm a\omega|{\cal R}_0|^2$ and therefore 
are real, such that they cancel out when summed) 
and the normalization of ${\cal R}_0$ 
was chosen such to yield the correct Wronskian normalization at the leading 
order in slow roll.

We shall solve Eq.~(\ref{Appendix: perturbative equation}) 
by the method of Green's functions. 
The retarded Green's function of the operator,
\begin{equation}
a^2\left(\!\begin{array}{cc}
           \partial_\tau^2-\frac{2}{\tau}\partial_\tau+k^2 & 0\cr
        0 & \partial_\tau^2-\frac{2}{\tau}\partial_\tau+k^2 \cr        
\end{array}
\!\right)G_{\rm ret}(\tau,\tau';k) = \delta(\tau-\tau')
\label{Appendix: Greens function equation}
\end{equation}
can be written as, 
\begin{equation}
G_{\rm ret}(\tau,\tau';k) = i\theta(\tau-\tau')\left[
   u(\tau,k) u^*(\tau',k)-u^*(\tau,k) u(\tau',k)\right]{1\!\!1}
\,,
\label{Appendix: Greens function solution}
\end{equation}
where $u(\tau,k)$ and $u^*(\tau,k)$ are the two fundamental solutions of 
the problem, 
\begin{equation}
\left( \partial_\tau^2-\frac{2}{\tau}\partial_\tau+k^2\right)u(\tau,k)=0
\,, 
\label{Appendix: fundamental de Sitter eom}
\end{equation}
{\it i.e.}
\begin{equation}
u(\tau,k) = \frac{H}{\sqrt{2k^3}}\left(1+ik\tau\right)e^{-ik\tau}
\,,\qquad
u^*(\tau,k) = \frac{H}{\sqrt{2k^3}}\left(1-ik\tau\right)e^{ik\tau}
\,,
\label{Appendix: fundamental de Sitter solutions}
\end{equation}
which satisfy the Wronskian,
\begin{equation}
 W[u,u^*] = u(\tau,k)\partial_\tau u^*(\tau,k)
        -[\partial_\tau u(\tau,k)] u^*(\tau,k) = \frac{i}{a^2}
\,.
\label{Appendix: Wronskian fund solution}
\end{equation}
With the Green's function in our hand, we can write the formal solution of 
Eq.~(\ref{Appendix: perturbative equation}) as,
\begin{equation}
{\cal R}^{(1)}(\tau,k) = 
\int_{\tau_0}^0d\tau' G_{\rm ret}(\tau,\tau';k)\cdot a^2(\tau'){\cal P}(\tau',k)
\,.
\label{Appendix: formal perturbative solution} 
\end{equation}
To make progress, note that the source ${\cal P}$ can be conveniently
broken into two parts,
\begin{equation}
{\cal P} = {\cal P}_2 + {\cal P}_4
\label{Appendix: breaking the source}
\end{equation}
where 

\begin{eqnarray}
{\cal P}_2 \!\!&=&\!\! 
\left(\!\begin{array}{cc}
       2\epsilon+\epsilon_2&  -2\eta_\omega\cr
        2\eta_\omega & 2\epsilon+\epsilon_2\cr        
\end{array}
\!\right)\!\frac{1}{\tau}\partial_\tau\!{\cal R}^{(0)}
= \frac{H}{2\sqrt{\epsilon k^3}M_{\rm P}}k^2e^{-ik\tau}
\tilde{\cal P}_2
,\quad 
\tilde{\cal P}_2 = \left(\!\begin{array}{cc}
       2\epsilon+\epsilon_2&  -2\eta_\omega\cr
        2\eta_\omega & 2\epsilon+\epsilon_2\cr        
\end{array}
\!\right)
,
\qquad
\label{Appendix: perturbative equation 2a}
\\
{\cal P}_4 \!\!&=&\!\!
  \frac{H}{2\sqrt{\epsilon k^3}M_{\rm P}}
   \frac{1}{\tau^2}\left(1+ik\tau\right)e^{-ik\tau} \tilde{\cal P}_4
\,,\qquad
\tilde{\cal P}_4 = 
\left(\!\begin{array}{cc}
      0 \;&\; 6\eta_\omega \cr
      0 \;&\;  \!-\!\left(\eta^2_{ss}\!-\! \eta_\omega^2\!+\!\Delta_\epsilon
                \!-\!\epsilon M_{\rm P}^2{\cal R}\right)\cr        
\end{array}
\!\right)
\,,
\label{Appendix: perturbative equation 2b}
\end{eqnarray}
where we took account of the fact that when $\partial_\tau$ acts on
$H$ and $\epsilon$ one generates higher order (quadratic) corrections
in slow-roll parameters, which are suppressed when compared with the terms 
we keep. 

Upon a careful look at the structure of the integrals 
in~(\ref{Appendix: formal perturbative solution}), one 
finds that the solution can be constructed in terms of 
two classes of integrals,
\begin{eqnarray}
I_n(a) \!\!&=&\!\! \int_{\tau_0=-1/H}^\tau \frac{d\tau'}{(\tau')^n}
\,,\qquad (n=0,1,2,3,4)
\label{Appendix: integrals In}\\
J_n(a) \!\!&=&\!\! \int_{\tau_0=-1/H}^\tau \frac{d\tau'}{(\tau')^n}
                  e^{-2ik\tau'}
\,,\qquad (n=0,1,2,3,4)
\,.
\label{Appendix: integrals Jn}
\end{eqnarray}
Inserting~(\ref{Appendix: perturbative equation 2a}--\ref{Appendix: perturbative equation 2b}) into~(\ref{Appendix: formal perturbative solution})
gives, 
\begin{eqnarray}
\left[{\cal R}^{(1)}\right]_2 
\!\!&=&\!\!
\frac{H}{2\sqrt{2\epsilon k^3}M_{\rm P}}\frac{i}{2k} \tilde{\cal P}_2
\left\{(1\!+\!ik\tau)e^{-ik\tau}\!\big[I_2\!-\!ikI_1\big]
    \!-\!(1\!-\!ik\tau)e^{ik\tau}\!\big[J_2\!+\!ikJ_1\big]
\right\}
,\qquad\;
\label{Appendix: formal perturbative solution 2}\\
\left[{\cal R}^{(1)}\right]_4 
\!\!&=&\!\!
\frac{H}{2\sqrt{2\epsilon k^3}M_{\rm P}}\frac{i}{2k^3} \tilde{\cal P}_4
\left\{(1\!+\!ik\tau)e^{-ik\tau}\!\left[I_4\!+\!k^2I_2\right]
    \!-\!(1\!-\!ik\tau)e^{ik\tau}\!\left[J_4\!+\!2ikJ_3\!-\!k^2J_2\right]
\right\}
.\qquad\;
\label{Appendix: formal perturbative solution 4}
\end{eqnarray}
The integrals~(\ref{Appendix: integrals In}--\ref{Appendix: integrals Jn})
are either elementary, or can be expressed in terms of 
the exponential integral (or equivalently integral since and cosine functions),
\begin{eqnarray}
 I_0(a) \!\!&=&\!\! -\frac{1}{aH}\!+\!\frac{1}{H}
\,,
\label{Appendix: integral I0}\\
 I_1(a) \!\!&=&\!\! -\ln(a)
\,,
\label{Appendix: integral I1}\\
 I_2(a) \!\!&=&\!\! aH\!-\!H
\,,
\label{Appendix: integral I2}\\
 I_3(a) \!\!&=&\!\! -\frac{(aH)^2}{2}\!+\!\frac{H^2}{2}
\,,
\label{Appendix: integral I3}\\
 I_4(a) \!\!&=&\!\! \frac{(aH)^3}{3}\!-\!\frac{H^3}{3}
\,,
\label{Appendix: integral I4}
\end{eqnarray}
and 
\begin{eqnarray}
 J_0(a) \!\!&=&\!\! \frac{i}{2k}e^{-2ik\tau}\!-\!\frac{i}{2k}e^{2ik/H}
\,,
\label{Appendix: integral J0}\\
 J_1(a) \!\!&=&\!\! -{\rm E}_1\left(\frac{2ik}{aH}\right)
     \!+\!{\rm E}_1\left(\frac{2ik}{H}\right)
= \ln\left(\frac{2k}{aH}\right)\!+\!\gamma_E+\frac{i\pi}{2}
  \!+\!{\rm E}_1\left(\frac{2ik}{H}\right) +{\cal O}\left(\frac{k}{aH}\right)
\,,\qquad
\label{Appendix: integral J1}\\
 J_2(a) \!\!&=&\!\! aHe^{-2ik\tau}\!-\!He^{2ik/H}
      -2ik J_1(a)
\,,
\label{Appendix: integral J2}\\
 J_3(a) \!\!&=&\!\! -\frac{(aH)^2}{2}e^{-2ik\tau}
                 \!+\!\frac{H^2}{2}e^{2ik/H} -ik J_2(a)
\,,
\label{Appendix: integral J3}\\
 J_4(a) \!\!&=&\!\! \frac{(aH)^3}{3}e^{-2ik\tau}
                 \!-\!\frac{H^3}{3}e^{2ik/H} -\frac23ik J_3(a)
\,,
\label{Appendix: integral J4}
\end{eqnarray}
where $E_1(iz)=\int_z^\infty dt e^{it}/t 
           = -\left[{\rm ci}(z)+i{\rm si}(z)\right]$ ($-\pi<{\rm Arg}(iz)<\pi $),
is the exponential integral and  ${\rm ci}(z)$ and ${\rm si}(z)$
are defined as, 
\begin{eqnarray}
{\rm ci}(z)  \!\!&=&\!\! -\int_z^\infty dt \frac{\cos(t)}{t}
                        = \int_0^z dt \frac{\cos(t)-1}{t} +\ln(z) +\gamma_E
\nonumber\\
{\rm si}(z)  \!\!&=&\!\! -\int_z^\infty dt \frac{\sin(t)}{t}
                        = \int_0^z dt \frac{\sin(t)}{t} -\frac{\pi}{2}
\,,
\label{Appendix: ci and si functions}
\end{eqnarray}
where the latter equalities can be used for getting the 
small argument expansion of $E_1(iz)$, $\rm ci(z)$ and $\rm si(z)$.
Upon inserting~(\ref{Appendix: integral I0}--\ref {Appendix: integral J4})
into Eqs.~(\ref{Appendix: formal perturbative solution 2}--\ref{Appendix: formal perturbative solution 4}) and
evaluating in the near super-Hubble limit $k/(aH)\ll 1$,
one obtains, 
\begin{eqnarray}
\left[{\cal R}^{(1)}\right]_2 
\!\!&=&\!\!
-\frac{H}{4\sqrt{\epsilon k^3}M_{\rm P}}\tilde{\cal P}_2
\left\{\ln\Big(\frac{2k}{H}\Big)\!+\!\gamma_E\!+\!\frac{i\pi}{2}
\!+\!{\rm E}_1\Big(\frac{2ik}{H}\Big)\!-\!2 
 \!-\!\frac{iH}{k}\Big(e^{2ik/H}\!-\!1\Big)
  \!+\! {\cal O}\Big(\frac{k}{aH}\Big)
\right\}
,\qquad\;
\label{Appendix: formal perturbative solution 2c}\\
\left[{\cal R}^{(1)}\right]_4 
\!\!&=&\!\!
-\frac{H}{6\sqrt{\epsilon k^3}M_{\rm P}}\tilde{\cal P}_4
\bigg\{\ln\Big(\frac{2k}{aH}\Big)\!+\!\gamma_E\!+\!\frac{i\pi}{2}
\!+\!{\rm E}_1\Big(\frac{2ik}{H}\Big)\!-\!\frac52
 \!-\!\frac{iH^3}{2k^3}e^{2ik/H}\Big(1\!-\!\frac{2ik}{H}
       \!+\!\frac{k^2}{H^2}\Big)
\nonumber\\
&&\hskip 2.7cm +\,\frac{iH^3}{2k^3}\Big(1+3\frac{k^2}{H^2}\Big)
  \!+\! {\cal O}\Big(\frac{k}{aH}\Big)
\bigg\}
\,,\qquad 
\label{Appendix: formal perturbative solution 4c}
\end{eqnarray}
such that the perturbed solution is of the form,
\begin{equation}
 {\cal R} = \frac{H}{2\sqrt{\epsilon k^3}M_{\rm P}}
 \Big[{1\!\!1}\!+\!\frac13\ln(a)\tilde{\cal P}_4
 \!-\!\frac13C_4(k)\tilde{\cal P}_4\!-\!\frac12C_2(k)\tilde{\cal P}_2\Big]
 \,,
\label{Appendix: perturbative solution structure}
\end{equation}
where $C_2(k)$ and $C_4(k)$ are complex $k-$dependent constants
(whose precise form can be read off from 
the curly brackets in
Eqs.~(\ref{Appendix: formal perturbative solution 2c}--\ref{Appendix: formal perturbative solution 4c})), which 
in general depend on the choice of the initial state at $\tau=\tau_0$,
but do not depend on time.

Several comments are in order concerning the 
result~(\ref{Appendix: perturbative solution structure}). 
Firstly, the answer in ~(\ref{Appendix: formal perturbative solution 2c}--\ref{Appendix: formal perturbative solution 4c}) was obtained 
by assuming 
a near super-Hubble limit, which means that, while $k/(aH)\ll 1$,
$\ln(aH/k)$ is still of the order of unity, such that when multiplied with 
a slow-roll parameter it is still much smaller than unity, {\it i.e.}
$\epsilon_i \ln(aH/k)\ll 1$, and the expansion in slow-roll
parameters we have used still applies. Secondly, 
the initial state deep in inflation (when $\tau_0=-1/H$)
has been chosen such that, at $\tau_0=-1/H$, no mixing between 
the curvature and entropy modes was present.
This can be easily seen by noticing that ${\cal R}^{(1)}$
vanishes at $\tau=\tau_0$, which immediately follows
from the observation that all 
integrals~(\ref{Appendix: integral I0}--\ref{Appendix: integral J4})
vanish at the initial time. Of course, this assumption
can be relaxed, and that will modify the solutions,
but the fundamental structure of the correction 
in~(\ref{Appendix: formal perturbative solution 2c}--\ref{Appendix: formal perturbative solution 4c}) will not change.
Thirdly, a non-trivial check of the validity 
of our approach can be made by checking that 
the next-to-leading corrections~(\ref{Appendix: formal perturbative solution 2c}--\ref{Appendix: formal perturbative solution 4c}) do not modify the normalization 
relation~(\ref{Appendix: normalization}), {\it i.e.} 
 the canonical commutation relation~(\ref{Appendix: canonical commutation})
 is still satisfied.
This is indeed so because the $\log(a)$ contributes with a real coefficient ,
such that it does not modify the Wronskian. The constant part is 
complex, but it is constant and thus does not contribute to the 
canonical momentum (at the next-to-leading order in slow-roll parameters).

From Eqs.~(\ref{Appendix: formal perturbative solution 4c}--\ref{Appendix: formal perturbative solution 4c}) we see that 
the structure of the next-to-leading solution is,
\begin{equation}
 {\cal R} ={\cal R}^{(0)}
\!+\!{\cal R}^{(1)}
\,,\qquad 
{\cal R}^{(1)}=\big[{\cal R}^{(1)}\big]_2\!+\!\big[{\cal R}^{(1)}\big]_4
%C_1(k)\tilde{\cal P}_4 \ln(a)
%\!+\! C_2(k)\tilde{\cal P}_4\!+\!C_3(k)\tilde{\cal P}_2
   \,,\qquad 
% \Pi_{\cal R} =\Pi_{\cal R}^{(0)}
 %  \!+\!2\epsilon M_{\rm P}^2a^3H C_1(k)\tilde{\cal P}_4
%\,,
\label{Appendix: structure of solution}
\end{equation}
%
%where $C_1$ is a real constant, and $C_2$ and $C_3$ are complex
%$k$-dependent (but time independent) constants.
 The scalar spectra are then,
 \begin{eqnarray}
 P_{\cal R} =\frac{k^3}{2\pi^2} {\cal R}\cdot {\cal R}^+
\label{Appendix: scalar spectrum}
\end{eqnarray}
which is matrix valued, whose individual entries (11), (12), (21) and (22) 
correspond to the spectra of the correlators, 
$\langle\hat {\cal R}_c(\tau,\vec x)\hat {\cal R}_c(\tau,\vec x\,')\rangle$,
$\langle\hat {\cal R}_c(\tau,\vec x)\hat {\cal S}(\tau,\vec x\,')\rangle$,
$\langle\hat {\cal S}(\tau,\vec x)\hat {\cal R}_c(\tau,\vec x\,')\rangle$, and
$\langle\hat {\cal S}(\tau,\vec x)\hat {\cal S}(\tau,\vec x\,')\rangle$,
respectively. Thus we have,
 %${\cal R}\cdot {\cal R}^+$ a $2\times 2$ matrix
%There is a logarithmically growing solution.
%
\begin{eqnarray}
{\cal P}_{{\cal R}_c} \!\!&\simeq&\!\!\frac{H^2}{8\pi^2\epsilon M_{\rm P}^2}
\left[1\!-\!(2\epsilon\!+\!\epsilon_2)\Re(C_2)
\!+\!4\eta_\omega^2\Big|\ln(a)\!-\!C_4\!+\!\frac12C_2\Big|^2\right]
\label{Appendix: spectrum RR}\\
{\cal P}_{{\cal R}_c{\cal S}} \!\!&\simeq&\!\!\frac{H^2}{8\pi^2\epsilon M_{\rm P}^2}
\left[2\eta_\omega\big(\ln(a)-C_4+i\Im(C_2)\big)\right]
={\cal P}^*_{{\cal S}{\cal R}_c}
\label{Appendix: spectrum RS}\\
{\cal P}_{{\cal S}} \!\!&\simeq&\!\!\frac{H^2}{8\pi^2\epsilon M_{\rm P}^2}
\left[1\!-\!(2\epsilon\!+\!\epsilon_2)\Re(C_2)
\!-\!\frac23\big(\eta^2_{ss}\!-\! \eta_\omega^2\!+\!\Delta_\epsilon
                \!-\!\epsilon M_{\rm P}^2{\cal R}\big)\Big(\!\ln(a)\!-\!\Re(C_4)\Big)\right]
\,,\qquad
\label{Appendix: spectrum SS}
\end{eqnarray}
where all quantities are evaluated at a near super-Hubble scale, 
$k_*=\mu (aH)_*$, with $\mu\ll 1$ and $|\ln(\mu)|={\cal O}(1)$.
Eqs.~(\ref{Appendix: spectrum RR}--\ref{Appendix: spectrum SS})
are the main results of this Appendix, and can be summarized as follows.
\begin{enumerate}
\item The corrections to the curvature spectrum~(\ref{Appendix: spectrum RR})
and the entropy spectrum~(\ref{Appendix: spectrum SS})
are suppressed by slow-roll parameters, as long as 
the fiducial scale $k_*$ is chosen such that $|\ln(k/aH)_*|= {\cal O}(1)$.
\item The cross-correlated spectrum between the curvature
and entropy perturbation~(\ref{Appendix: spectrum RS})
is dynamically created even if it was not there initially in inflation.
Its amplitude is however slow-roll suppressed when compared with
that of the curvature and entropy perturbations.
\item There are also higher order corrections, 
but -- without an explicit calculation -- their precise form 
remains unclear.
Nevertheless, one may be tempted 
to resum the first order corrections. For example,
resuming the $[{\cal R}^{(1)}]_4$ correction~(\ref{Appendix: formal perturbative solution 4c}), 
one obtains a contribution to the entropy spectrum of the 
type $(k/aH)^{\delta_4^{(1)} n_{\cal S}}$, with 
$\delta_4 n_{\cal S} = \frac23\left(\eta_{ss}^2-\eta_\omega^2
+\Delta_\epsilon-\epsilon M_{\rm P}^2{\cal R}\right)$.
The other part may come from the logrithmic correction
of $[{\cal R}^{(1)}]_2$ 
in~(\ref{Appendix: formal perturbative solution 2c}), 
and amounts to $\delta_2^{(1)} n_{\cal S}=-2\epsilon-\epsilon_2$.
Summing the two contributions yields, 
 $\delta^{(1)} n_{\cal S}=-2\epsilon+\frac23\left(\eta_{ss}^2-\eta_\omega^2
-\epsilon M_{\rm P}^2{\cal R}\right)$,
 where we took account of, $\Delta_\epsilon\simeq 3\epsilon_2/2$.
This correctly captures a part of the spectral index
in Eq.~(\ref{spectral index: entropy perturbation}),
which was obtained by solving the equations to the leading order 
in spatial gradients, but it does not capture the contribution
from the mixing between the curvature and entropy modes.
While this is encouraging, 
as far as we know this resummation
-- also known as the dynamical renormalization group -- 
is not guaranteed to give correct answers and therefore ought 
to be used with a caution.

\end{enumerate}
With these remarks in mind, the principal result of this Appendix 
is that -- as long as one is interested in the leading order results 
in slow-roll parameters -- it is legitimate to work with the leading 
order spectra~(\ref{spectra LO}) used in the main text. If one were interested 
in subleading results, one can quite easily employ the results of this appendix 
to incorporate the subleading 
corrections~(\ref{Appendix: spectrum RR}--\ref{Appendix: spectrum SS})
 into the spectra both during inflation
as well as after inflation.

\section*{Appendix B: Postinflationary evolution of field condensates}
\label{Appendix B: Postinflationary evolution of field condensates}

 In this appendix we model the evolution of the condensates
 $\phi$ and $\chi$ in a simple model in which inflation
 is terminated by an inflaton mass term and for simplicity 
 we consider the case where after inflation,
 \begin{equation}
 f\! -\! 1\ll 1\qquad 
 (\chi^ 2\ll M_{\rm P}^2/|\xi|)
 \,. 
 \label{Appendix: condition on f}
 \end{equation}
 The potential in the transformed frame is then 
 ({\it cf.} Eq.~(\ref{potential in Einstein frame})),
\begin{equation}
V(\phi,\chi)=\frac{1}{f^2}\left[\frac 12 m_\phi^2\phi^2
  + V_0{\rm e}^{-\lambda\phi/M_{\rm P}}
  +\frac12 m_\chi^2 \chi^2
 \right]
\,,
\label{Appendix: potential in Einstein frame}
\end{equation}
where we set $\lambda_\chi=0$. This potential exhibits a minimum
given by $(\phi_0,\chi_0=0)$, where $\phi_0$ is the solution of
the equation,
$m_\phi^2\phi_0=\frac{\lambda V_0}{M_{\rm P}}
{\rm e}^{-\lambda\phi/M_{\rm P}}$, and  it can be expressed 
in terms of the Lambert $W$ function, 
\begin{equation}
\phi_0=\frac{M_{\rm P}}{\lambda}\,
 W\!\left(\frac{\lambda^2V_0}{m_\phi^2M_{\rm P}^2}\right)
 \,,
 \label{Appendix: Lambert W function}
\end{equation}
which simplifies in the two limits, 
when its argument is large and small, 
\begin{eqnarray}
\phi_0&=&\frac{M_{\rm P}}{\lambda}
\left\{
 \ln\left(\frac{\lambda^2V_0}{m_\phi^2M_{\rm P}^2}\right)
 -\ln\left[\ln\left(\frac{\lambda^2V_0}{m_\phi^2M_{\rm P}^2}\right)\right]
 +{\cal O}\left(\frac{\ln(\ln(x_0))}{\ln(x_0)}\right)
 \right\}
 \,,\quad (x_0\gg 1)
 \nonumber\\
 \phi_0&=&\frac{M_{\rm P}}{\lambda}
\left\{
 \left(\frac{\lambda^2V_0}{m_\phi^2M_{\rm P}^2}\right)
 -\left(\frac{\lambda^2V_0}{m_\phi^2M_{\rm P}^2}\right)^{\!\!2}
 +{\cal O}\left(x_0^3\right)
 \right\}
 \,,\quad (x_0\ll 1)
 \,,
 \label{Appendix: Lambert W function 2}
\end{eqnarray}
where $x_0=(\lambda^2V_0)/(m_\phi^2M_{\rm P}^2)$.

Now taking account of 
Eqs.~(\ref{configuration space metric})
and~(\ref{Christoffel symbols}), 
the field condensate equations~(\ref{background field equations}),
 reduce to, 
\begin{eqnarray}
\ddot \phi+\frac{2\xi\chi}{fM_{\rm P}^2}\dot\phi\dot\chi
   + 3H\dot\phi +\frac{1}{f}
\left[m_\phi^2\phi-\frac{\lambda V_0}{M_{\rm P}}{\rm e}^{-\lambda\phi/M_{\rm P}}
  \right] &=& 0
\label{Appendix: EOM for phi}
\\
\ddot \chi
  -\frac{\xi\chi}{M_{\rm P}^2-(1\!-\!6\xi)\chi^2}\dot\phi^2
  +\xi\chi\left(\frac{2}{M_{\rm P}^2-\xi\chi^2}
       -\frac{1\!-\!6\xi}{M_{\rm P}^2-(1\!-\!6\xi)\xi\chi^2}
      \right)\dot\chi^2
&&
      \nonumber\\
   + 3H\dot\chi +\frac{-2\xi\chi}{M_{\rm P}^2-(1\!-\!6\xi)\xi\chi^2}
\left[\frac{m_\chi^2M_{\rm P}^2}{-2\xi}
      -\frac{1}{f}\left(m_\phi^2\phi^2
    +2V_0{\rm e}^{-\lambda\phi/M_{\rm P}}
      +m_\chi^2\chi^2\right)
  \right] &=& 0
\,,
\label{Appendix: EOM for chi}
\end{eqnarray}
where we moved back to the variable $\chi^2=(f\!-\!1)/(-\xi)$,
which is now more convenient. These equations are hard to solve 
and moreover do not properly reflect the physics of preheating
since they do not capture the field fluctuations dynamics,
which can be created after inflation at a fast rate, thereby 
inducing a significant backreaction which is not captured by~(\ref{Appendix: EOM for phi}--\ref{Appendix: EOM for chi}).

Inflation ends in this  model when 
$\phi$ is sufficiently close to $\phi_0$, which will be 
typically the case when $|\phi-\phi_0|\ll M_{\rm P}$,
after which $\phi$ starts   oscillating and decaying 
around $\phi_0$. Analogously, $\chi$ will oscillate around 
its minimum $\chi=0$ with an amplitude $|\chi|\ll M_{\rm P}$
and $\chi^2\ll M_{\rm P}^2/|\xi|$, such that all complicated terms 
in the denominators of~(\ref{Appendix: EOM for phi}--\ref{Appendix: EOM for chi}) can be dropped.
For notational convenience, we shift the inflaton field $\phi$ by 
$-\phi_0$, 
\begin{equation}
\phi \rightarrow \phi-\phi_0
\,,
\label{Appendix: shifting the inflaton field}
\end{equation}
such that after inflation both $\phi$ and $\chi$ oscillate around 
their minimum at zero. Since the amplitude of fluctuations 
is small when compared with the Planck scale, one can 
linearize~(\ref{Appendix: EOM for phi}--\ref{Appendix: EOM for chi}). This is obviously true for all the terms, except for those 
which potentially contain resonant decay. Since the amplitude 
of these terms is small, only narrow parametric resonance is possible, and this resonance will occur only if the 
ratio of the amplitude and frequency of the oscillations
is to a good approximation an integer squared (see {\it e.g.}  Ref.~\cite{Shtanov:1994ce}). For simplicity, we leave 
a more complete analysis of preheating for future, and here
assume that parametric resonance is absent and 
study the linearized version of Eqs.~(\ref{Appendix: EOM for phi}--\ref{Appendix: EOM for chi}),~\footnote{Somewhat more general
equations from~(\ref{Appendix: EOM for phi:2}--\ref{Appendix: EOM for chi:2}) 
can be obtained if one works with energy densities, $\rho_\phi$ and $\rho_\chi$, 
\begin{eqnarray}
\dot\rho_\phi + 2\epsilon_\phi H\rho_\phi 
 &=& -\Gamma_\phi (\rho_\phi - \rho_{\phi,0})
        -\Gamma_{\phi\chi}(\rho_\phi-\rho_\chi)
\label{conservation equation for phi}
\\
\dot\rho_\chi + 2\epsilon_\chi H\rho_\chi 
 &=& -\Gamma_\chi (\rho_\chi - \rho_{\chi,0})
        +\Gamma_{\phi\chi}(\rho_\phi-\rho_\chi)
\label{conservation equation for chi}
\,,
\end{eqnarray}
where  $\rho_{\phi,0}$ and $\rho_{\chi,0}$ denote the equilibrium
densities for $\phi$ and $\chi$, respectively.
 Note that Hubble damping for the two fields 
 can be different if the equation of state parameters
 differ,  
 $w_\phi={\cal P}_\phi/\rho_\phi\neq w_\chi={\cal P}_\phi/\rho_\phi$. In this case the damping rates are determined by, 
 $2\epsilon_\phi H=3(1+w_\phi)H$ and  
 $2\epsilon_\chi H=3(1+w_\chi)H$.
Equations~(\ref{conservation equation for phi}--\ref{conservation equation for chi}) are more general 
from~(\ref{Appendix: EOM for phi:2}--\ref{Appendix: EOM for chi:2}) as they  
also include the transfer of energy between 
the two fields and indicate that -- due to the backreaction 
from the created field fluctuations -- the decays stop when 
the fields get thermalized.
We do not attempt to solve these equations here,
but instead work with the simpler 
ones~(\ref{Appendix: EOM for phi:2}--\ref{Appendix: EOM for chi:2}).
} 
\begin{eqnarray}
\ddot \phi + (3H+\Gamma_\phi)\dot\phi 
  + m_{\phi,\rm eff}^2\phi &\simeq& 0
  \,,\qquad m_{\phi,\rm eff}^2 
   \equiv \left(1+\frac{\lambda\phi_0}{M_{\rm P}}\right)m_\phi^2
\label{Appendix: EOM for phi:2}
\\
\ddot \chi
   +  (3H+\Gamma_\chi)\dot\chi +m_{\chi,\rm eff}^2\chi 
 &=& 0
  \,,\qquad 
  m_{\chi,\rm eff}^2 
   \equiv m_\chi^2 +\frac{-2\xi m_\phi^2\phi_0}
             {\lambda M_{\rm P}}
                \left(1+\frac{\lambda\phi_0}{M_{\rm P}}\right)
\,,
\label{Appendix: EOM for chi:2}
\end{eqnarray}
where $\phi_0$ is defined in~(\ref{Appendix: Lambert W function})
and $\Gamma_\phi$ and $\Gamma_\chi$ denote the decay
rates of $\phi$ and $\chi$, respectively. {To keep our discussion as general as possible,
we shall not discuss here in great length the precise microscopic origin of the decay rates.~\footnote{Tree-level perturbative 
decay rates can be found {\it e.g.} in Ref.~\cite{Abbott:1982hn,Shtanov:1994ce,Linde:2005ht}.
Consider, for example, the following interaction Lagrangian,
\begin{equation}
{\cal L}_{\rm int}=-\frac{g}{2}\phi^2\chi^2-(y_\phi\phi+y_\chi\chi)\bar\psi\psi
-\frac{h_\phi}{2}\phi^2\theta^2-\frac{h_\chi}{2}\chi^2\theta^2
\,,
\label{Appendix interactions} 
\end{equation}
where $\psi$ and $\theta$ are some light fermionic and scalar fields,
{\it i.e.} their masses satisfy, $m_\psi\ll m_\chi,m_\phi$ and $m_\theta \ll m_\chi,m_\phi$.
The corresponding tree level decay rates 
are then given by, 
$\Gamma(\phi\rightarrow\chi\chi)\simeq [g^2\phi_0^2/(8\pi m_\phi)]\theta(m_\phi-2m_\chi)$, 
$\Gamma(\chi\rightarrow\phi\phi)\simeq [g^2\chi_0^2/(8\pi m_\chi)]\theta(m_\chi-2m_\phi)$, 
$\Gamma(\chi\rightarrow\bar\psi\psi)\simeq y_\chi^2m_\chi/(8\pi)$, 
$\Gamma(\phi\rightarrow\bar\psi\psi)\simeq y_\phi^2m_\phi/(8\pi)$,
$\Gamma(\chi\rightarrow\theta\theta)\simeq h_\chi^2\chi_0^2/(8\pi m_\chi)$, 
$\Gamma(\phi\rightarrow\theta\theta)\simeq h_\phi^2\phi_0^2/(8\pi m_\phi)$,
respectively. To make a connection with the model studied here, note that one can read off the coupling $g$ from 
Eqs.~(\ref{Appendix: EOM for phi}--\ref{Appendix: EOM for chi}), 
$g\simeq 2\xi m_{\phi,\rm eff}^2/M_{\rm P}^2$.
Inserting this into the above rate results in, 
$\Gamma(\phi\rightarrow\chi\chi)\simeq (\xi H_{\rm e}/M_{\rm P})^2m_{\phi,\rm eff}$,
where $H_{\rm e}$ is the Hubble rate at the end of inflation, 
$H_{\rm e}^2\simeq m_{\phi,\rm eff}^2\phi_0^2/(6M_{\rm P}^2)
<10^{-10}~M_{\rm P}^2$,
and for definiteness we have assumed $m_{\phi ,\rm eff}>2m_{\chi ,\rm eff}$.
This rate is at most,  $\Gamma(\phi\rightarrow\chi\chi)
  < 10^{-10}\xi^2 m_{\phi,\rm eff}$, 
  and since $\xi^2$ 
cannot be much larger than unity, this perturbative decay channel is very slow and will be 
most likely dominated by resonant decay channels, which are the subject of 
a forthcoming publication.
More generally, the perturbative decay rate is given by 
$\Gamma=\Im[\Sigma(k^0=\omega,\vec k^2=0)]/(2\omega)$, 
where $\Sigma$ denotes the self-energy and $\omega$ is the frequency of the zero mode,
see {\it e.g.} Refs.~\cite{Abbott:1982hn} and~\cite{Linde:2005ht}.
}
Since the main goal of our analysis of the postinflationary dynamics is to study
how the non-minimal coupling of the inflaton affects postinflationary dynamics,
keeping the decay rates phenomenological as  
in~(\ref{Appendix: EOM for phi:2}--\ref{Appendix: EOM for chi:2}) suffices for that purpose.
The decay rates
may be of a perturbative origin or due to a parametric resonance. 
If tree level perturbative decay rates do not vanish, they will typically dominate the condensate decays, and their time dependence 
will be mild and therefore adiabatic.
If quantum loops 
or parametric resonance dominates the decay rates,
they will be generally time dependent and 
may or may not be changing adiabatically in time.
We postpone a more detailed analysis of the microscopic origin of the 
decay rates in the postinflationary epochs, and how that can affect
postinflationary dynamics for a later publication.

Since in the approximation used to 
get~(\ref{Appendix: EOM for phi:2}--\ref{Appendix: EOM for chi:2}) 
the field  condensates $\phi$ and $\chi$ decouple, 
Eqs.~(\ref{Appendix: EOM for phi:2}--\ref{Appendix: EOM for chi:2}) are relatively easily solved under the assumptions that 
$\epsilon$ is approximately constant and equal to either 
$3/2$ (matter era) or $2$ (radiation era).
The field masses and decay rates can be either larger on smaller 
than the expansion rate. Since the expansion rate is dropping 
as $H\propto 1/t$, they may be larger only for a 
(brief) period after inflation. In that case we have 
for rescaled fields, 
$\tilde \phi = {\rm e}^{\Gamma_\phi t/2}\phi$, 
$\tilde \chi = {\rm e}^{\Gamma_\chi t/2}\chi$, 
\begin{eqnarray}
\tilde \phi &=& \phi_0 + \frac{\dot\phi_0}{H_0}
\left(1-a^{-(3-\epsilon)}\right)
\,,\qquad 
  \left| m_{\phi,\rm eff}^2-\frac{\Gamma_\phi^2}{4}\right|
  \ll H^2
\label{Appendix: phi early times after inflation}
\\
\tilde \chi &=& \chi_0 + \frac{\dot\chi_0}{H_0}
\left(1-a^{-(3-\epsilon)}\right)
\,,\qquad 
  \left| m_{\chi,\rm eff}^2-\frac{\Gamma_\chi^2}{4}\right|
  \ll H^2
\,,
\label{Appendix: chi early times after inflation}
\end{eqnarray}
where $H_0=H(t_0)$, $\phi(t_0)=\phi_0$, 
$\dot\phi_0=(d\phi/dt)(t=t_0)$ and $a_0=a(t_0)=1$.
These solutions are approximately correct only for a relatively 
short period of time, for which $m_{\phi,\rm eff}t\ll 1$
and $m_{\chi,\rm eff}t\ll 1$. When (one of) these conditions
are violated, the fields will start oscillating. 

After the expansion rate has sufficiently dropped, the conditions
in~(\ref{Appendix: phi early times after inflation}) and 
(\ref{Appendix: chi early times after inflation}) will get violated
and the field condensates will begin oscillating 
around their minima. To study 
that case, the equations of motion for 
the rescaled fields, 
$\tilde\phi = a^{3/2}{\rm e}^{\Gamma_\phi t/2}\phi$
and  
$\tilde\chi = a^{3/2}{\rm e}^{\Gamma_\chi t/2}\chi$
become those of harmonic oscillators with frequencies that 
adiabatically vary in time (they do not change much 
in one expansion time), implying that the approximate 
solutions can be written in the form, 
\begin{eqnarray}
 \phi(t) &=& \phi_0\frac{{\rm e}^{-\Gamma_\phi t/2}}{a^{3/2}}
 \cos\left[\int_{0}^t \Omega_\phi(t')dt'+\varphi_\phi\right]
\,,\quad \Omega_\phi^2(t) = m_{\phi,\rm eff}^2 
   -\frac14(\Gamma_\phi+3H)^2-\frac32\dot H
\label{Appendix: solution for phi}
\\
 \chi(t) &=& \chi_0\frac{{\rm e}^{-\Gamma_\chi t/2}}{a^{3/2}}
 \cos\left[\int_{0}^t \Omega_\chi(t')dt'+\varphi_\chi\right]
\,,\quad \Omega_\chi^2(t) = m_{\chi,\rm eff}^2 
   -\frac14(\Gamma_\chi+3H)^2-\frac32\dot H
\,,\qquad 
\label{Appendix: solution for chi}
\end{eqnarray}
where $\phi_0=\phi(0)$, $\chi_0=\chi(0)$ and 
$\varphi_\phi$ and $\varphi_\chi$ are unimportant phases
that can be set to zero and for notational simplicity we choose the time at the end of inflation,
$t_e=0$.
If $\Gamma_\phi$ and $\Gamma_\chi$ vary (adiabatically) with time, then one should exact replacements, 
$\Gamma_\phi t\rightarrow \int_0^t \Gamma_\phi(t')dt'$
and 
$\Gamma_\chi t\rightarrow \int_0^t \Gamma_\chi(t')dt'$
in~(\ref{Appendix: solution for phi}--\ref{Appendix: solution for chi}). On the other hand, if $m_{\phi,\rm eff},m_{\chi,\rm eff}\gg H$, then the integrals in the oscillatory functions simplify to,
$\int_{0}^t \Omega_\phi(t')dt'\approx \Omega_\phi t$
and 
$\int_{0}^t \Omega_\chi(t')dt'\approx \Omega_\chi t$,
where in the last relations, $\Omega_\phi\approx 
\sqrt{m_{\phi,\rm eff}^2-\Gamma_\phi^2/4}$
and $\Omega_\chi\approx 
\sqrt{m_{\chi,\rm eff}^2-\Gamma_\chi^2/4}$.

The approximate solutions~(\ref{Appendix: solution for phi}--\ref{Appendix: solution for chi})
 are used in the main text to analyze the postinflationary 
 evolution of the scalar cosmological perturbations.

%\bibliography{mybibfile}

\end{document}